# A Study of an N Molecule-Quantized Radiation Field-Hamiltonian

BY

MICHAEL THOMAS TAVIS

DOCTOR OF PHILOSOPHY, GRADUATE PROGRAM IN PHYSICS

UNIVERSITY OF CALIFORNIA, RIVERSIDE, DECEMBER 1968

PROFESSOR FREDERICK W. CUMMINGS, CHAIRMAN

## ABSTRACT OF THE DISSERTATION

In this dissertation a simple Hamiltonian for a system of interacting molecules and radiation field is developed from a model of N Two-Level Molecules interacting, via a dipole approximation, with a single mode, quantized radiation field. The exact eigenvalues and eigenvectors of this Hamiltonian are compared with analytical results obtained from several different approximation schemes applied to the simple model. The consequence of this comparison is the development of validity criteria for the utilization of these approximation schemes to more difficult problems. For example, an analysis done here indicates that using an independent molecule approximation in an explanation of the "build up" of coherent radiation in a gas laser (maser) may be inappropriate.

# Table of Contents

# Table of Figures

# List of Tables

# CHAPTER I: INTRODUCTION

An insight into the collective interactions of a system of atoms or molecules with an electromagnetic field is important for a proper understanding of such devices as a laser (maser), since these interactions lead to interesting phenomena such as the generation of coherent radiation. Many authors[1,2,3,4] have attempted to explain the complicated laser phenomena by using various approximations in dealing with the molecule plus field interactions. Although these attempts have been fairly successful, the validity of these approximations has not been fully established.

In this dissertation, we do not attempt to solve a problem as complicated as that of a laser system. Instead, we have treated the simpler problem of N Two-Level Molecules (TLMs) interacting with but a single mode, quantized radiation field. Pumping and radiation loss from the cavity, which characterize the interaction of the laser system with the "outside," have been ignored; the system considered here is assumed to be completely self-contained. Since we have used an idealized model for the molecule plus field interactions, it has been possible to find an exact solution. Several approximations are also used to solve the same problem, and in this manner we have determined the validity of the approximations, at least for this system. These validity criteria are likely to be applicable to similar but more realistic systems. For example, it has been claimed[5] that the approximation of treating each molecule as independent of one another when interacting with a radiation field can explain the buildup of coherent radiation in a laser. However, the analysis presented here indicates that this approximation may not be valid.

It is hoped that this development will lead to a more complete understanding of the cooperative effects existing in the molecular plus field interactions, such as the generation of coherent light. Indeed it is found that under certain conditions the exact ground states of this simple problem show a strong resemblance to the states of the coherent radiation field[6] in that both states exhibit a Poisson photon number distribution and near-classical dispersion in photon number, i.e. ,the mean square deviation in photon number is equal to the average number of photons.

Finally, the states developed here may be used as a basis set for which states of more realistic models may be expanded, although this is not attempted in this dissertation.

Similar problems have been treated before. Nearly fifteen years ago, Dicke[7] considered a similar problem. He was one of the first authors to realize the importance of treating the assembly of radiating molecules as a single quantum system on the grounds that the molecules, which are interacting with a common radiation field, should not be treated as independent. Even though his results were obtained through the use of first order perturbation theory (Fermi's Golden Rule),which is valid for times short compared to the transition time, he was able to demonstrate cooperative phenomena between the molecules. He showed that the spontaneous emission rate of a group of radiating molecules is proportional to $N^2$ where N is the number of molecules. As is well known, an independent molecule model leads to a transition rate always proportional to N. Although not of immediate interest to this dissertation, except in a peripheral way, a number of interesting results stated by Dicke are derived in Appendix D. As stated before, the model problem is solved exactly, without the use of perturbation theory. More recently K. W. H. Stevens and B. Josephson, Jr.[(8)], have considered a similar though slightly more complex problem. Their results, though exact, were obtained for only a small number of molecules.

A summary of the main body of the dissertation follows: In Chapter II the complex Hamiltonian of a molecule plus field interaction is reduced to a simple Hamiltonian. The steps used are given. First the classical Hamiltonian for a charged particle interacting with an electromagnetic field is written. The change from classical to quantum physics is made and the single particle Hamiltonian is generalized to one for N neutral molecules, each molecule consisting of M particles. Approximations are made to simplify this Hamiltonian. The main simplifications are in the use of the dipole approximation in the interaction terms of the Hamiltonian, in considering that each molecule has only two energy levels, that the molecules interact with only one mode of the radiation field, and that each molecule interacts with exactly the same electric field. The reason for this development is to point out these many approximations necessary to reach the final form of the Hamiltonian (2.37), and in the use of operators and states which are formally identical to those developed for treating a system of spin-1/2 particles[7].

In Chapter III the Hamiltonian (2.37) is further simplified by ignoring terms which describe the effect of a simultaneous absorption of a photon and de-excitation of one TLM or, conversely, the effect of a photon emission and TLM excitation. This new Hamiltonian (3.1) still contains all of the degeneracy of the problem which has made the perturbation treatment of the problem so difficult[9]. However, the problem is amenable to exact treatment by again applying the spin-1/2 formalism, namely that of the algebra of the group SU(2). Since the Hamiltonian for N spin-1/2 particles interacting with a single mode radiation field factors into the irreducible representations of the group SU(2), the original problem of diagonalizing a matrix of dimension $2^N$ reduces to diagonalization of matrices of order 2r+1, where $r \leq N/2$ is called the "cooperation number" and is the analogue of the total angular momentum, S, for a system of spins. The diagonalization of these smaller matrices is reduced to a finite difference equation with specified boundary conditions. A discussion of most of the assumptions used in obtaining eq. (3.1) is also given in Chapter III.

In Chapter IV this difference equation is solved exactly in closed form for the eigenvalues and vectors for the correlated system (molecule plus field). Unfortunately, the expressions developed there are very difficult to understand except graphically and they present a tedious numerical calculation without the use of a high-speed computer. Exact solutions also afford a standard for approximation. In Chapter V and Appendix B several approximations are made to the Hamiltonian (3.1). These approximations were made in order to understand the exact solutions of the problem more fully and to have a basis for analyzing the approximations used by other authors in the treatment of laser and related problems.

In Chapter VI a discussion of the numerical results, including comparisons between the exact solutions and the approximations of Section V, is given. Since the Hamiltonian (3.1) has been solved exactly for the case where the radiation field frequency, $\omega$, is not necessarily equal to the energy level spacing, $\Omega$, of the TLM, the effect of having $\omega$ far from $\Omega$ is discussed.

Finally, the main results of this dissertation are summarized in Chapter VII. The validity of various approximations is reviewed and their relevance to more complicated problems is discussed.

The formulae for the ensemble averages of the positive frequency part of the electric field, E(t), and the photon number, $EE^{+}(t)$, are developed in Appendix F. This development follows the analysis used by Cummings[(10)] for the single Two-Level Molecule case. Without many simplifying approximations, the formulae are very tedious to evaluate, even with the facilities of a high-speed computer.

# CHAPTER II: DEVELOPMENT OF THE HAMILTONIAN

## A. General Development

In this chapter the Hamiltonian for N Two-Level Molecules (TLMs) interacting with a quantized electromagnetic field via a dipole approximation is developed. The classical form of the non-relativistic Hamiltonian for a single particle of mass, m, and charge, e, moving in an electromagnetic field is given by

$$H - \frac{1}{2m}\left(\vec{p} - \frac{e}{c}\vec{A}\right)^2 + e\varphi, \tag{2.1}$$

where p is the momentum conjugate to the position of the particle, $\vec{A}$ is the magnetic vector potential, and $\varphi$ is the scalar field. The Hamiltonian is derived in the usual way[(11)], i.e. , from a classical Lagrangian which includes a generalized velocity-dependent potential obtained from the Lorenz force on the particle due to the electromagnetic fields (E, B). E and B, the electric and magnetic fields, obey Maxwell's equations

$$\vec{\nabla}\, x\, \vec{E} + \frac{1}{c}\frac{\partial \vec{B}}{\partial t} = 0, \tag{2.2a}$$

$$\vec{\nabla} \cdot \vec{B} = 4\,\pi\,\rho, \tag{2.2b}$$

$$\vec{\nabla}\, x\vec{B} - \frac{1}{c}\frac{\partial \vec{D}}{\partial t} = \frac{(4\pi\bar{\mathrm{j}})}{c}, \tag{2.2c}$$

$$\vec{\nabla} \cdot \vec{B} = 0. \tag{2.2d}$$

Equation (2.2d) implies that

$$\vec{B} = \vec{\nabla}\, x\, \vec{A}\,, \tag{2.3a}$$

and Eq. (2.2a) that

$$\vec{E} = -\vec{\nabla}\varphi - \frac{1}{c}\frac{\partial \vec{A}}{\partial t}. \tag{2.3b}$$

The conversion from classical physics to quantum physics is accomplished by the replacement of the conjugate momentum, $\vec{p}$, by $\frac{\hbar}{i}\vec{\nabla}$. For convenience $\vec{p}$ will be retained wherever possible. An additional potential energy, V, is added to the Hamiltonian in Eq. (2.1). This energy may be regarded as the potential energy that binds the particles (of electrostatic origin if the particle is an electron)[(12)]

Finally, the Hamiltonian for the single particle plus field is written in the form

$$H = H_m + H_F + H_I, \tag{2.4}$$

where

$$H_m = \frac{1}{2m}\vec{p}^2 + V, \tag{2.5a}$$

$$H_I = \frac{-e}{mc}\vec{A}\cdot\vec{p} + \frac{i\,e\,\hbar}{2\,mc}(\vec{\nabla}\cdot\vec{A}) + \frac{e^2}{2mc^2}\left|\vec{A}\right|^2 + e\varphi, \tag{2.5b}$$

and the final term $H_f$ is the energy content of the electromagnetic field

$$H_f = \frac{1}{2\pi}\int(\vec{E}^2 + \vec{H}^2)d^3x. \tag{2.5c}$$

The Hamiltonian is now generalized to N electrically neutral molecules each containing M particles. The interaction energy of the j[th] molecule with the electromagnetic field can then be written[(13)]

$$H_{I_j} = \sum_{k=1}^{M} \left\{ \frac{-e_k}{m_k c} \vec{A}(\overrightarrow{x_k}) \cdot \overrightarrow{p_k} + \frac{i e_k \hbar}{2 m_k c} \left[ \vec{\nabla} \cdot \vec{A}(\overrightarrow{x_k}) \right] + \frac{e_k^2}{2 m_k c} \left| \vec{A}(\overrightarrow{x_k}) \right|^2 + e_k \varphi(\overrightarrow{x_k}) \right\} \tag{2.6}$$

By performing a gauge transformation

$$\vec{A}(\overrightarrow{x_k}) \rightarrow \vec{A}(\overrightarrow{x_k}) + \vec{\nabla}_k \Lambda \tag{2.7a}$$

$$\varphi(\overrightarrow{x_k}) \rightarrow \varphi(\overrightarrow{x_k}) - \frac{1}{c} \frac{\partial \Lambda}{\partial t}, \tag{2.7b}$$

The interaction Hamiltonian for one molecule becomes

$$\begin{aligned} H_{I_j} = \sum_{k=1}^{M} \Bigg\{ & \frac{-e_k}{m_k c} \left( \vec{A}(\overrightarrow{x_k}) + \vec{\nabla}_k \Lambda \right) \cdot \vec{p}_k + \frac{i e_k \hbar}{2 m_k c} \left[ \vec{\nabla} \cdot \left( \vec{A}(\overrightarrow{x_k}) + \vec{\nabla}_k \Lambda \right) \right] \\ & + \frac{e_k^2}{2 m_k c} \left[ \vec{A}(\overrightarrow{x_k}) + \vec{\nabla}_k \Lambda \right]^2 + e_k \varphi(\overrightarrow{x_k}) - \frac{1}{c} \frac{\partial \Lambda}{\partial t} \Bigg\} \end{aligned} \tag{2.8}$$

In order to simplify this expression, A is chosen such that

$$\vec{\nabla}_k \Lambda = -\vec{A}(\overrightarrow{x_J}) \tag{2.9}$$

The dipole approximation is used at this point. This means that over the dimensions of the molecule the vector potential may be assumed constant. Therefore

$$\vec{A}(\overrightarrow{x_k}, t) = \vec{A}(\overrightarrow{x_J}, t), \tag{2.10}$$

where $\overrightarrow{x_J}$ is the center of mass of the molecule, and $\Lambda$ is found to be

$$\Lambda = -\vec{A}(\vec{x_j}, t) \cdot \vec{x_k}. \tag{2.11}$$

When the Coulomb gauge is applied (no sources for the electromagnetic field $\varphi(\vec{x_k})$ is set equal to zero; therefore,

$$\vec{E} = -\frac{1}{c}\frac{\partial \vec{A}}{\partial t}, \tag{2.12}$$

And the interaction Hamiltonian is given by

$$H_{I_j} = \vec{E}(\vec{x_j}) \cdot \sum_{k=1}^{m} e_k \vec{x_k} \tag{2.13}$$

The summation over k is easily recognized as being the dipole moment of the molecule and

$$H_{I_j} = -\vec{E}(\vec{x_j}) \cdot \vec{\mu_j}. \tag{2.14}$$

The general Hamiltonian for the N molecules follows from the above

$$H_I = -\sum_j \vec{E}(\vec{x_j}) \cdot \vec{\mu_j}. \tag{2.15}$$

## B. Molecular Quantization

As a further simplification of the problem, it will be assumed that the N molecules are confined to a cavity, that the internal coordinates of the individual molecules are unaffected by collisions, that but two internal states are involved for each molecule, and that the wave functions of the separate molecules do not overlap. The energy level spacing for each molecule is taken as $\hbar\Omega$ with the zero of energy half way between the energy levels. With these assumptions, the Hamiltonian for the molecules is given by

$$H_m = H_o + \hbar\Omega \sum_{j=1}^{N} R_{j3}\,, \tag{2.16}$$

where $H_o$ acts on the center of mass coordinates and represents the translational and the non electromagnetic field interaction energies of the molecules. The internal energy of the j[th] molecule, $\hbar\Omega R_{j3}$, has the eigenvalues $\pm\frac{1}{2}\hbar\Omega.$ Since $H_o$ and all the $R_{j3}$'s commute with each other, the energy eigenstates of the molecules could be chosen to be simultaneous eigenstates of $H_o$, $R_{13}$, $R_{23}$,⋯, $R_{n3}$, and is given by a product of a state representing the center of mass of the molecules and a state representing the internal energies of the molecules

$$= |CM > \{+ + - + \cdots > . \tag{2.17}$$

If the number of + (molecule in upper energy state) and - molecule in lower energy state) is denoted by $n_+$ and $n_-$, respectively, then m is defined as

$$m = \frac{1}{2}(n_+ - n_-), \tag{2.18a}$$

and

$$N = n_+ + n_-. \tag{2.18b}$$

The total energy of the molecular system, given by

$$E_{CMm} = E_g + \hbar\Omega m\,. \tag{2.19}$$

has a degeneracy of order

$$\frac{N!}{n_+!\,n_-!}, \tag{2.20}$$

due to the various permutations of + and - molecules. Since the molecules have non-overlapping wave functions, questions of symmetry for the center of mass state are not relevant. It will also be assumed that C.M. motion of the molecules is insignificant so that questions concerning Doppler effects may be ignored. With these assumptions, that part of the Hamiltonian and states of the molecules concerning C.M. motion and interactions will be discarded leaving the internal states of the molecules as the significant part of the problem.

It is advantageous to employ the analogy between the $\pm$ molecular system and a system of spin-1/2 particles and define the operators $R_{j1}$and $R_{j2}$along with $R_{j3}$as

$$\begin{aligned} R_{j1}\left|\cdots \pm_j \cdots\right> &= \frac{1}{2}\left|\cdots \mp_j \cdots\right>, \\ R_{j2}\left|\cdots \pm_j \cdots\right> &= \pm\frac{1}{2}i\left|\cdots \mp_j \cdots\right>, \\ R_{j3}\left|\cdots \pm_j \cdots\right> &= \pm\frac{1}{2}\left|\cdots \pm_j \cdots\right>. \end{aligned} \tag{2.21}$$

The operators

$$R_{j\pm} = R_{j1} + iR_{j2} \tag{2.22}$$

are raising the lowering operators of the internal states of the molecule. Also, it is convenient to define operators

$$R_k = \sum_{j=1}^{N} R_{jk}, k = 1,2,3 \tag{2.23a}$$

and the operators

$$R_{\pm} = R_1 \pm R_2 \tag{2.23b}$$

$$R^2 = R_1^2 + R_2^2 + R_3^2. \tag{2.23c}$$

The interaction Hamiltonian for all the molecules, Eq. (2.15), may be expressed in terms of the operators just developed. Since the dipole operator is odd, a general form for $\vec{\mu}_j$ is expressed by

$$\vec{\mu_j}_{op} = \langle\vec{\mu_j}\rangle\left(R_{j+} + R_{j-}\right) \tag{2.24}$$

and

$$H_I = -\sum_{j=1}^{N} \langle\vec{\mu_j}\rangle\vec{E}\left(\vec{x_j}\right)\left(R_{j+} + R_{j-}\right). \tag{2.25}$$

## C. Field Quantization

Jaynes and Cummings[(14)] have discussed a type of field quantization appropriate to the problem considered here. Their basic starting point is that the electromagnetic field should be expanded in terms of the resonant modes of the cavity under consideration instead of the usual plane wave expansion. The cavity may be represented by a volume, V, bounded by a closed surface, S. Let

$$\vec{E}_\lambda(\vec{x}), \qquad k_\lambda^2 = \frac{\omega_\lambda^2}{c^2}$$

be the eigenfunctions and eigenvalues of the boundary-value problem

$$\begin{gathered} \vec{\nabla}\, x\, \vec{\nabla}\, x\, \vec{E}_\lambda - k^2 \vec{E}_\lambda = 0 \; in \; V, \\ \vec{n}\, x\, \vec{E}_\lambda = 0 \; on \; S, \end{gathered} \tag{2.26}$$

where $\vec{n}$ is a unit vector normal to S. The normal modes $\vec{E}_\lambda(\vec{x})$ and $\vec{H}_\lambda(\vec{x})$, are related to one another through Maxwell's equations (2.2)

$$\begin{gathered} \vec{\nabla}\, x\, \vec{E}_\lambda = k_\lambda \vec{H}_\lambda \,, \\ \vec{\nabla}\, x\, \vec{H}_\lambda = -k_\lambda \vec{E}_\lambda \,, \end{gathered} \tag{2.27}$$

are normalized to unity

$$\begin{gathered} \int_V \left(\vec{E}_\lambda \cdot \vec{E}_{\lambda'}\right) dV = \delta_{\lambda\lambda'}, \\ \int_V \left(\vec{H}_\lambda \cdot \vec{H}_{\lambda'}\right) dV = \delta_{\lambda\lambda'} \,. \end{gathered} \tag{2.28}$$

Then the electric and magnetic fields may be expanded in the forms:

$$\vec{E}_\lambda(\vec{x}, t) = \sqrt{4\pi} \sum_\lambda \omega_\lambda q_\lambda \vec{E}_\lambda(\vec{x}), \tag{2.29a}$$

and

$$\vec{H}_\lambda(\vec{x}, t) = -\sqrt{4\pi} \sum_\lambda p_\lambda \vec{E}_\lambda(\vec{x}), \tag{2.29b}$$

Using this expansion, the Hamiltonian for the field alone Eq. (2.5c)

is written

$$H_f = \frac{1}{2}\sum_{\lambda}({p_\lambda}^2 + {\omega_\lambda}^2{q_\lambda}^2), \tag{2.30}$$

This is the familiar harmonic oscillator Hamiltonian for which the quantization technique is well known namely that the canonically conjugate coordinates and momenta satisfy the commutation relations

$$[q_\lambda, q_{\lambda'}] = [p_\lambda, p_{\lambda'}] = 0, \tag{2.31a}$$

and

$$[q_\lambda, p_{\lambda'}] = i\hbar\delta_{\lambda\lambda'}. \tag{2.31b}$$

The $q_\lambda$, $p_\lambda$ are now canonically transformed to $a_\lambda^\dagger$ and $a_\lambda$ , the creation and destruction operators for a photon in the mode $\lambda$. These operators are given by

$$a_\lambda^\dagger = \frac{p_\lambda + i\omega_\lambda q_\lambda}{\sqrt{2\hbar\omega_\lambda}} \tag{2.32a}$$

$$a_\lambda = \left(a_\lambda^\dagger\right)^\dagger = \frac{p_\lambda - i\omega_\lambda q_\lambda}{\sqrt{2\hbar\omega_\lambda}} \tag{2.32b}$$

The electric field and the electromagnetic field Hamiltonian in terms of these operators are

$$\vec{E}_\lambda(\vec{x}, t) = \sum_\lambda \sqrt{2\pi\hbar\omega_\lambda}\,\vec{E}_\lambda(\vec{x})\left[e^{-\frac{i\pi}{2}}a_\lambda^\dagger + e^{\frac{i\pi}{2}}a_\lambda\right] \tag{2.33a}$$

and

$$H_f = \sum_{\lambda} \hbar\omega_{\lambda}\left(\hat{n}_{\lambda} + \frac{1}{2}\right), \tag{2.33b}$$

where

$$\hat{n}_{\lambda} = a_{\lambda}^{\dagger} a_{\lambda} \tag{2.34}$$

is the number operator which acting on the state $|n_{\lambda}>$ gives the number of photons in mode $\lambda$ times the same state: $\hat{n}_{\lambda}\,|n_{\lambda}> = n_{\lambda}\,|n_{\lambda}>$.

The complete Hamiltonian expressed in terms of the operators developed above is written as

$$H = \hbar\Omega \sum_{j} R_{j3} + \sum_{\lambda} \hbar\omega_{\lambda}\left(\hat{n}_{\lambda} + \frac{1}{2}\right) - \sum_{\lambda,j}\left(\gamma_{\lambda j} a_{\lambda} + \gamma_{\lambda j}^{*} a_{\lambda}^{\dagger}\right)\left(R_{j+} + R_{j-}\right), \tag{2.35}$$

where

$$\gamma_{\lambda j} = \sqrt{2\pi\hbar\omega_{\lambda}}\,\vec{E}_{\lambda}(\vec{x}) \cdot \langle\vec{\mu}_j\rangle e^{\frac{i\pi}{2}}. \tag{2.36}$$

This is the Hamiltonian that many authors[1,2,15,16] use in a discussion of the laser and related problems. Even Eq. (2.35) is too difficult to treat exactly and further assumptions are required. These assumptions are that each molecule has exactly the same dipole moment and that all TLMs are assumed to be located at equivalent fixed mode positions or to be confined to a container whose dimensions are small compared to the radiation wavelength, i. e. , each molecule sees exactly the same $\vec{E}_{\lambda}$ field. This is experimentally feasible. Also it is assumed that only one mode of the cavity interacts significantly with the molecules. Therefore the energy of the field for the other modes may be considered as a constant of motion and may be dropped.

The Hamiltonian with these assumptions becomes

$$H = \hbar\Omega R_3 + \hbar\omega\hat{n} - (\gamma a + \gamma^* a^\dagger)(R_+ + R_-) \tag{2.37}$$

Here all mode subscripts have been dropped and the summation over the index j for the separate molecules performed. In the next chapter the problem to be treated exactly will be developed.

## CHAPTER III: FORMULATION OF THE PROBLEM

The Hamiltonian which will be taken to describe the interaction of N Two-Level Molecules with a single-mode radiation field is given by

$$H = \Omega R_3 + \omega a^\dagger a - \gamma a R_+ - \gamma^* a^\dagger R_- \tag{3.1a}$$

$$= H_o - \gamma a R_+ - \gamma^* a^\dagger R_-. \tag{3.1b}$$

To recapitulate, the identical Two-Level Molecules (TLMs) are assumed to have non-overlapping space functions and the energy separation of each TLM *($\Omega$)* is not necessarily equal to the mode frequency ($\omega$) of the electromagnetic field. For convenience, ħ has been set equal to unity. The complex coupling constant, $\gamma = |\gamma| e^{i\varphi_1}$, is given by Eq. (2.36). In the case of a cylindrical cavity[14], for example, with only the lowest TM mode excited, the eigenfunction, E, of the cavity is

$$\left|\vec{E}(\vec{x})\right| = \frac{1}{J_1\sqrt{V}} J_o\left(\frac{\omega}{c} r\right) \tag{3.2}$$

where $J_1 = J_1(u) = 0.5191$, and $u = 2.405$ is the first root of $J_o(u) = 0$. V is the volume of the cavity. The TLMs are coupled to the radiation field in the dipole approximation, and all TLMs are further assumed to be located at equivalent fixed mode positions, or to be confined to a container whose dimensions are small compared to the radiation wavelength. This means that $\vec{E}(\vec{x})$ in Eq. (2.36) may be taken as a constant. (In the case of the cylindrical cavity $\vec{E} = \frac{1}{J_1\sqrt{V}}$ if the TLMs are assumed to lie along the axis of the cylinder. In terms of frequency, $\gamma = \frac{\langle\mu\rangle}{J_1}\sqrt{\frac{2\pi\omega}{\hbar V}}e^{i\pi/2}$. By substituting specific values of $\langle\mu\rangle$ and V into Eq. (2.36), the value of $\gamma$ may be found. For example, a value for $\gamma$ of approximately 5 cps has been found for an ammonia beam maser[(14)] with a frequency of $2 \times 10^{10}$ cps. Terms in Eq. (2.27) have been ignored in the dipole coupling in Eq. (3.1) which do not conserve particle number, c, (i.e. , those terms in Eq. (2.37) which do not commute with $a^\dagger a + R_3$).

The validity of each of the assumptions used to obtain the final form of the Hamiltonian (3.1) is somewhat difficult to ascertain. The use of the dipole approximation is quite sound for most systems of interest since the wavelength of the relevant radiation is at least $10^3$ times the linear dimensions of most molecules. The assumptions that the TLMs interact with only a single mode and that each TLM sees exactly the same electric field strength are realizable in the laboratory. This is easily seen since the single mode radiation field may be chosen by constructing a resonant cavity that admits only a single mode near the resonant frequency of the TLMs. The second assumption that each TLM sees the same electric field strength has already been discussed. The assumption that each TLM is identical is not rigorously justified since collision for the case of a gas or wave function overlap for the case of a solid will cause a spread in the TLM frequency, $\Omega$. The effect of this continuum of TLM energies will not be considered here. Dicke(7) has discussed, within the framework of his approximation scheme, the effects of Doppler broadening. In the case of an ammonia beam maser, the TLMs will essentially interact with three different frequencies. Since all the molecules are moving at the same velocity, these frequencies are (1) the basic frequency of the molecules and (2) the two frequencies obtained from adding or subtracting the translational energies of the molecules. The final approximation of dropping the doubling terms has been discussed for the case of a single molecule interacting with a radiation field by E. T. Jaynes and F. W. Cummings[14]. The validity of dropping these terms for N TLMs is not too readily apparent and will be discussed in Appendix E. The contribution to the energy due to these terms is believed to be small except for very high intensity fields[7,14]. Their contribution to the energy may be determined by a perturbative technique after the eigenvalues and eigenstates are found.*[1] (Note that the model above describes many of the conditions existing in an ammonia beam maser[14].)

A form of H which will be more amenable to treatment is obtained by rearranging some of the terms in Eq. (3.1) and dividing by $\Omega$.

$$H = (a^\dagger a + R_3) + a^\dagger a \frac{\omega - \Omega}{\Omega} - \kappa a R_+ - \kappa^* a^\dagger R_- \quad (3.3)$$

$$= H_o - \kappa a R_+ - \kappa^* a^\dagger R_-. \quad (3.4)$$

$$\kappa = \frac{\gamma}{\Omega}. \quad (3.5)$$

States of the non-interacting system are defined[7] such that

$$H_o|n>|r,m> = \left(m + n + n\frac{\omega - \Omega}{\Omega}\right)|n>|r,m> \quad (3.6a)$$

[1] See Appendix E

$$R_3|r,m> = \sum_{j=1}^{N} R_{j3}|r,m> = m|r,m> \tag{3.6b}$$

$$R_{\pm}|r,m> = \sum_{j=1}^{N} R_{j\pm}|r,m>$$

$$= e^{\pm i\varphi_2}\big(r(r+1) - m(m\pm 1)\big)^{1/2}|r,m\pm 1> \tag{3.6c}$$

and

$$a|n> = e^{i\varphi_3}\sqrt{n}|n-1> \tag{3.6d}$$

The states $|r,m>$ are formally identical to states of total angular momentum and total z-component of momentum for a system of spins. These $|r,m>$states are formed in the same way, from single molecule states $|1/2,\pm 1/2>$, as the total spin states are formed from the individual spinor states. The "cooperation number," r, analogous to the total angular momentum of a spin system, satisfies

$$R^2|r,m> = r(r+1)|r,m>, \tag{3.7}$$

with m$\leq r \leq \frac{N}{2}$ where r and m are either integer of half-integer. The term "cooperation number" has significance in the fact that the projection of the vector operator on the “1-2” plane gives essentially the total dipole moment of the molecular system while the third component, $R_3$, gives the energy. For larger "cooperation number," the possible interaction with the electromagnetic field also becomes larger. On the other hand, a value of r = 0 implies no interaction with the electromagnetic field at all. The operators in Eq.(3.6) satisfy the commutation relations

$$[R_3, R_{\pm}] = \pm R_{\pm} \tag{3.8a}$$

$$[R_+, R_-] = 2R_3 \tag{3.8b}$$

$$[a, a^\dagger] = 1. \tag{3.8c}$$

Since both $R^2$ and $c = a^t a + R_3$ commute with H, the eigenstates may be chosen to be eigenstates of these two operators as well, and the eigenstates may be labeled by the eigenvalues r and c. The symbol c represents the conservation of the number of the photons plus the number of TLMs in the excited state (c is not a good quantum number for the general Hamiltonian (2.37)). If $\omega=\Omega$[17], c is the eigenvalue of $H_o$ and is given by Eq. (3.4). For a given r and c, there will be (in general) 2r+1 energy eigenvalues and for $\omega=\Omega$ these eigenvalues will be symmetrically displaced about the constant c, where $-r < c < \infty$ .

It is helpful to display the elements

$$< n| < r, m|H|r', m' > |n' >$$

as a matrix where there is a grouping into an infinite number of blocks of dimensions $2^N$ along the main diagonal. Each $2^N$ dimensional block in turn breaks up into smaller blocks along the main diagonal, their number and dimension determined according to the irreducible representations of the group SU(2). This is shown schematically in Figure 1. Only the shaded blocks have nonzero elements. There are

$$\frac{N!\,(2r+1))}{\left(\frac{N}{2}+r+1\right)!\left(\frac{N}{2}-r\right)!},$$

identical blocks for each value of r, and each smaller block is of dimension 2r+1. The values of r range from N/2 for the largest single shaded block at the top left-hand corner of the figure down to either r = 0, or r = 1/2 depending on whether N is even or odd. For a given value of r, an integer change in c corresponds to a change to an adjacent block of dimension $2^N$.

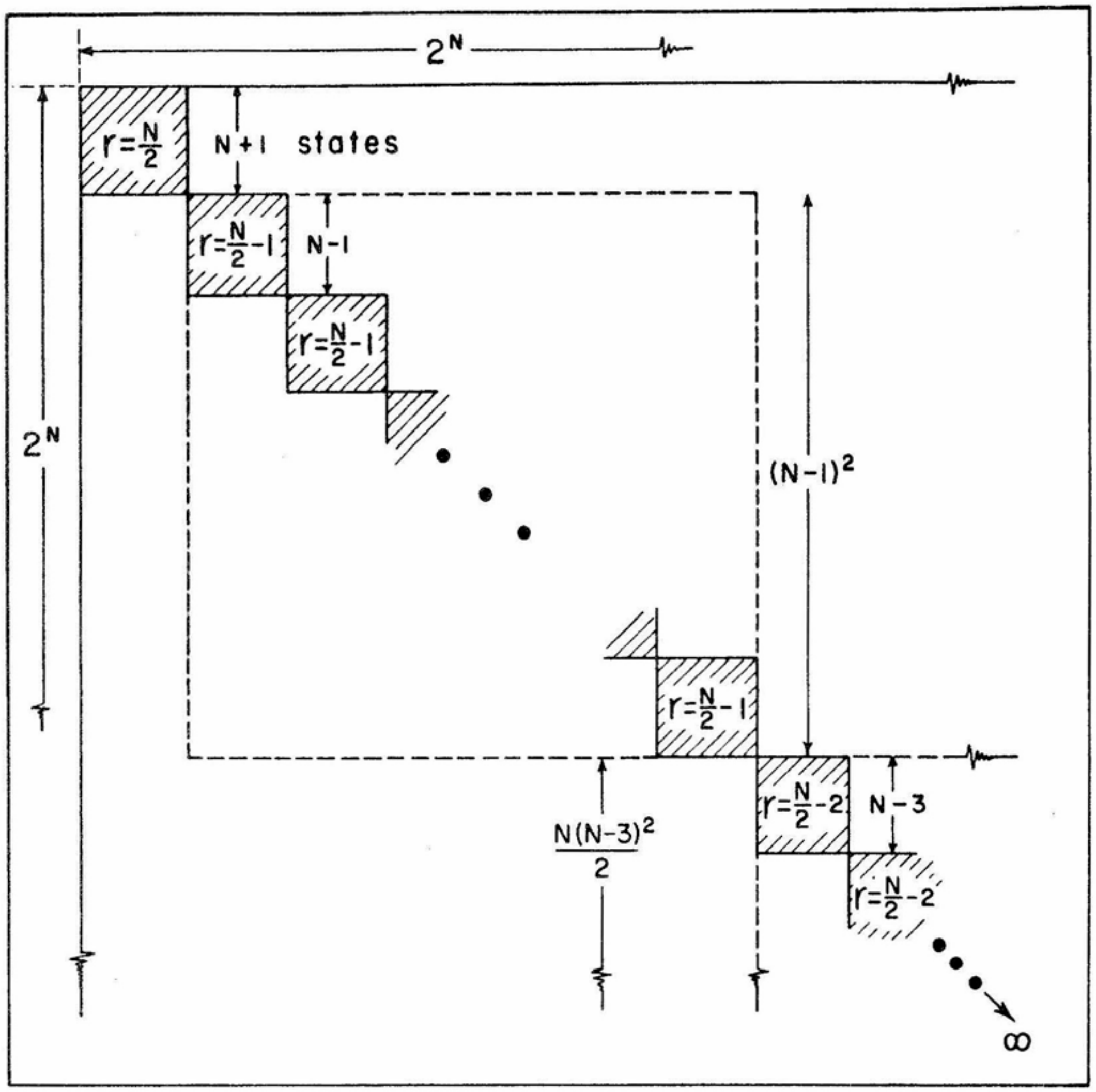

Figure 1: Schematic of Matrix Representation of H

Figure 2 shows the elements of one of the shaded blocks of Figure 1 for a particular r value. The block has nonzero entries only along the diagonal and immediately adjacent to it. If the value of c is such that c $\leq$ r, then the block will have dimension less than 2r + 1 and will be of dimension c+ r + 1 instead; thus, besides having an infinite number of blocks of dimension $2^N$, there will also be N large blocks of dimension smaller than $2^N$.

Denote the eigenstates of H as |r, c, j > .

$$H|r,c,j> = \lambda_{r,c,j}|r,c,j> \tag{3.9}$$

where j takes on the 2r + 1 values 0,1,2, . . . , 2r, if c $\geq$ r, or the c + r + 1 values 0,1,⋯,c+r if c < r.

Recalling that the states |r, c, j > are eigenstates of H, c, and $R^2$, and that m = c-n varies between r and -r,

$$|r, c, j> = \sum_{n=max(0,c-r)}^{c+r} A_n^{(r,c,j)} |n> |r, c-n>, \tag{3.10}$$

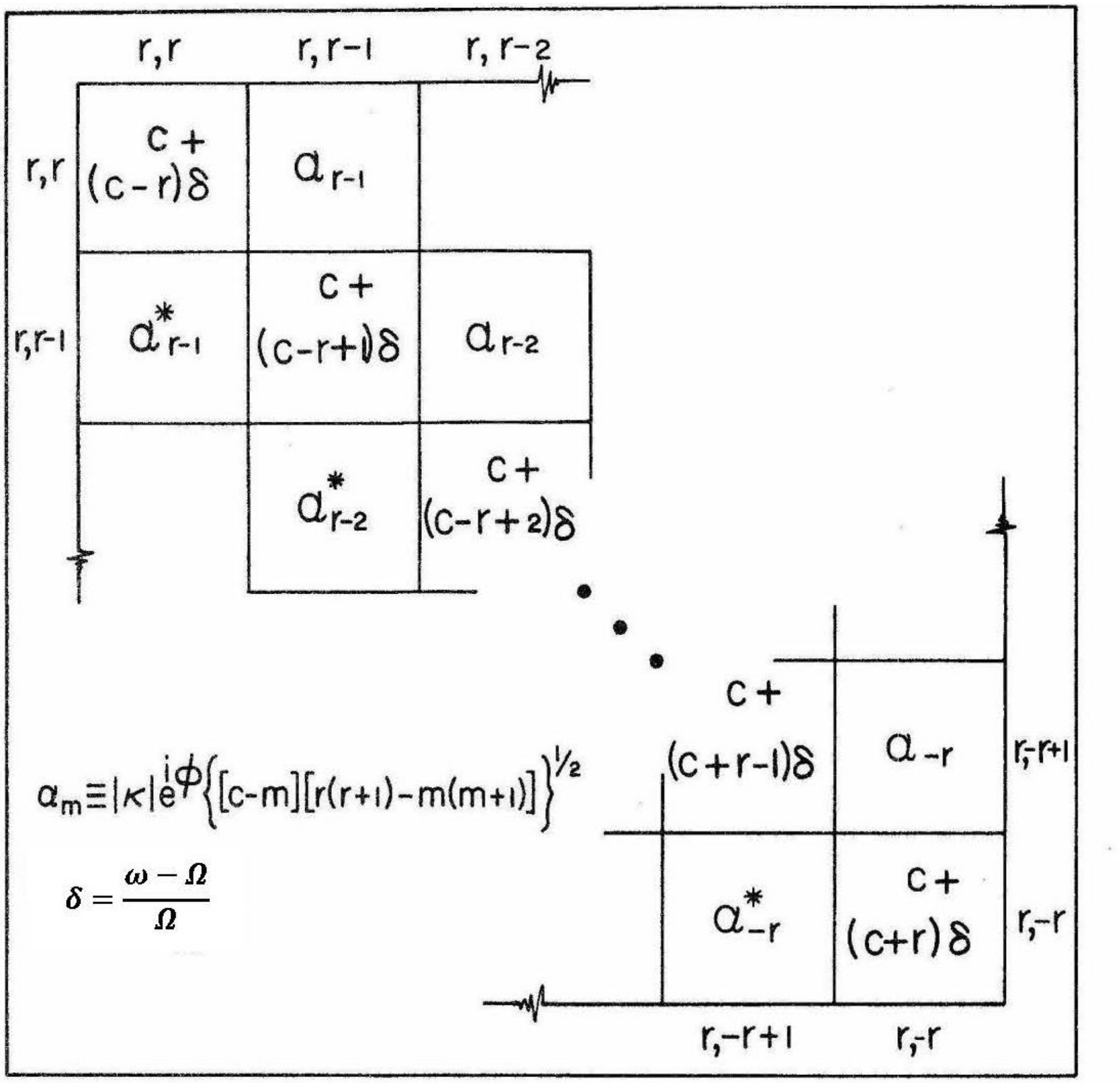

**Figure 2: Sub-matrix for given r and c.**

where max(0, c-r) is an abbreviation for the maximum value of 0 or c-r. From Eqs. (3.4), (3.6), and (3.8), the $A_n^{(r,c,j)}$ satisfy the difference equation

$$-|\kappa|e^{-i\varphi}\sqrt{n}C_{r,c-n}A_{n-1}^{(r,c,j)} + \left(c + \frac{\omega-\Omega}{\Omega}n - \lambda_{r,c,j}\right)A_n^{(r,c,j)} - |\kappa|e^{i\varphi}\sqrt{n+1}C_{r,c-n-1}A_{n+1}^{(r,c,j)} = 0, \tag{3.11}$$

where

$$\varphi = \varphi_1 + \varphi_2 + \varphi_3, \tag{3.12a}$$

and

$$C_{r,c-n} = [r(r+1) - (c-n)(c-n+1)]^{1/2}. \tag{3.12b}$$

The $A^{(r,c,j)}$ satisfy the boundary conditions

$$A_{r+c+1}^{(r,c,j)} = A_{max(-1,c-r-1)}^{(r,c,j)} = 0. \tag{3.13}$$

It is convenient to define $B_n$'s so that

$$A_n = \frac{\left(e^{-i\varphi}\right)^n B_n}{\sqrt{n!}\,C_{min(c-1,r-1)}C_{min(c-2,r-2)}\cdots C_{c-n}} \tag{3.14}$$

The superscripts (subscripts) (r, c, j) have been dropped for simplicity whenever this does not cause confusion. Only the r subscript will be suppressed in the $C_{r,c-n}$ to conform with Figure 2. If an effective eigenvalue

$$q = \frac{c-\lambda}{|\kappa|} \tag{3.15}$$

and relative tuning parameter

$$\beta = \frac{\omega - \Omega}{|\kappa|\Omega} \tag{3.16}$$

is defined, $B_n$ will satisfy the difference equation

$$B_{n+1} - (q + \beta n)B_n + nC_{c-n}^2 B_{n-1} = 0. \tag{3.17}$$

The largest value of q for a given value of r and c corresponds to the ground state of the system.

# CHAPTER IV: EXACT SOLUTION

The exact solution of Eq. (3.17) (non-normalized) can be obtained by "unraveling" from one end. That is, by starting from

$$B_{max(-1,c-r-1)} = B_{c+r+1} = 0, \tag{4.1}$$

one may obtain a solution for $B_n$ in the form

$$B_n = \sum_{\ell=0}^{[t/2]} (-1)^{\ell} \mathcal{S}_{\ell}^{(t-1)}, \tag{4.2}$$

where

$$n = t + \alpha \tag{4.3a}$$

$$\alpha = max(0, c - r), \tag{4.3b}$$

and $[t/2]$ is the first integer equal to or less than t/2.

The $\mathcal{S}_{\ell}^{t}$ are given by

$$\mathcal{S}_{\ell}^{t} = \sum_{m_1=1}^{t} \sum_{m_2=m_1+2}^{t} \sum_{m_3=m_2+2}^{t} \cdots \sum_{m_\ell=m_{\ell-1}+2}^{t} \left\{ \prod_{\substack{y=0 \\ y\neq[m_i,m_i-1]}}^{t} (q + (y+\alpha)\beta) \right\} \prod_{j=\{m_i\}} \mathcal{C}_j \tag{4.4a}$$

where $\mathcal{C}_{m_i} = (m_i + \alpha) C^2_{c-(m_i+\alpha)}$. In Eq. (4.4a) the first product cannot contain terms with y equal to any of the $m_i$ or $m_i$-1 and the second product contains only terms with j equal to one of the $m_i$.[2] Equation (4.4a) and the following explanation may be somewhat clarified by writing a few

[2] Note that Eq. (4.4a) consists of the sum of products of $\mathcal{C}_j$ taken $\ell$ at a time with no nearest neighbors.

examples of $\mathcal{S}_\ell^t$.

$$\mathcal{S}_0^t = (q+\alpha\beta)(q+(\alpha+1)\beta)(q+(\alpha+2)\beta)\cdots(q+(\alpha+t)\beta)$$

$$\mathcal{S}_3^6 = (q+(\alpha+6)\beta)\mathcal{C}_1\mathcal{C}_3\mathcal{C}_5 + (q+(\alpha+4)\beta)\mathcal{C}_1\mathcal{C}_3\mathcal{C}_6 + (q+(\alpha+2)\beta)\mathcal{C}_1\mathcal{C}_4\mathcal{C}_6 + (q+\alpha\beta)\mathcal{C}_2\mathcal{C}_4\mathcal{C}_6$$

$$\begin{aligned}\mathcal{S}_2^5 &= (q+(\alpha+4)\beta)(q+(\alpha+5)\beta)\mathcal{C}_1\mathcal{C}_3 \\ &+ (q+(\alpha+2)\beta)(q+(\alpha+5)\beta)\mathcal{C}_1\mathcal{C}_4 + (q+(\alpha+2)\beta)(q+(\alpha+3)\beta)\mathcal{C}_1\mathcal{C}_5 \\ &+ (q+\alpha\beta)(q+(\alpha+5)\beta)\mathcal{C}_2\mathcal{C}_4 + (q+\alpha\beta)(q+(\alpha+3)\beta)\mathcal{C}_2\mathcal{C}_5 \\ &+ (q+\alpha\beta)(q+(\alpha+1)\beta)\mathcal{C}_3\mathcal{C}_5\end{aligned} \qquad (4.4b)$$

$$\mathcal{S}_3^5 = \mathcal{C}_1\mathcal{C}_3\mathcal{C}_5$$

$$\mathcal{S}_0^{-1} = 1, and\ \mathcal{S}_\ell^t = 0\ if\ \ell > \frac{1}{2}(t+1), e.g., \mathcal{S}_3^4 = 0.$$

The above equations (4.4) are much too difficult to use in a practical way; however, a recursion relation for the "$\mathcal{S}_\ell^t$" exists which makes the use of Eq. (4.2) practical. This recursion relation

$$\mathcal{S}_\ell^t = (q+(\alpha+t)\beta)\mathcal{S}_\ell^{t-1} + \mathcal{C}_t\mathcal{S}_{\ell-1}^{t-2} \qquad (4.5)$$

is found by induction or inspection.

The exact eigenvalues, or equivalently the q's, are determined from Eq. (4.2) for $B_{r+c+1}$ namely

$$B_{r+c+1} = \sum_{\ell=0}^{[t'/2]} (-1)^{\ell} \mathcal{S}_{\ell}^{(t'-1)} = 0, \tag{4.6}$$

where t' = r+c+1-$\alpha$. These are polynomials in q of degree 2r+1 if c≥ r, and of degree r+c+l if c<r. If 2r+1 or r+c+1 is even and $\beta$=0, one of the roots of Eq. (4.6) is q=0. For this case, the eigenvector can be found directly from the equation for the $A_n$ and is given by

$$A_n = (-1)^{t/2} e^{-it\varphi} \sqrt{\frac{\mathcal{C}_{t-1}\mathcal{C}_{t-3}\cdots\mathcal{C}_1}{\mathcal{C}_t\mathcal{C}_{t-2}\cdots\mathcal{C}_2}}, \text{ t even}$$
$$= 0 \qquad , t\ odd \tag{4.7}$$

Also in the special case ($\beta = 0$) the q's are such that $q_{2r}=-q_0$, $q_{2r-1}=-q_1$, and so on symmetrically placing the q-values about zero, and the states $A_n^{(j)}$, j =2r, 2r-1,⋯, r-1, if r is an integer (or r+1/2 if r is a half-integer) are found from the states $A_n$, j = 0, 1,⋯, r-1 (or r-1/2 if r is a half-integer) by replacing $\varphi$ by $\varphi$ + $\boldsymbol{\pi}$· in Eq. (3.14). Care must be taken with these latter statements since $0 \leq j \leq r + c$ when c < r.

A high speed computer was used to calculate the effective eigenvalues "q" from Eq. (4.6) as well as the normalized eigenvectors $A_n^{(r,c,j)}$ and values of average atom energy, or inversion <m> , and the dispersion in photon number $< (n - <n> )^2 >$ for each eigenvector. A brief discussion of the numerical technique used is given in Appendix A.

# CHAPTER V: APPROXIMATE ANALYTIC SOLUTIONS FOR $\beta = 0$

In this chapter approximate expressions for the eigenvalues and eigenvectors of Eq. (3.1) will be obtained for the special case $\omega = \Omega$.

## A. Differential Equation Approach

The approximation obtained here is accurate for the ground and low-lying excited states (and, because of symmetry about q = 0 for the special case $\beta = 0$, for the most highly excited states and eigenvalues as well). Toward this end Eq. (3.11) is rewritten by defining $E_n$ such that

$$A_n = \frac{\left(\bar{q}e^{-i\varphi}\right)^n E_n}{\sqrt{n!}\, C_{c-1}C_{c-2}\cdots C_{c-n}}\,, c < r \tag{5.1a}$$

and

$$A_n = \frac{\left(\bar{q}e^{-i\varphi}\right)^n E_n}{\sqrt{n!}\, C_{r-1}C_{r-2}\cdots C_{c-n}}\,, c \geq r \tag{5.1b}$$

where

$$\bar{q} = \frac{(c-\lambda)}{2|\kappa|} = \frac{1}{2}q. \tag{5.2}$$

$E_n$ then satisfies the difference equation

$$\bar{q}^2 E_{n+1} - 2\bar{q}^2 E_n + nC_{c-n}^2 E_{n-1} = 0. \tag{5.3}$$

This is rewritten in the form

$$E_{n+1} + E_{n-1} - 2E_n + \left\{\frac{(4nC_{c-n}^2 - q^2)}{q^2}\right\} E_{n-1} = 0. \tag{5.4}$$

If $\Delta E_n \equiv E_{n+1} - E_n$, then $\Delta^2 E_n = E_{n+2} + E_n - 2E_{n+1}$

which implies that

$$\Delta^2 E_{n-1} + \left\{\frac{(4nC_{c-n}^2 - q^2)}{q^2}\right\} E_{n-1} = 0. \qquad (5.5)$$

The first approximation made to this equation is the replacement of $\Delta^2$ by $\frac{d^2}{dx^2}$. This will be accurate for large enough values of r, when n takes on a sufficient number of values to make this replacement sensible; clearly it will not be accurate for r = 1/2, for example, when there are only two values of n for a given c. Also, this approximation is not expected to be accurate for large j when $A_n$ changes rapidly as n varies. The number of "zeros" of the vector is in fact given by j and for j = r (r interger) the $A_n$ have zero value every other term. This may be seen from Eq. (4.7). The differential equation for E(n) is then

$$E'' = \frac{(4nC_{c-n}^2 - q^2)}{q^2} E = 0 \qquad (5.6)$$

The second approximation, which is made to Eq. (5.6), is most easily seen by reference to Figure 3. This figure shows the cubic equation, considered as a continuous function of n:

$$F(n) = q^2 - 4nC_{c-n}^2 = q^2 + 4[n^3 - 2n^2 x - n(y^2 - x^2)], \qquad (5.7)$$

where

$$x = c + \frac{1}{2} \qquad (5.8a)$$

$$y = r + \frac{1}{2} \qquad (5.8b)$$

The minimum of F(n) is at $n_o$ , obtained from $F'(n_o) = 0$, namely

$$n_o = \frac{2}{3}x + \frac{1}{3}\sqrt{3y^2 + x^2} \tag{5.9}$$

Note that the minimum position does not depend on the effective eigenvalue q. Writing F(n) as a function of $(n - n_o)$ gives

$$F(n_o) = -\alpha_1 + q^2 + \alpha_2(n - n_o)^2 + 4(n - n_o)^3\,, \tag{5.10}$$

where the $\alpha's$ are defined by

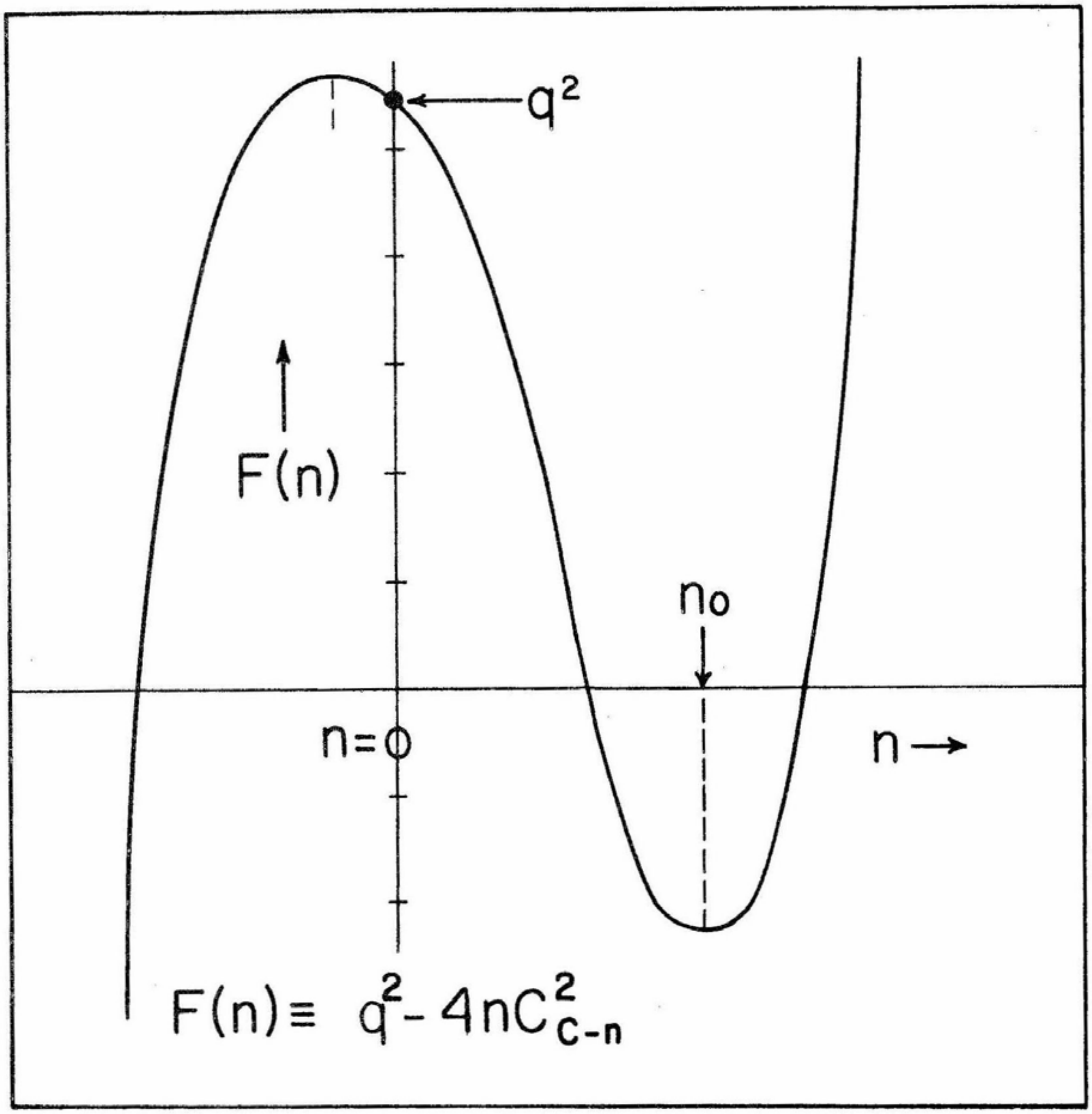

Figure 3: Schematic of cubic F(n) vs. n.

$$\alpha_1 = 4n_o C_{c-n}^2 \tag{5.11a}$$

$$\alpha_2 = 4\sqrt{3y^2 + x^2} \tag{5.11b}$$

For all $(n\text{-}n_o)$ such that $|n\text{-}n_o| << \alpha_2$, discarding the cubic term will be a good approximation. For example, it turns out that the dispersion of the ground state is less than $n_o$, which means that the maximum relevant $|n\text{-}n_o|$ is about $\sqrt{n_o}$ ; but $\boldsymbol{\alpha}_2$ is of the order of $n_o$ (or greater) so that for large $n_o$ discarding the cubic is clearly a valid approximation, at least for the ground state, and as it develops, for a number of the first excited states as well.

With these two approximations, Eq. (5.4) becomes

$$E''(n) + \left\{\frac{\alpha_1 - q^2}{q^2} - \frac{\alpha_2}{q^2}(n - n_o)^2\right\} E(n) = 0. \tag{5.12}$$

Making a change of variable,

$$t = \left(\frac{\alpha_2}{q^2}\right)^{1/4} (n - n_o) \tag{5.13}$$

In Eq. (5.12) gives

$$\left\{\frac{d^2}{dt^2} + \left[\frac{\alpha_1 - q^2}{(\alpha_2 q^2)^{\frac{1}{2}}} - t^2\right]\right\} E(t) = 0. \tag{5.14}$$

This is in the standard form for the harmonic oscillator[12] and if we apply the usual boundary condition that E vanishes as $t \to \pm\infty$ (third assumption) then

$$\frac{\alpha_1 - q^2}{(\alpha_2 q^2)^{\frac{1}{2}}} = 2j + 1,$$

$$j = 0,1,\cdots\cdots, r - 1; r\ intger$$

$$j = 0,1,\cdots\cdot, r - \frac{1}{2}; r\ half - integer. \tag{5.15a}$$

Solving the quadratic equation for q,

$$q_j = -(\alpha_2)^{1/2}\left(j + \frac{1}{2}\right) + \left[\left(j + \frac{1}{2}\right)^2 \alpha_2 + \alpha_1\right]^{1/2} \tag{5.15b}$$

which gives the approximate effective eigenvalues for j positive ≤ r. For j > r, the eigenvalues are again given as explained below Eq. (4. 7); they are the negative of the values for j < r. Also

$$E_j(t) = e^{-t^2/2} H_j(t), \tag{5.17}$$

where $H_j(t)$ is the Hermite polynomial of order j. Then

$$(E_{n-1})_j = \{exp[-(\alpha_2)^{1/2}(n - n_o)^2/2q_j]\} H_j\left[\left(\frac{\alpha_2}{q^2}\right)^{1/4}(n - n_o)\right] \tag{5.18}$$

## A.1 Dispersion in Photon Number in the Ground State within the Differential Equation Approximation

Table 1 lists limiting forms for ready reference of several previously defined quantities as well as $\sigma^2$, the dispersion in the ground state. From the table it is readily seen for all four limiting regions considered, that $\bar{q}_o/C_{c-n}$ is nearly equal to $\sqrt{n_o}$. This is also the statement that $\alpha_2 \ll \alpha_1$ for these limits, and so $\bar{q}_o = -\frac{\sqrt{\alpha_2}}{2} + \sqrt{\frac{\alpha_2}{4} + \alpha_1} \cong \sqrt{\alpha_1}$. Therefore $\bar{q}_o = \frac{q_o}{2}$ can be written as $\bar{q}_o = \sqrt{2} n_o \left| \sqrt{y^2 + n_o^2} - n_o \right|^2$ from which it is clear that $\bar{q}_o \leq \sqrt{2y} n_o$. For the first two columns of Table 1 this is a good approximation. For c>>r>>1, $C_{c-n_o} \to r$ and $\bar{q}_o \to \sqrt{n_o} r$. The product of the $C_n$'s which appear in the denominator of Eq. (5. 1) may be written as

$$\left.\begin{array}{ll} c < r & C_{c-1}C_{c-2}\cdots C_{c-n} \\ c > r & C_{r-1}C_{r-2}\cdots C_{c-n} \end{array}\right\} = Const. \sqrt{\frac{(r-c+n)!}{(r+c-n)!}}, \tag{5.19}$$

where the constant (Const.) is independent of n. In the ground state the non-normalized $A_n^{(r,c,0)}$ (since $H_o(t) = 1$) are

$$A_n^{(0)} \cong \frac{(\bar{q}_o)^n e^{in\varphi}}{\sqrt{n!}\sqrt{\frac{(r-c+n)!}{(r+c-n)!}}} e^{\frac{-(n-n_o)^2\sqrt{\alpha_2}}{2q_o}} \tag{5.20a}$$

Since $\bar{q}_o \cong \sqrt{n_o}\sqrt{(r-c+n_o)(r+c-n_o)} > 1$ and using Sterling's approximation for the factorial terms

$$n! = \sqrt{2\pi n} n^n e^{-n}, \tag{5.20b}$$

Eq. (5. 20a) becomes, after some simple algebra,

$$A_n^{(0)} \cong \frac{exp\left\{\frac{n}{2}[ln(n_o)+1] + \frac{(r-c+n)}{2} ln\left(\frac{r-c+n_o}{r-c+n}\right) - \frac{(r+c-n)}{2} ln\left(\frac{r+c-n_o}{r+c-n}\right) - n + c - \frac{(n-n_o)^2\sqrt{\alpha_2}}{2q_o} + in\varphi\right\}}{\left\{\left[(2\pi n)\frac{r-c+n}{r+c-n}\right]^{1/4} exp\left[\frac{r-c}{2} ln(r-c+n_o) - \frac{r+c}{2} ln(r+c-n_o)\right]\right\}} \tag{5.21}$$

## Table 1: Some Limiting Cases

| | $c = -r + \epsilon$ <br> $r \gg \epsilon \gg 1$ | $r \gg c \gg 1$ | $r = c \gg 1$ | $c \gg r \gg 1$ |
|---|---|---|---|---|
| $\frac{\alpha_2}{4}$ | $2r\left(1 - \frac{\epsilon}{4r}\right)$ | $\sqrt{3}r\left(1 + \frac{c^2}{6r^2}\right)$ | $2c$ | $c\left(1 + \frac{3r^2}{2c^2}\right)$ |
| $n_o$ | $\frac{\epsilon}{2}$ | $\frac{r}{\sqrt{3}}\left(1 + \frac{2c}{\sqrt{3}r}\right)$ | $\frac{4c}{3}$ | $c\left(1 + \frac{r^2}{2c^2}\right)$ |
| $C_{c-n_o}$ | $\sqrt{\epsilon\left(r - \frac{\epsilon}{3}\right)}$ | $\sqrt{\frac{2}{3}}r\left(1 + \frac{c}{2\sqrt{3}r}\right)$ | $\frac{2\sqrt{2}}{3}c$ | $r$ |
| $\frac{\alpha_1}{4}$ | $\frac{\epsilon^2(r - \epsilon/3)}{2}$ | $\frac{2}{3\sqrt{3}}r^3\left(1 + \frac{\sqrt{3}c}{r}\right)$ | $\frac{32}{27}c^3$ | $r^2c\left(1 + \frac{r^2}{2c^2}\right)$ |
| $\bar{q}_o$ | $\epsilon\sqrt{\left(r - \frac{\epsilon}{3}\right)/2}$ | $\sqrt[4]{3}\frac{\sqrt{2}}{3}r^{3/2}$ | $\frac{4}{3}\sqrt{\frac{2}{3}}c^{3/2}$ | $r\sqrt{c}$ |
| $\sigma^2$ | $\frac{n_o}{2}$ | $\frac{n_o}{\sqrt{6}}$ | $\frac{n_o}{2\sqrt{3}}$ | $\frac{r}{2}$ |

Replace n by $n_o$ in the denominator of Eq. (5.21). These terms in the denominator may then be included in the normalization and can be ignored. If the various *ln* terms in the numerator of Eq. (5. 21) are expanded, for example

$$ln\left(\frac{n_o}{n}\right) = ln\left(\frac{n_o - n + n}{n}\right) \cong \frac{n_o - n}{n} - \frac{1}{2}\frac{(n_o - n)^2}{n^2}, \tag{5.22}$$

and then collected, and if the constant term $e^{\left[\frac{n_o}{2} + n_o - c\right]}$ is dropped, then Eq. (5.21) becomes

$$A_n^{(0)} \propto exp\left\{-\frac{(n_o - n)^2}{4}\left[\frac{1}{n_o} + \frac{1}{r - c + n_o} - \frac{1}{r + c - n_o} + \frac{2\sqrt{\alpha_2}}{q_o}\right]\right\} e^{in\varphi} \tag{5.23a}$$

Terms have been ignored in Eq. (5.23a) which resulted from the substitution of $n_o$ for n in the inner square bracket of that expression. For all four limiting regions of Table 1, the sum of the first three terms in the inner bracket of Eq. (5.23a) is very small in comparison to the last term; therefore,

$$A_n^{(0)} \approx exp\left\{\frac{-(n-n_o)^2}{4(q_o)/2\sqrt{\alpha_2}}\right\} e^{in\varphi}. \tag{5.23b}$$

The dispersion in photon number in the ground state is

$$\sigma^2 = \langle (n-n_o)^2 \rangle = \frac{\sum \left| A_n^{(0)} \right|^2 (n-n_o)^2}{\sum \left| A_n^{(0)} \right|^2}$$

or using Eq. (5.23)

$$\sigma^2 = \frac{q_o}{2\sqrt{\alpha_2}}. \tag{5.24}$$

The bottom row of Table 1 shows $\sigma^2$ for the four limiting cases. The dispersion (5.24) is in every case smaller than the "classical" dispersion obtained from a Poisson distribution, namely $\sigma^2 = n_o$ [(6)], although for r ≥ c the dispersion is of order $n_o$, the average photon number. When the amount of energy in the electromagnetic field greatly exceeds the amount of energy available to the uncoupled atoms, that is c>>r, then the dispersion is much less than the average photon number $n_o \cong$ c and is $\sim r/2$ instead.

In a recent paper by the author[(17)] an error was made in deriving the results of the corresponding section, namely the error was in replacing the product of terms in Eq. (5.19) by $C^n_{c-n_o}$. The wave function and dispersion given in that paper were therefore incorrect.

Note that for the excited states, j ≠ 0, a similar analysis is possible and

$$A_n^j \cong exp\frac{-j(n-n_o)}{r}exp\left[\frac{-\alpha_2^{1/2}(n-n_o)^2}{2q_j}\right]H_j\left[\left(\frac{\alpha_2}{q_j^2}\right)^{1/4}(n-n_o)\right]$$

where j<<r. If j=2r-I where I<<r, then

$$A_n^j \cong (-1)^n exp\frac{-I(n-n_o)}{r}exp\left[\frac{-\alpha_2^{1/2}(n-n_o)^2}{2q_j}\right]H_j\left[\left(\frac{\alpha_2}{q_j^2}\right)^{1/4}(n-n_o)\right]$$

The second exponential dependence predominates over the first. This is the reason for the resemblance between exact states and the states of the harmonic oscillator.

The next two approximations are based on modifying the magnitude of either TLM or field part of the interaction terms in the Hamiltonian. This modification is performed only on the interaction terms and in such a way that $R^2$ and ($a^\dagger a$+ $R_3$) still commute with the Hamiltonian; that is, r and c remain good quantum numbers.

## B. Average Field Approach

The first of these modification processes is to average over the magnitude of the photon creation and destruction operators in the interaction terms of Eq. (3.4). Towards this end define operators $\mathcal{L},\mathcal{L}^\dagger$ (18) for which the following hold

$$\mathsf{L}|n>=|n-1>, n>0$$
$$= 0 \qquad , n = 0, \tag{5.25a}$$

$$\mathsf{L}^{\dagger}|n>=|n+1>, \tag{5.25b}$$

$$[n,\mathsf{L}] = -\mathsf{L}, \tag{5.25c}$$

$$[n,\mathsf{L}^{\dagger}] = \mathsf{L}^{\dagger}, \tag{5.25d}$$

$$and\ [\mathsf{L},\mathsf{L}^{\dagger}] = |0><0|. \tag{5.25e}$$

Eq. (5.25e) implies that $\mathcal{L}$ is not unitary. For a full discussion of these operators see reference (18). The photon creation and destruction operators can now be defined in terms of $\mathcal{L}$ and $\mathcal{L}^{\dagger}$ as

$$a = \mathsf{L}\sqrt{\hat{n}} \tag{5.26a}$$
$$a^{\dagger} = \sqrt{\hat{n}}\mathsf{L}^{\dagger}, \tag{5.26a}$$

where $\hat{n}$ is simply the number operator. In terms of these new operators, Eq. (3.4) for $\omega = \Omega$ becomes

$$H = n + R_3 - \kappa\mathsf{L}\sqrt{\hat{n}}\,R_+ - \kappa^*\sqrt{\hat{n}}\mathsf{L}^{\dagger}R_-, \tag{5.27}$$

and performing the indicated average

$$H = n + R_3 - \kappa\sqrt{n_o}\,\mathsf{L}R_+ - \kappa^*\sqrt{n_o}\mathsf{L}^{\dagger}R_-. \tag{5.28}$$

Two ways of approaching Eq. (5.28) are possible. The first is to again form the difference equation, as was done in the case of the exact solution, by defining a "field" averaged solution

$$\overline{|rc\mathrm{j}>}_f = \sum_{n=max(0,c-r)}^{c+r} F_n|n>|r,m> \tag{5.29}$$

so that

$$-|\kappa|e^{-i\varphi}\sqrt{n_o}\,C_{c-n}F_{n-1} + (c-\lambda_f)F_n - |\kappa|e^{i\varphi}\sqrt{n_o}\,C_{c-n-1}F_{n+1} = 0\,, \tag{5.30}$$

where all understood subscripts (superscripts) have been dropped. Following the procedure given for Chapters 3 and 4, define

$$F_n = \frac{e^{-in\varphi}G_n}{C_{min(c-1,r-1)}\cdots C_{c-n}}\,, \tag{5.31a}$$

so that

$$G_{n+1} - q_f G_n + C_{c-n}^2 G_{n-1} = 0\,, \tag{5.31b}$$

where

$$q_f = \frac{c-\lambda_f}{|\kappa|\sqrt{n_o}}. \tag{5.31c}$$

The same boundary conditions apply as previously, Eq. (4. 1),

$$G_{max(-1,c-r-1)} = G_{c+r+1} = 0\,, \tag{5.32a}$$

implying that

$$G_n = \sum_{l=0}^{[t/2]} (-1)^l U_l^{(t-1)}\,, \tag{5.32b}$$

and

$$U_l^t = q_f^{t-2l} \sum_{m_1=1}^{t}\ \sum_{m_2=m_1+2}^{t}\ \sum_{m_3=m_2+2}^{t} \cdots \sum_{m_l=m_{l-1}+2}^{t} \left\{ \prod_{j=\{m_i\}} C_{c-(j+\alpha)}^2 \right\}. \tag{5.32c}$$

Again the effective eigenvalues $q_f$ are found from the roots of $G_{c+r+1}=0$ and all statements about their number and symmetries are the same as given after Eq. (4.7) for the special case $\omega = \Omega$. Also for the case that (2r+1, r+c+1) is odd, the exact non-normalized solution when $q_f=0$ and $\beta=0$ for even t is

$$F_n = (-1)^t e^{-it\varphi} \sqrt{\frac{C_{c-(t-1+\alpha)}^2 \cdots C_{c-(1+\alpha)}^2}{C_{c-(t+\alpha)}^2 \cdots C_{c-(2+\alpha)}^2}}. \tag{5.33}$$

Numerical solutions for the $F_n$ and $q_f^{(j)}$ were obtained as outlined in Appendix A as in the case of the exact solution. For c>r, the fact that the $q_f$'s are found to lie along a straight line leads to another formal approach to Eq. (5.28) valid for c > r. Rewriting Eq. (5. 28) gives formally

$$H = \sum_{j=1}^{N} \frac{(a^{\dagger}a + R_3)}{N} - \kappa\sqrt{n_o}\,\mathsf{L}R_{j+} - \kappa^{*}\sqrt{n_o}\,\mathsf{L}^{\dagger}R_{j-}\,. \tag{5.34}$$

Recalling that c is a good quantum number, replace the first term by

$$c_j = \frac{c}{N} \tag{5.35}$$

This may be interpreted as the Hamiltonian of N separate TLMs interacting with the electromagnetic field. Each interaction conserves the average energy c of the system but the TLMs "see" each other only through an average field. Therefore this problem can be treated by solving exactly for the eigenvalues and vectors for just one TLM, and the total state and energy of the system is formed from the single TLM in such a way as to give states of total r, c, and j, where j is an index representing the energy of the state.

The matrix for one TLM interacting with the field in terms of the basis states $|c_j - \frac{1}{2} > |\frac{1}{2}, \frac{1}{2} > and\ |c_j + \frac{1}{2} > |\frac{1}{2}, -\frac{1}{2} >$ is displayed in Figure 4.

| | |
|---|---|
| $c_j$ | $-\boldsymbol{\kappa}\sqrt{\boldsymbol{n_o}}$ |
| $-\boldsymbol{\kappa}^*\sqrt{\boldsymbol{n_o}}$ | $c_j$ |

**Figure 4 :**

The eigenvalues and eigenvectors for the ground and excited states, respectively, for this (single interacting) TLM case are easily found to be

$$\lambda_o = c_j - |\kappa|\sqrt{n_o} \tag{5.36a}$$

$$\overline{|\frac{1}{2}, c_j, 0 >}_f = \frac{1}{\sqrt{2}}\left[|c_j - \frac{1}{2} > |\frac{1}{2}, \frac{1}{2} > e^{i\varphi} + |c_j + \frac{1}{2} > |\frac{1}{2}, -\frac{1}{2} >\right] \tag{5.36b}$$

$$\lambda_1 = c_j + |\kappa|\sqrt{n_o} \tag{5.36c}$$

$$\overline{|\frac{1}{2}, c_j, 1 >}_f = \frac{1}{\sqrt{2}}\left[|c_j - \frac{1}{2} > |\frac{1}{2}, \frac{1}{2} > - |c_j + \frac{1}{2} > |\frac{1}{2}, -\frac{1}{2} > e^{-i\varphi}\right] \tag{5.36d}$$

These eigenvectors are generated from the non-interacting basis states by a rotation of $45^\circ$ in the internal space of TLM plus field. The states $|\overline{r, c, j} >_f$ are constructed in a manner similar to that used by Dicke(7) and make full use of the definitions of the |r, m> states. Construct a state from

the single TLM states such that r and c are constants of motion and that this state has $n_+$ TLMs in the excited state and $n_-$ TLMs in the ground state. The total energy of the system is then

$$\lambda_j^{(f)} = (n_+ + n_-)c_j + (n_+ - n_-)|\kappa|\sqrt{n_o} = c - 2|\kappa|\sqrt{n_o}(r - j) \tag{5.36e}$$

The ground state is represented by j=0 and $n_-$=N=2r. The index m is not used here for $n_+ - n_-$ since m is reserved to represent the internal energy of the N non-interacting TLMs .

If the ground and excited TLM states are represented by - and +, then the state $|\overline{r,c,j}>_f$ is constructed from these states in exactly the same way that the |r,m> states are constructed from $|\uparrow>=|\frac{1}{2},\frac{1}{2}>$ and $|\downarrow> = |\frac{1}{2},-\frac{1}{2}>$ states. It is more difficult, however, to give the $|\overline{r,c,j}>_f$ in terms of the |n>|r,m> states hence these constructions were found by inspection. For example, consider the two TLM cases. States are given by

$$\overline{|1,c,2>}_f = |+>|+>,$$
$$\overline{|1,c,1>}_f = \frac{1}{\sqrt{2}}(|+>|-> + |->|+>),$$
$$\overline{|1,c,0>}_f = |->|->, \tag{5.37}$$

and

$$\overline{|0,c,0>}_f = \frac{1}{\sqrt{2}}(|+>|-> - |->|+>).$$

In terms of the |r, m > states, upon inserting the photon states, these are

$$|\mathbf{1},\boldsymbol{c},\mathbf{2}>_f=\frac{\mathbf{1}}{\mathbf{2}}\{|\uparrow>|\uparrow>+\boldsymbol{e}^{\boldsymbol{i}(\boldsymbol{\pi}-\boldsymbol{\varphi})}[|\uparrow>|\downarrow>+|\downarrow>|\uparrow>]+\boldsymbol{e}^{\mathbf{2}\boldsymbol{i}(\boldsymbol{\pi}-\boldsymbol{\varphi})}|\downarrow>|\downarrow>\}$$
$$=\frac{\mathbf{1}}{\mathbf{2}}\{|\boldsymbol{c}-\mathbf{1}>|\mathbf{1},\mathbf{1}>+\sqrt{\mathbf{2}}\boldsymbol{e}^{\boldsymbol{i}(\boldsymbol{\pi}-\boldsymbol{\varphi})}|\boldsymbol{c}>|\mathbf{1},\mathbf{0}>+\boldsymbol{e}^{\mathbf{2}\boldsymbol{i}(\boldsymbol{\pi}-\boldsymbol{\varphi})}|\boldsymbol{c}+\mathbf{1}>|1,-1>\};$$

$$|\mathbf{1},\boldsymbol{c},\mathbf{1}>_f=\frac{\mathbf{1}}{\mathbf{2}\sqrt{\mathbf{2}}}\{|\uparrow>|\uparrow>\boldsymbol{e}^{\boldsymbol{i\varphi}}+|\uparrow>|\downarrow>+e^{\boldsymbol{i}(\boldsymbol{\pi}-\boldsymbol{\varphi})}|\downarrow>|\uparrow>\boldsymbol{e}^{\boldsymbol{i\varphi}}+\boldsymbol{e}^{\boldsymbol{i}(\boldsymbol{\pi}-\boldsymbol{\varphi})}|\downarrow>|\downarrow>\}$$
$$=\frac{\mathbf{1}}{\mathbf{2}\sqrt{\mathbf{2}}}\{|\uparrow>|\uparrow>\boldsymbol{e}^{\boldsymbol{i\varphi}}+|\uparrow>|\downarrow>\boldsymbol{e}^{\boldsymbol{i\pi}}+|\downarrow>|\uparrow>+\boldsymbol{e}^{\boldsymbol{i}(\boldsymbol{\pi}-\boldsymbol{\varphi})}|\downarrow>|\downarrow>\}$$
$$=\frac{\mathbf{1}}{\mathbf{2}\sqrt{\mathbf{2}}}\{\mathbf{2}|\boldsymbol{c}-\mathbf{1}>|\mathbf{1},\mathbf{1}>\boldsymbol{e}^{\boldsymbol{i\varphi}}+\sqrt{\mathbf{2}}(\mathbf{1}+\boldsymbol{e}^{\boldsymbol{i\pi}})|\boldsymbol{c}>|\mathbf{1},\mathbf{0}>+\mathbf{2}\boldsymbol{e}^{\boldsymbol{i}(\boldsymbol{\pi}-\boldsymbol{\varphi})}|\boldsymbol{c}+\mathbf{1}>|1,-1>\}$$
$$=\frac{\mathbf{1}}{\sqrt{\mathbf{2}}}\{|\boldsymbol{c}-\mathbf{1}>|\mathbf{1},\mathbf{1}>\boldsymbol{e}^{\boldsymbol{i\varphi}}+\boldsymbol{e}^{\boldsymbol{i}(\boldsymbol{\pi}-\boldsymbol{\varphi})}|\boldsymbol{c}+\mathbf{1}>|1,-1>\}; \tag{5.38}$$

$$|\mathbf{1},\boldsymbol{c},\mathbf{0}>_f=\frac{\mathbf{1}}{\mathbf{2}}\{|\uparrow>|\uparrow>\boldsymbol{e}^{\mathbf{2}\boldsymbol{i\varphi}}+\boldsymbol{e}^{\boldsymbol{i\varphi}}(|\uparrow>|\downarrow>+|\downarrow>|\uparrow>)+|\downarrow>|\downarrow>\}$$
$$=\frac{\mathbf{1}}{\mathbf{2}}\{|\boldsymbol{c}-\mathbf{1}>|\mathbf{1},\mathbf{1}>\boldsymbol{e}^{\mathbf{2}\boldsymbol{i\varphi}}+\sqrt{\mathbf{2}}\boldsymbol{e}^{\boldsymbol{i\varphi}}|\boldsymbol{c}>|1{,}0>+|c+1>|1,-1>\}$$

and

$$|\mathbf{0},\boldsymbol{c},\mathbf{0}>_f=\frac{\mathbf{1}}{\mathbf{2}\sqrt{\mathbf{2}}}\{(\mathbf{1}-\boldsymbol{e}^{\boldsymbol{i\pi}})(|\uparrow>|\downarrow>-|\downarrow>|\uparrow>)\}=|\boldsymbol{c}>|0{,}0>$$

The general state $|r,c,j>_f$ for r = N/2 is now given. This state may be generalized for states with r <N/2 since the form of these states, with r≤N/2, in terms of |r, m> states is exactly the same as for states where $r=r', r'=N'/2$, $N'\neq$N.

$$|\overline{r,c,\mathrm{j}}>_f=\frac{1}{2^{N/2}\sqrt{\frac{N!}{n_+!\,n_-!}}}\sum_\rho\left\{\prod_{k=1}^{n_+}\left(|c_j-\frac{1}{2}>|\frac{1}{2},\frac{1}{2}>+|c_j+\frac{1}{2}>|\frac{1}{2},-\frac{1}{2}>e^{i(\pi-\varphi)}\right)_k\right.$$
$$\left.\times\prod_{k'=1}^{n_-}\left(|c_j-\frac{1}{2}>|\frac{1}{2},\frac{1}{2}>e^{i\varphi}+|c_j+\frac{1}{2}>|\frac{1}{2},-\frac{1}{2}>\right)_{k'}\right\},r=\frac{N}{2} \tag{5.39}$$

where $\rho$ represents the various permutations on the ground and excited states and k and $k'$ represent the different excited and ground state TLMs interacting with the field. This state may be represented by a double sum

$$|\overline{r,c,j}>_f = \frac{1}{2^{N/2}\sqrt{\frac{N!}{n_+!\,n_-!}}}\sum_{k=0}^{n_+}\sum_{k'=0}^{n_-} e^{ik(\pi-\varphi)}e^{i(n_- - k')\varphi}$$

$$\times\left[\frac{n_+!}{k!\,(n_+ - k)!}\right]\times\left[\frac{n_-!}{k'!\,(n_- - k')!}\right]\times\left[\frac{N!}{n_+!\,n_-!}\right]\div\left[\frac{N!}{(k+k')!\,(N-k-k')!}\right] \qquad (5.40)$$

$$\times\sqrt{\frac{N!}{(k+k')!\,(N-k-k')!}}\,|c-(r-(k+k'))>|r,r-(k+k')>.$$

The first phase factor and first factorial term comes from the phase and number of terms encountered in the product over the individual excited state TLMs interacting with the field. The second phase factor and second factorial term comes from the product over individual ground states. The third factorial term comes from the number of permutations over the $n_+$ and $n_-$ excited and ground states. The fourth factorial term comes from the number of terms necessary to form the |r,m> states and the square root term is necessary so that the |r,m> states are normalized. Performing a cancellation of terms in Eq. (5.40) gives

$$|\overline{r,c,j}>_f = \frac{1}{2^{N/2}\sqrt{\frac{N!}{n_+!\,n_-!}}}\sum_{k=0}^{n_+}\sum_{k'=0}^{n_-} e^{ik(\pi-\varphi)}e^{i(n_- - k')\varphi}$$

$$\times\frac{(k+k')!\,(N-(k+k')!)}{k!\,(n_+ - k)!\,k'!\,(n_- - k')!}\sqrt{\frac{N!}{(k+k')!\,(N-(k+k'))!}} \qquad (5.41)$$

$$\times|c-(r-(k+k'))>|r,r-(k+k')>.$$

Define L=k+$k'$; m=r-L, $L' = k - k'$.Also note that $n_+ = j, n_- = N - j$, and $N = 2r$. Therefore

$$|\overline{r,c,j}>_f = \frac{1}{2^r\sqrt{\frac{(2r)!}{j!\,(2r-j)!}}}\sum_{L=0}^{2r}\sum_{L'=down}^{up'} e^{\frac{(L+L')\pi i}{2}}[L!\,(2r-L)!]$$

$$\div\left[\left(\frac{L+L'}{2}\right)!\left(\frac{L-L'}{2}\right)!\left(j-\left(\frac{L+L'}{2}\right)\right)!\left(n_- - \left(\frac{L-L'}{2}\right)\right)!\right] \quad (5.42a)$$

$$\times\, e^{(n_- - L)\varphi}\sqrt{\frac{(2r)!}{L!\,(2r-L)!}}\,|c-m>|r,m>\,,$$

where

$$n_- = 2r - j\,, \quad (5,42b)$$

$$m = r - L\,, \quad (5.42c)$$

$$down = max[-L, L - 2n_-]\,, \quad (5.42d)$$

$$up = min[L, 2j - L]\,. \quad (5.42e)$$

and the prime on the second sum indicates that only every other value starting with "down" is used in the sum. Equation (5.42a) is the general $|\overline{r,c,j}>_f$ state expressed in terms of |c-m>|r, m> states even for r≠N/2. In these terms the ground state is

$$|\overline{r,c,0}>_f = \frac{1}{2^r}\sum_{L=0}^{2r} e^{i(2r-L)\varphi}\sqrt{\frac{(2r)!}{L!\,(2r-L)!}}\,|c-m>|r,m> \quad (5.43a)$$

with energy

$$\lambda_o^{(f)} = c - 2r|\kappa|\sqrt{n_o}\,, \tag{5.43b}$$

and the most excited state is

$$|\overline{r,c,2r}>_f = \frac{1}{2^r}\sum_{L=0}^{2r} e^{iL(\pi-\varphi)}\sqrt{\frac{(2r)!}{L!\,(2r-L)!}}|c-m>|r,m>\,, \tag{5.43c}$$

with energy

$$\lambda_{2r}^{(f)} = c + 2r|\kappa|\sqrt{n_o}\,. \tag{5.43d}$$

In the present approximation, which is valid when c ≥ 5r, it is seen that <m>≡0. This may be understood by realizing that the average of "m" in both the excited and ground states for the single TLM interacting with the field is zero; therefore, the constructed states $|\overline{r,c,j}>_f$ must also have <m> = 0. This means that $n_o \cong c$ and that

$$\lambda_j^{(f)} = c - 2(r-j)|\kappa|\sqrt{n_o}\,. \tag{5.44}$$

It is of interest to note that a differential equation may also be obtained from a difference equation within the present approximation. The technique is identical to that used in approximation "A" of this section. The eigenvalues obtained in this way are given by

$$\left(\lambda_f\right)_{diff} = c - 2|\kappa|\sqrt{n_o}\left[-\left(j+\frac{1}{2}\right) + \sqrt{\left(j+\frac{1}{2}\right)^2 + \left(\frac{r+1}{2}\right)^2}\right]. \tag{5.45}$$

By comparing the ground state energies calculated by Eq. (5.45), Eq. (5.44), or by Eq. (5.16), (the last of which is obtained within the framework of the differential equation approximation for the exact Hamiltonian), it is seen that all three results agree for $c \geq r$ but disagree markedly for $c<r$. When $c \geq 5r$ the extent of agreement between the exact eigenvalues and those calculated from Eq. (5.44) is to at least three significant figures, and the "exact" and "approximate" eigenvectors are almost indistinguishable. The above approximations for $c>r$ could be interpreted as the N-TLMs being bathed in a large photon flux such that the TLMs are in thermal equilibrium, with exactly half the TLMs excited (infinite positive temperature.) Of course, in this approximation all questions of spontaneous emission effects are completely ignored. It is also of interest that $\sigma^2$ for the ground state and most excited state equals r/2 as in Table 1.

## C. Modified TLM Approach

In Section B an approximation valid for $c > r$ has been discussed. An approximation valid for $c < 0$ would also be convenient. Toward this end rewrite Eq. (3.4) as

$$H = R_3 + a^{\dagger}a - \kappa\sqrt{R^2 - {R_3}^2 + R_3}\Gamma_+ a - \kappa^* \Gamma_- \sqrt{R^2 - {R_3}^2 + R_3} a^{\dagger} \,, \tag{5.46}$$

where

$$\Gamma_+ |r, m > = |r, m+1 > \,, \tag{5.47a}$$

$$\Gamma_+ |r, r > = 0 \,, \tag{5.47b}$$

$$\Gamma_- |r, m > = |r, m-1 > \,, \tag{5.47c}$$

$$\Gamma_- |r, -r > = 0 \,. \tag{5.47d}$$

In order to approximate the TLM part of the interaction, namely to modify the operator $\sqrt{R^2 - {R_3}^2 + R_3}$, it is necessary to define an operator $\hat{L}$ =$R_3$+r such that $\hat{L}\,|r, m> = (m+r)|r, m> = \ell|r, m>$. Since c < 0, m can be assumed to be near -r so that $\ell << r$. Replace the operator $R^2$ by r(r+1) in $\sqrt{R^2 - {R_3}^2 + R_3}$ obtaining$\sqrt{(\mathrm{r} - \mathrm{R}_3)(\mathrm{r} + \mathrm{R}_3 + 1)}$. In terms of $\hat{L}$ this becomes $\sqrt{(2\mathrm{r} - \hat{L})(\hat{L} + 1)}$ and the approximation consists of dropping the terms due to $\hat{L}^2$ and $\hat{L}$ so that $\sqrt{R^2 - {R_3}^2 + R_3} \to \sqrt{2\mathrm{r}(\hat{L} + 1)}$ Then Eq. (5.46) becomes

$$H_T = R_3 + a^\dagger a - \kappa\sqrt{2r\big(\hat{L} + 1\big)}\Gamma_+ a - \kappa^* \Gamma_- \sqrt{2r\big(\hat{L} + 1\big)}a^\dagger\,. \tag{5.48}$$

There exists a one-to-one correspondence between the interaction terms of this Hamiltonian expressed in terms of the states $|0>|r,c>;|1>|r,c-1>;\cdots|c-r>\ |r,-r>$ and the interaction terms of the Hamiltonian

$$H_f = R_3 + a^\dagger a - \kappa\sqrt{2r}R_+\mathcal{L} - \kappa^*\sqrt{2r}R_-\mathcal{L}^\dagger\,. \tag{5.49}$$

expressed in terms of the states $|0>\ |r'\ r'>;\ |1>|r', r'-1>;\cdots|\ r'>|\ r',- r'>$ where $r'=\frac{c+r}{2}$. From the correspondence to Section B of this chapter, it can be seen that the deviation from the average energy c for the Hamiltonian in Eq. (5.49) is $-2(r'-j)\sqrt{2r}$ where j runs from 0 to (2r') and 2r replaces $n_o$ in Eq. (5.36). Therefore

$$\lambda_j^{(T)} = c - (c + r - 2j)|\kappa|\sqrt{2r}\,. \tag{5.50}$$

where j = 0,1,⋯c+r.

Also from this analogy the states $|\overline{r, c, j}>_T$ are given

$$|\overline{r,c,j}>_T = \frac{1}{2^{(c+r)/2}\sqrt{\frac{(c+r)!}{j!\,(c+r-j)!}}} \sum_{L=0}^{c+r} \sum_{L'=down_T}^{up_T'} e^{\frac{(L+L')\pi i}{2}} [L!\,(c+r-L)!]$$

$$\div \left[\left(\frac{L+L'}{2}\right)!\left(\frac{L-L'}{2}\right)!\left(j-\left(\frac{L+L'}{2}\right)\right)!\left(n_- - \left(\frac{L-L'}{2}\right)\right)!\right] \qquad (5.51)$$

$$\times e^{(n_- - L)\varphi} \sqrt{\frac{(c+r)!}{L!\,(c+r-L)!}}\,|L>|r,c-L>,$$

where

$$n_- = c + r - j\,, \qquad (5.52a)$$

$$down_T = max[-L, L-2n_-]\,, \qquad (5.52b)$$

$$up_T = min[L, 2j-L]\,, \qquad (5.52c)$$

and the prime on the second sum again means sum over only every other term starting with $L' = down_T$.

The ground state is

$$|\overline{r,c,0}>_T = \frac{1}{2^{(c+r)/2}} \sum_{L=0}^{r+c} e^{i(c+r-L)\varphi} \sqrt{\frac{(c+r)!}{L!\,(c+r-L)!}}\,|L>|r,c-L>, \qquad (5.53a)$$

with energy

$$\lambda_o^{(T)} = c - (c+r)|\kappa|\sqrt{2r}\,. \qquad (5.53b)$$

By consulting Table 1 for the case c < 0 and recalling that $\boldsymbol{\epsilon}$ = c+r, it is easily seen that the ground state energies agree. Also for

c< 0 the effective eigenvalues lie along a straight line in agreement with Eq.(5.50). Of course, for this case the average value of "m" $\neq 0$. In Eq. (5.53) L represents the number of photons so that $<L>=\frac{c+r}{2}$ agreeing with the results in Table 1; also, $\langle L^2\rangle - \langle L\rangle^2 = \frac{c+r}{4}$, again agreeing with the result from the table.

The above approximation may be understood as individual photons interacting with the N-TLMs. Each photon acts independently of the others as if the average energy of the entire system, c, for one photon is -r+1.

In Appendix B other approximations to the Hamiltonian (3.4) for $\omega = \Omega$ will be discussed. The general conclusions are that the approximations discussed there are in poor agreement with the exact results.

# CHAPTER VI: DISCUSSION OF NUMERICAL RESULTS

This chapter contains a discussion of the numerical results obtained from the exact solution and a comparison with the results given by the approximations of the preceding section.

Figures 5 through 9 show $A_n^{(r,c,j)}$ as a function of n for several representative values of r,c,j for the special case of $\beta$ = 0. The value j = 0 corresponds to the ground state, j = 1 to the first excited state, and j = 2r the most highly excited state. The value $q_o$ corresponds to the ground state and is the largest value of q. From these figures it is seen that the states resemble, in form, the harmonic oscillator states[12] .This resemblance led to the differential equation approximation discussed in Chapter V. In these figures the dispersion calculated from the exact results is given. If these values are compared to those calculated from Eq. (5.24) and given in Table 1, it will be seen that the two results agree very well for all cases considered. Figure 10 is a plot showing a comparison between the exact q's and $q_j' s$ given by Eq. (5.16) vs. j for $\beta$= 0. It is easily seen that for the ground and first few excited states the results are in good agreement. As j becomes larger and approaches r, the exact and differential solution results disagree markedly. The reason for this is due to the use of the harmonic oscillator boundary conditions (below Eq. (5. 14)) and the replacement of the second difference by the second differential, rather than neglect of the cubic term in Eq.

(5.12). A second order perturbation calculation on the cubic term indicates that inclusion of this term will correct the eigenvalues in the wrong direction and by a negligible amount when this perturbation calculation is valid.[3]

[3] See Appendix C

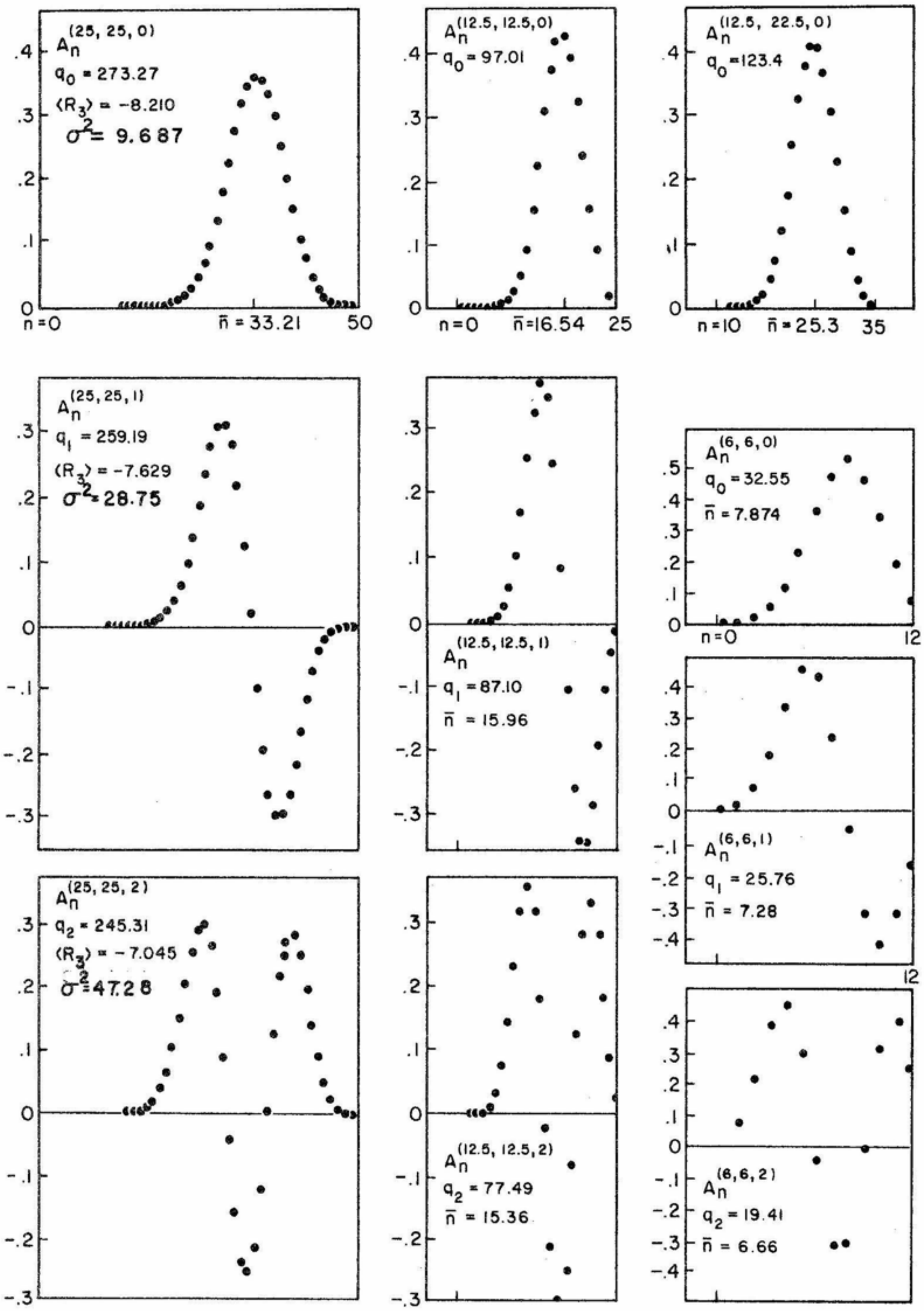

Figure 5: Selected Eigenvectors $\boldsymbol{A_n^{(r,c,j)}}$

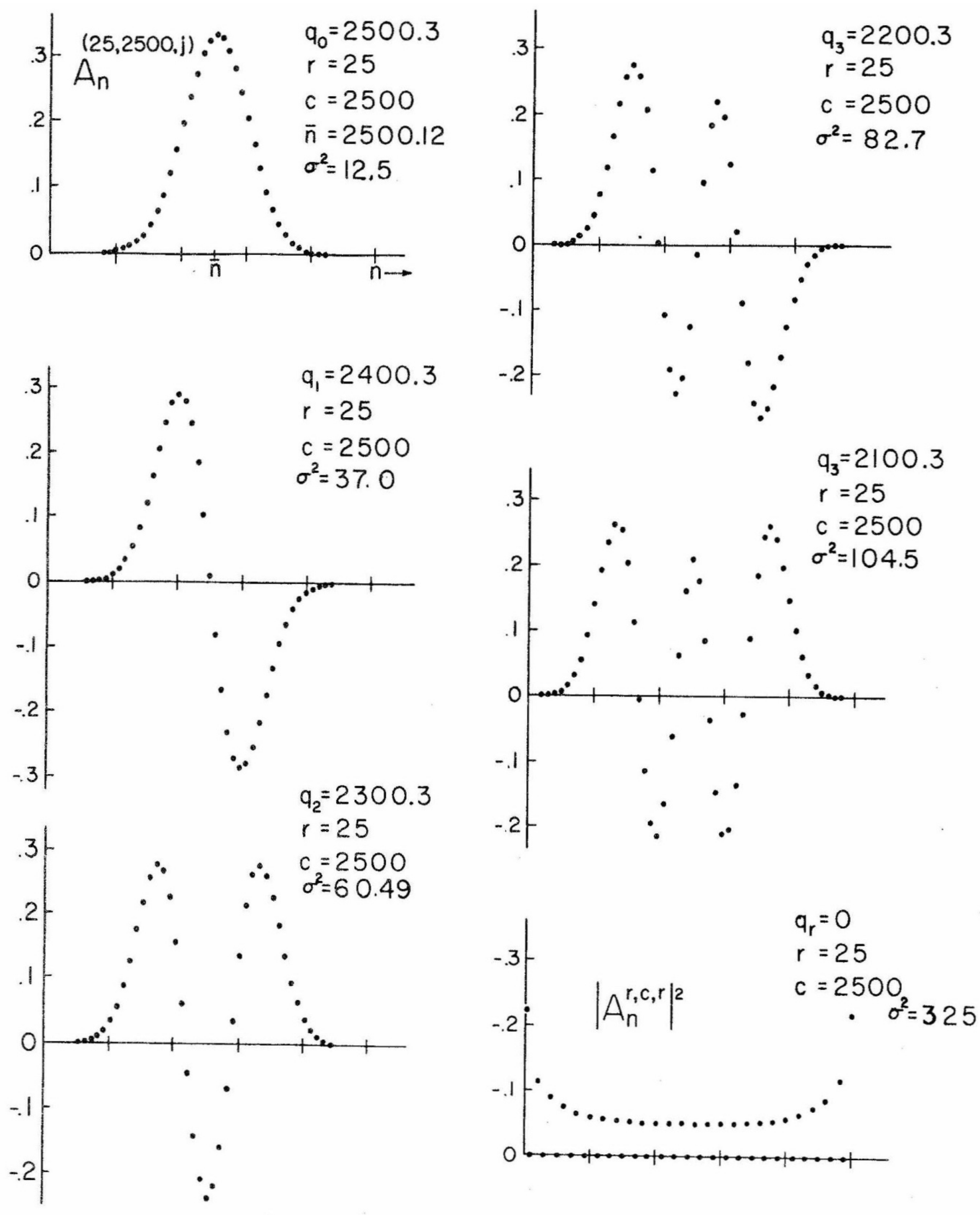

Figure 6: $A_n^{(25,2500,j)}$,j=0, 1, ⋯,4 and $\left|A_n^{(25,2500,25)}\right|^2$

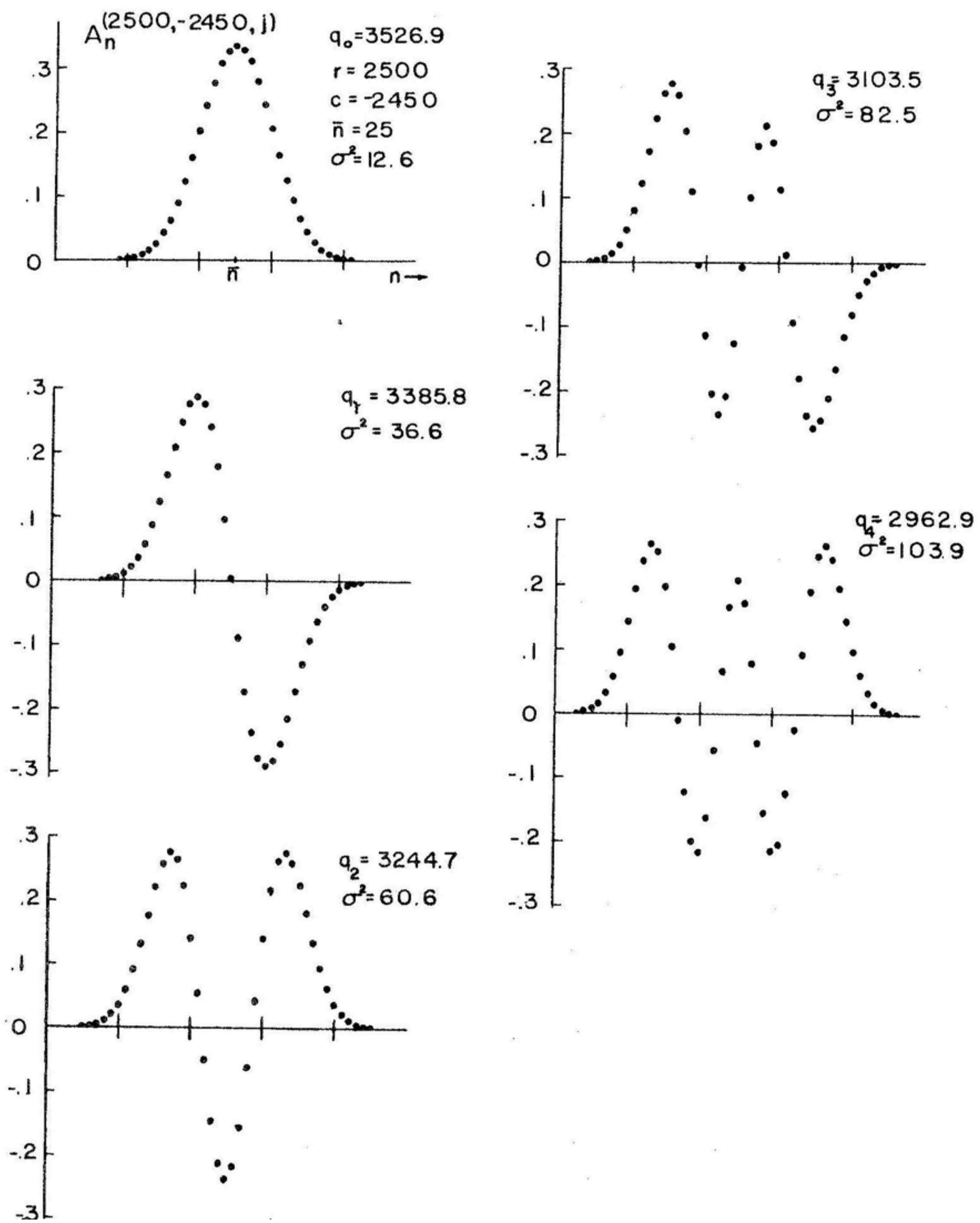

Figure 7: $\boldsymbol{A_n^{(2500,2450,j)}}$,j=0,1,2,3,4

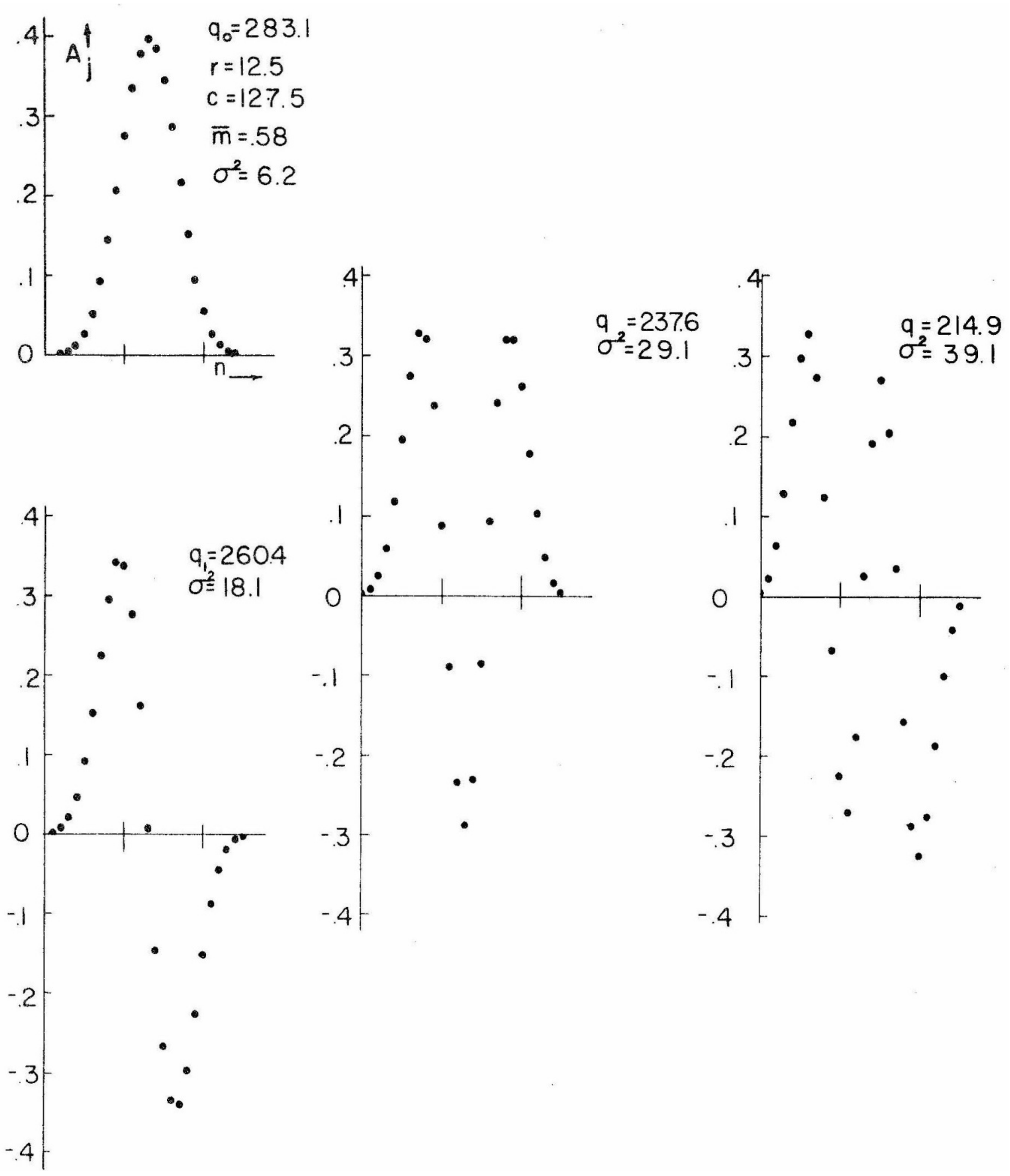

Figure 8: $\mathbf{A_n^{(12.5,127.5,j)}}$,j=0,1,2,3

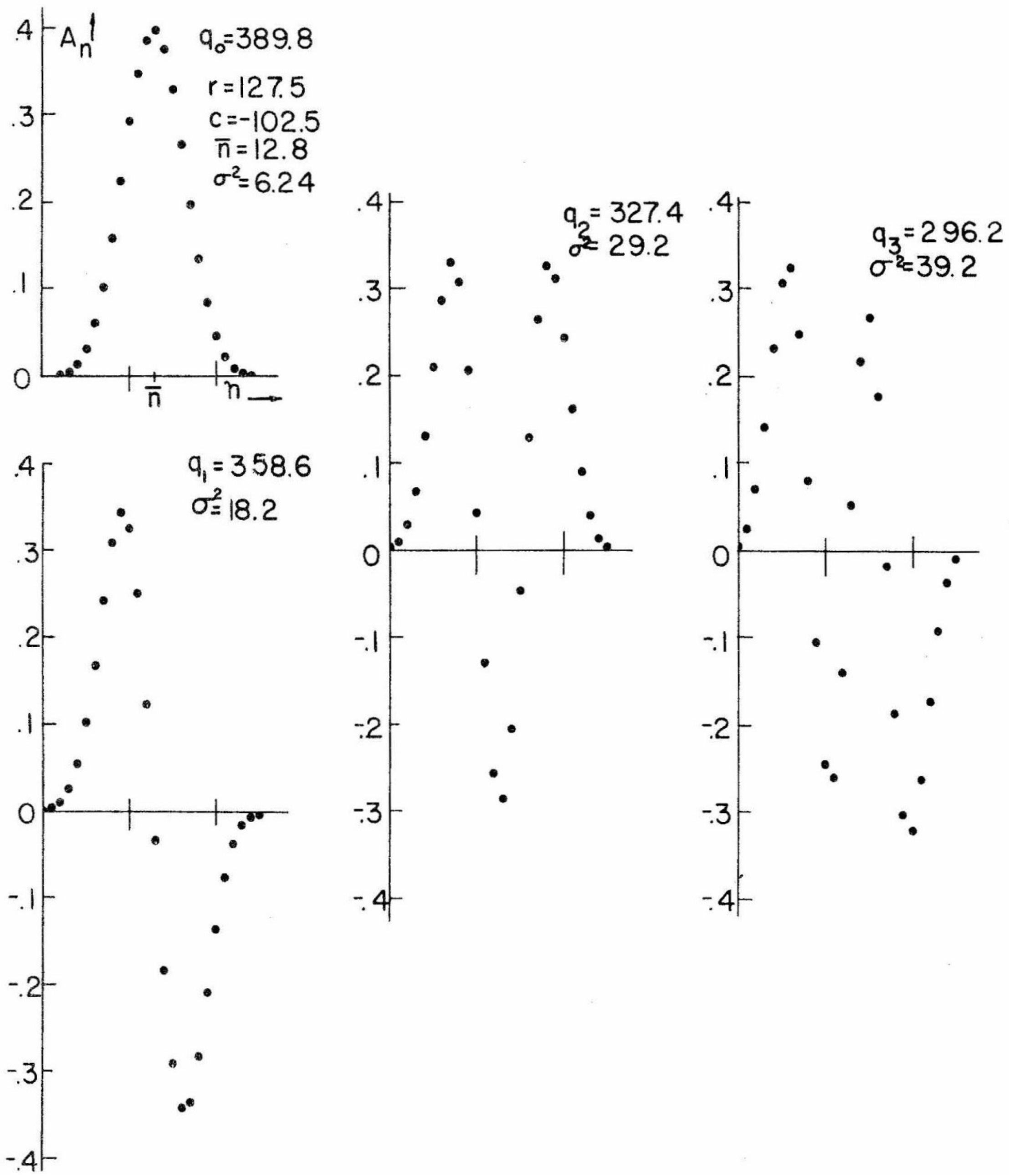

Figure 9: $\mathbf{A_n^{(127.5,-102.5,j)}}$,j=0,1 2,3

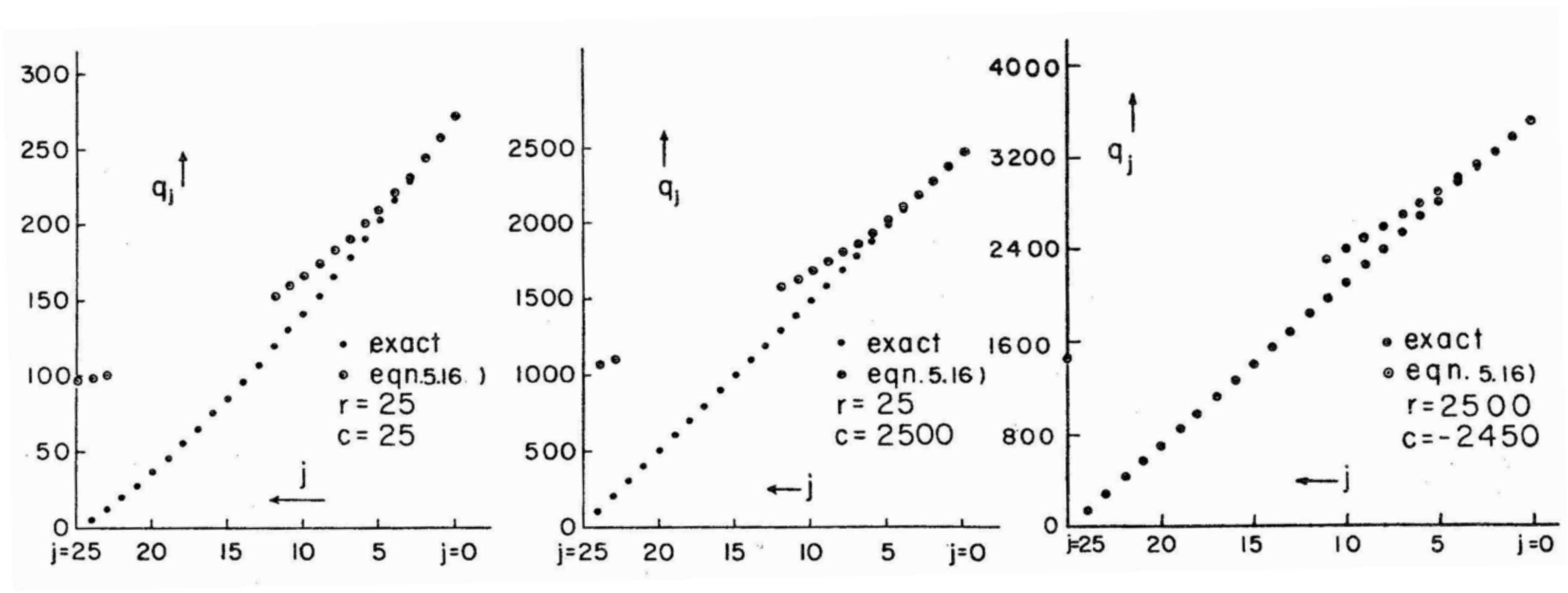

Figure 10: Eigenvalues

In Figure 11 values of q versus c for representative values of j and fixed r are given. Again $\beta = 0$ for this figure. It is seen from Figure 11 that q is proportional to $c^{1/2}$ for c > r. This agrees with the results given in Table 1. Also from Figures 10 and 11 it is noted that except for $c \cong r$, the eigenvalues are very closely linear in the index j, and the eigenvectors resemble the harmonic-oscillator eigenfunctions for all states (Figure 6, especially the lower right-hand corner which gives the squared eigenvector for j = 25). Since the "differential equation approach" did not give a linear eigenvalue spectrum the last two approaches of Chapter V were conceived. The "average field approach," which is essentially treating the problem as if each TLM were independent, gives excellent agreement with the exact solution for c > r. In fact, when $c \geq 5r$ and above the results are nearly identical. This agreement includes both eigenvectors and eigenvalues. For c << r the exact solution again gives a linear effective eigenvalue spectrum. The "modified TLM approach," discussed above, gives good agreement with the exact results for $c = -r + \epsilon$ where $\epsilon \ll r$. As $\epsilon \to r$ or $c \to 0$ the results for this approach do not agree with the exact solution as well as for $\epsilon$ small. The reason for this can be understood in a heuristic manner. For c < 0, the approximation is equivalent to considering each photon interacting independently with the TLM system. That is, the system, as far as each photon sees it, is just one photon and N TLMs in the down state or no photons and just one excited TLM, i. e. , |1>|r,-r> and |0>|r,-r+1> . This is the reason for the term$\sqrt{2r}$ appearing in Eq. (5.50). But as more and more actual photons interact with the system, it would be unreasonable to expect that each photon have N TLMs or equivalently 2r (TLMs)'s with which to interact. Equation (5.50) could then be heuristically modified to take this into account and an approximate formula is given by

$$\lambda = c - (c + r - 2j)\sqrt{2r - \alpha}\,|\kappa| \tag{6.1}$$

where j = 0, 1,⋯,c+r and $\alpha$ is a number representing the number of TLMs in the excited state. If the case c = 0 were considered, then for Eq. (6.1) to agree with the results in Table 1

$$r\sqrt{2r - \alpha} = \frac{2\sqrt{2}(3)^{1/4}r^{3/2}}{3} \tag{6.2}$$

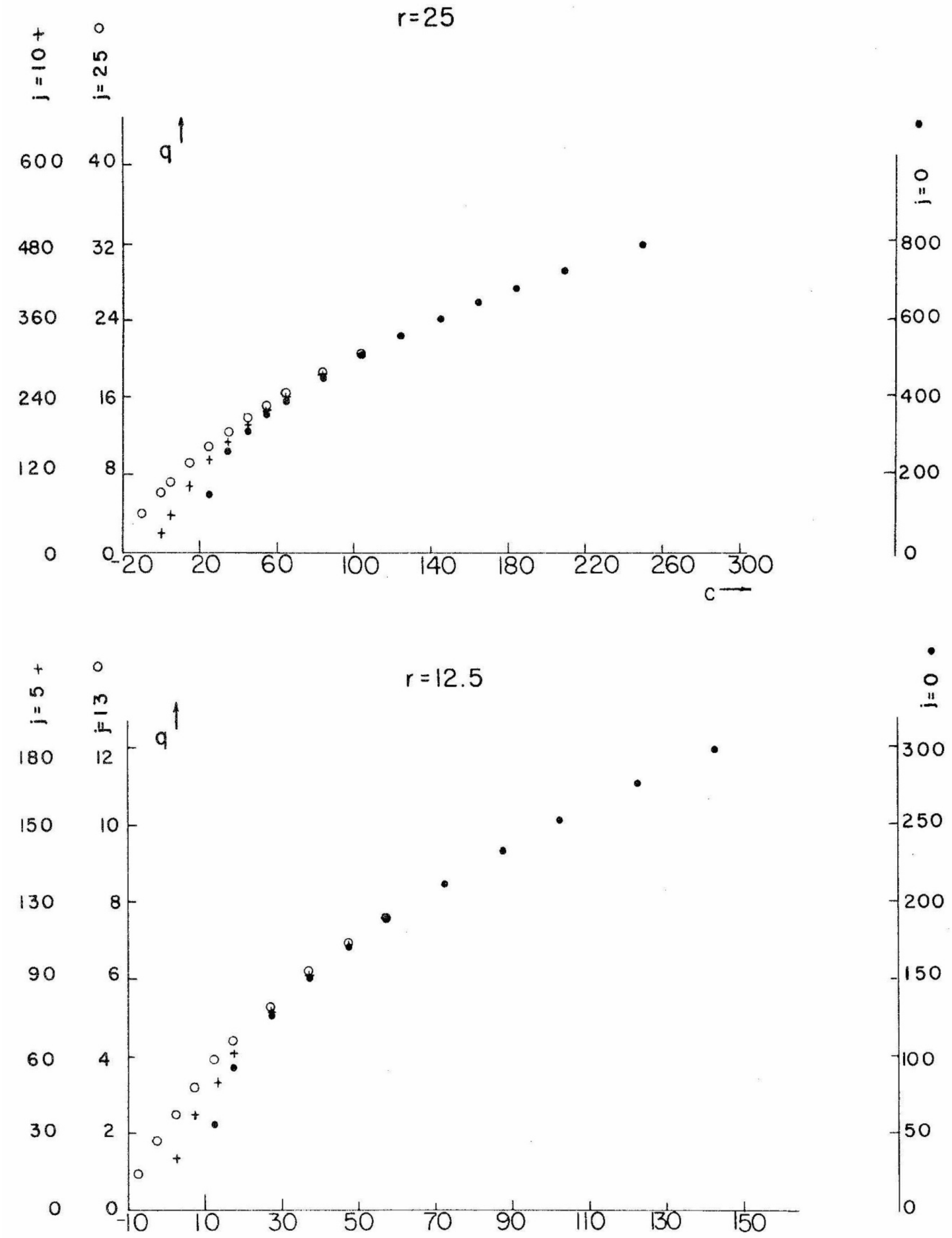

Figure 11: q vs c

This implies that $\alpha$= 0.45r ≅ r/2 or that for c = 0 there is a maximum of r free photons in the system and that approximately half of them are absorbed as excitations on the TLMs. Figures 12 and 13 are eigenvectors calculated from Eq. (5.42) and/or Eq. (5.51). Figure 12 is to be compared with Figures 6 and 7 and Figure 13 with Figures 8 and 9. As is seen, the figures have almost exactly the same form, it is noted however, that the discrepancy between the exact results, displayed in Figures 8 and 9 where c = 10r or

10(r+c) = r, and the results in Figure 13 is

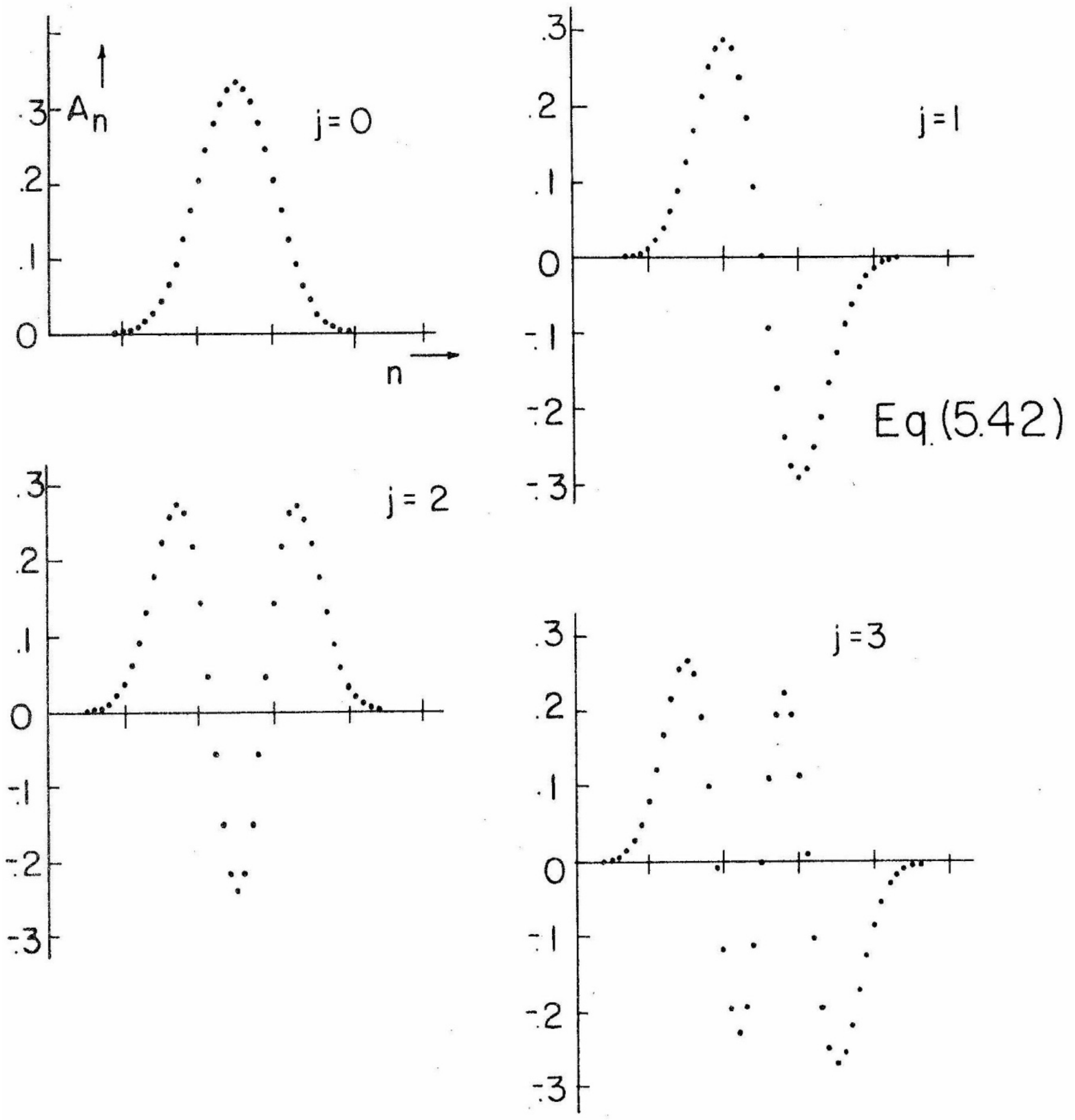

Figure 12: Approximate $\boldsymbol{A_n}$ for r or $\frac{(r+c)}{2}$ =25

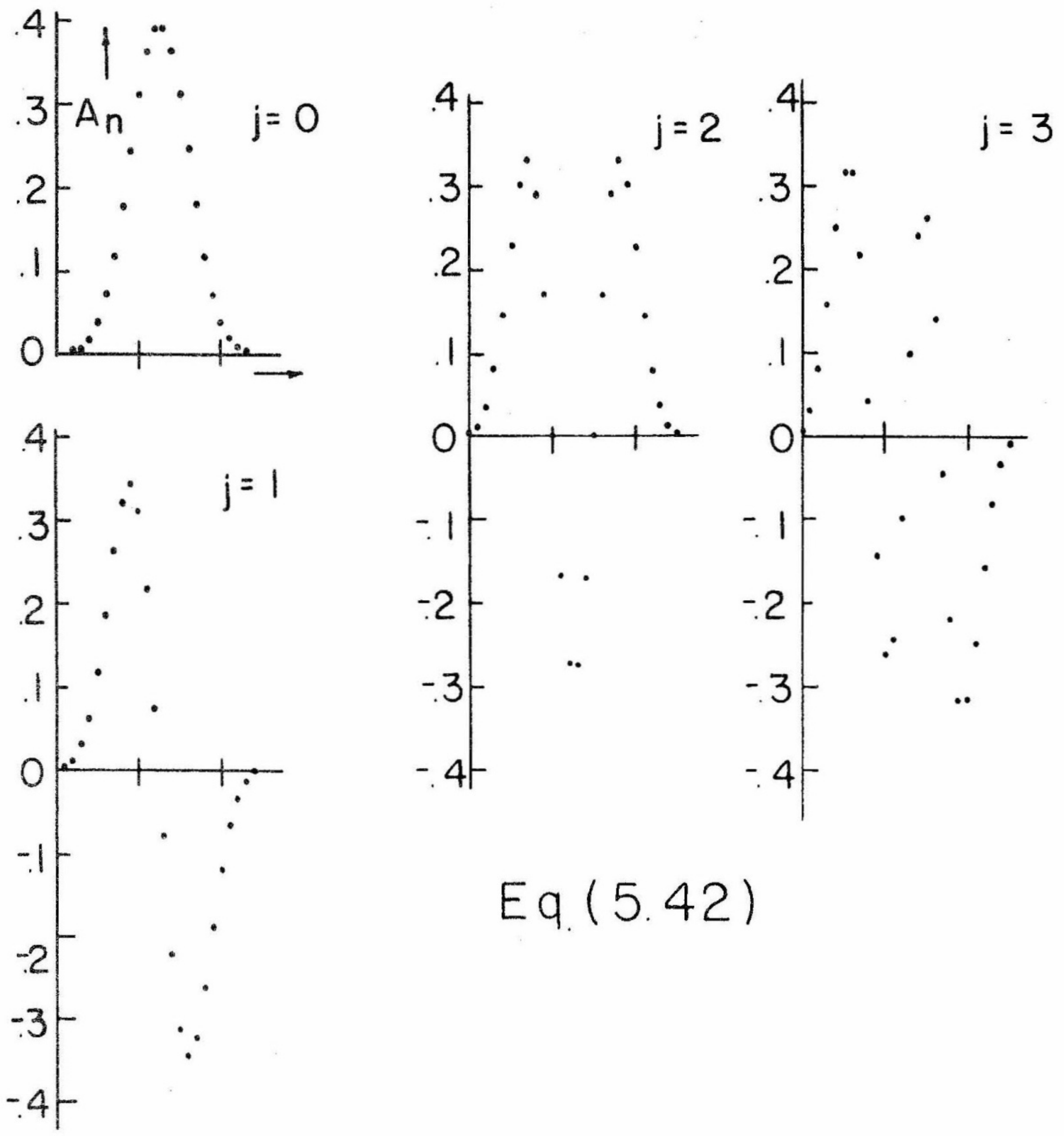

Figure 13: Approximate $\boldsymbol{A_n}$ for r or $\frac{(\boldsymbol{r}+\boldsymbol{c})}{\boldsymbol{2}}$ =12.5

larger than the discrepancy between Figure 12 and Figures 6 and 7, where c = 100r or 100(r+c) =r. The eigenvalues calculated from Eq. (5.44) and Eq. (5.50) agree with the exact results to at least three significant figures for the cases considered. That is, the "average field approach" is very good for c≥5r and the "modified TLM approach" for c = -r+$\epsilon$ where 5$\epsilon$ < r.

For comparison purposes Figures 14 and 15 are given. Figure 14 shows eigenvectors calculated using the "average TLM approach" given in Appendix B for c < r, and Figure 15 shows eigenvectors from the "average field approach" for c < 0. These figures disagree so markedly from the exact results, Figures 8 and 9, that it indicates that this approximation of averaging over the TLM part of the interaction terms of the Hamiltonian is completely invalid, and that for c < r the "average field approach" is also invalid. Numerical results for the "average TLM approach" for c > r are even worse.

Figures 16 and 17 contain values of $A_n^{(r,c,j)}$ versus n and $q_j$ versus j for $\beta \neq 0$ calculated from the exact solution for several values of $\beta$. As $\beta$ becomes large, the N-TLM system and the quantized radiation field slowly decouple. This is seen by comparing Figure 16, for large $\beta$, with Figures 5 through 9. In Figure 16 only one term of each vector is large while the others are very small (in the case of large $\beta$), while in Figures 5 through 9 all components of the vector are relevant. By rewriting Eq. (3.4), the Hamiltonian may be given by

$$H = \omega c + (\Omega - \omega)R_3 - \gamma a R_+ - \gamma^* a^\dagger R_- \,. \tag{6.3}$$

Approximations identical to those applied in Chapter V for c > r and c<0 where $\beta$=0 may be applied to Eq. (6.3). In the case c > r, Eq. (6.3)

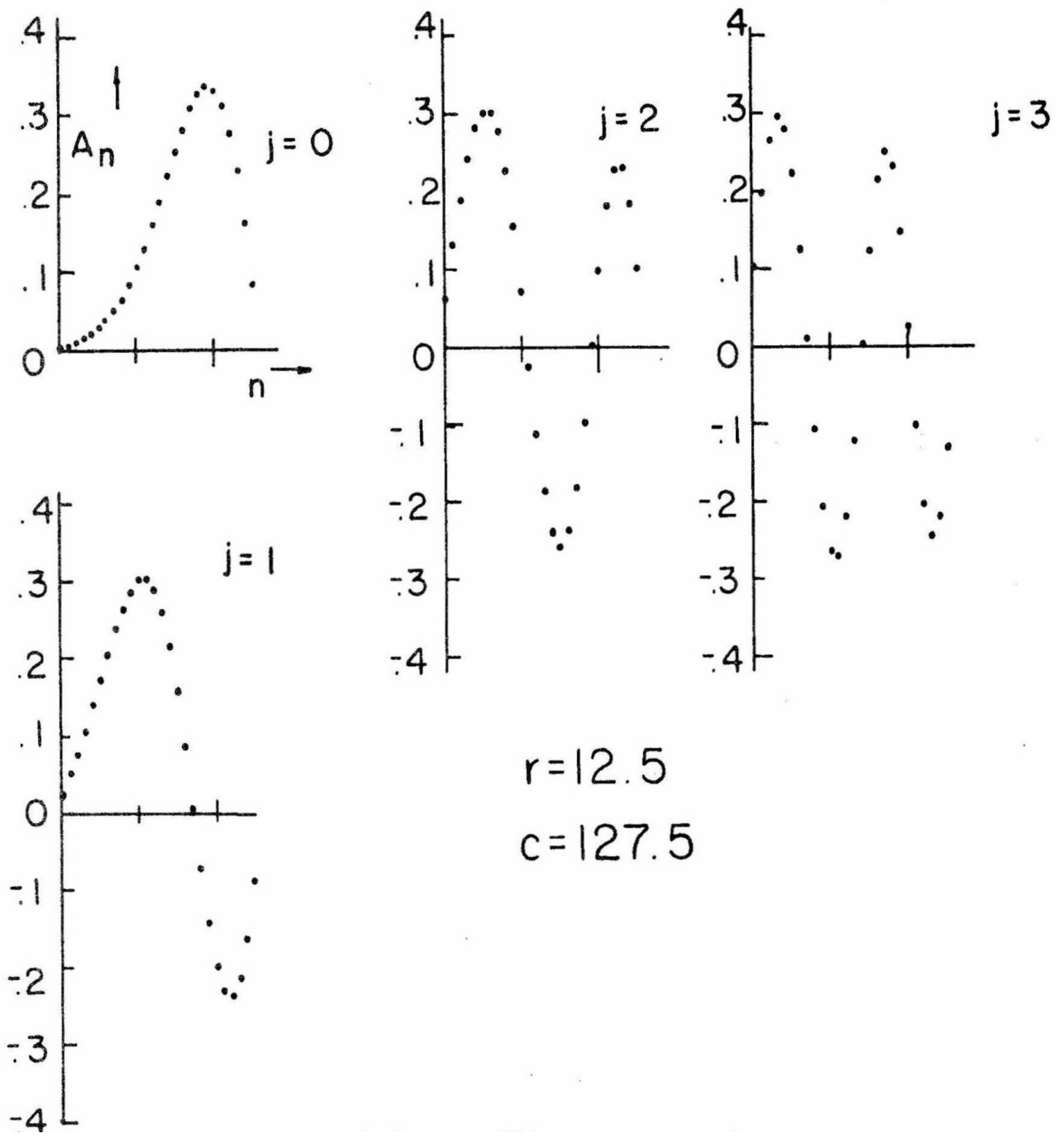

Figure 14: Eigenvectors Average TLM Approach

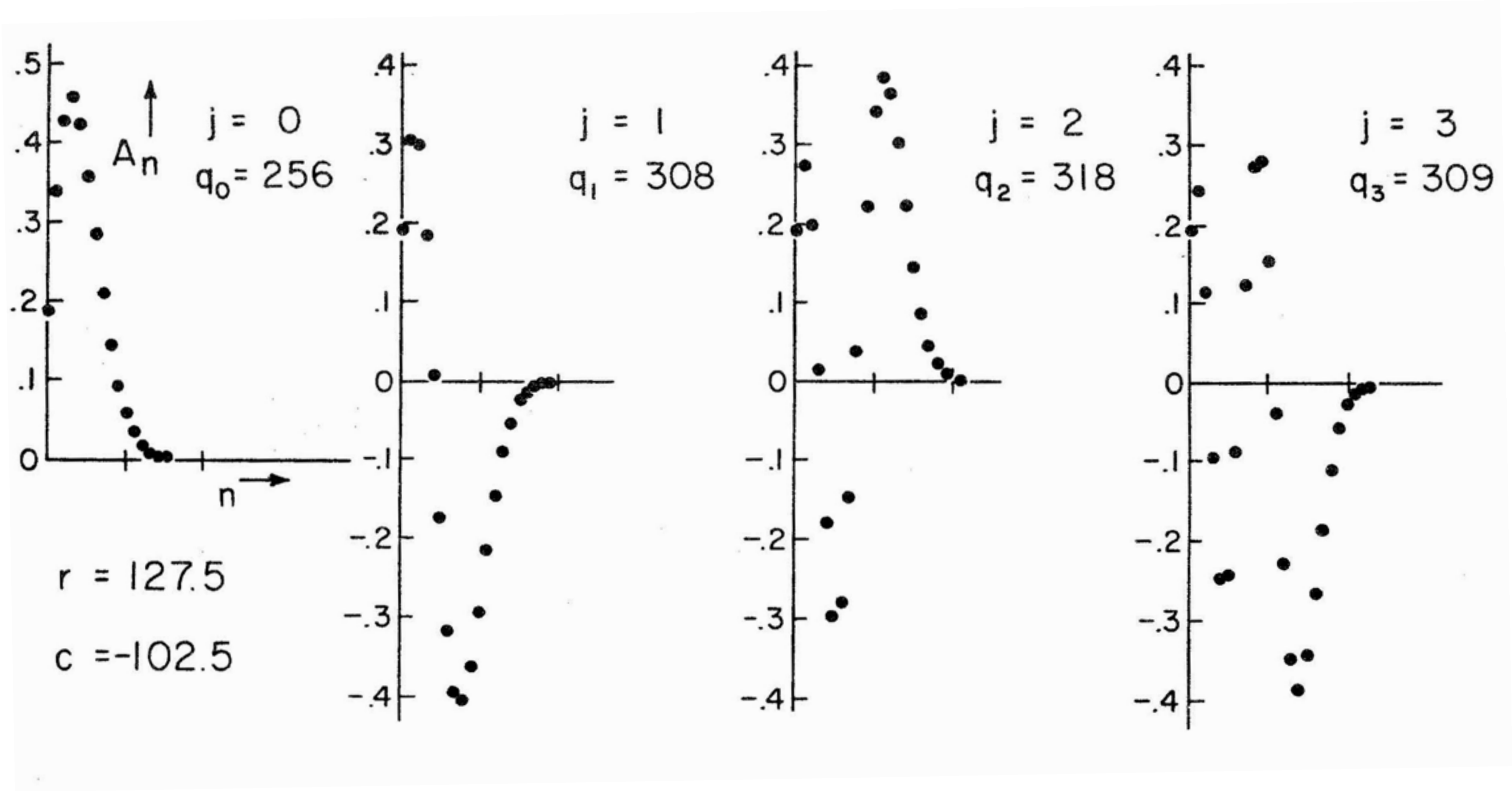

Figure 15: Eigenvectors Average Field Approach

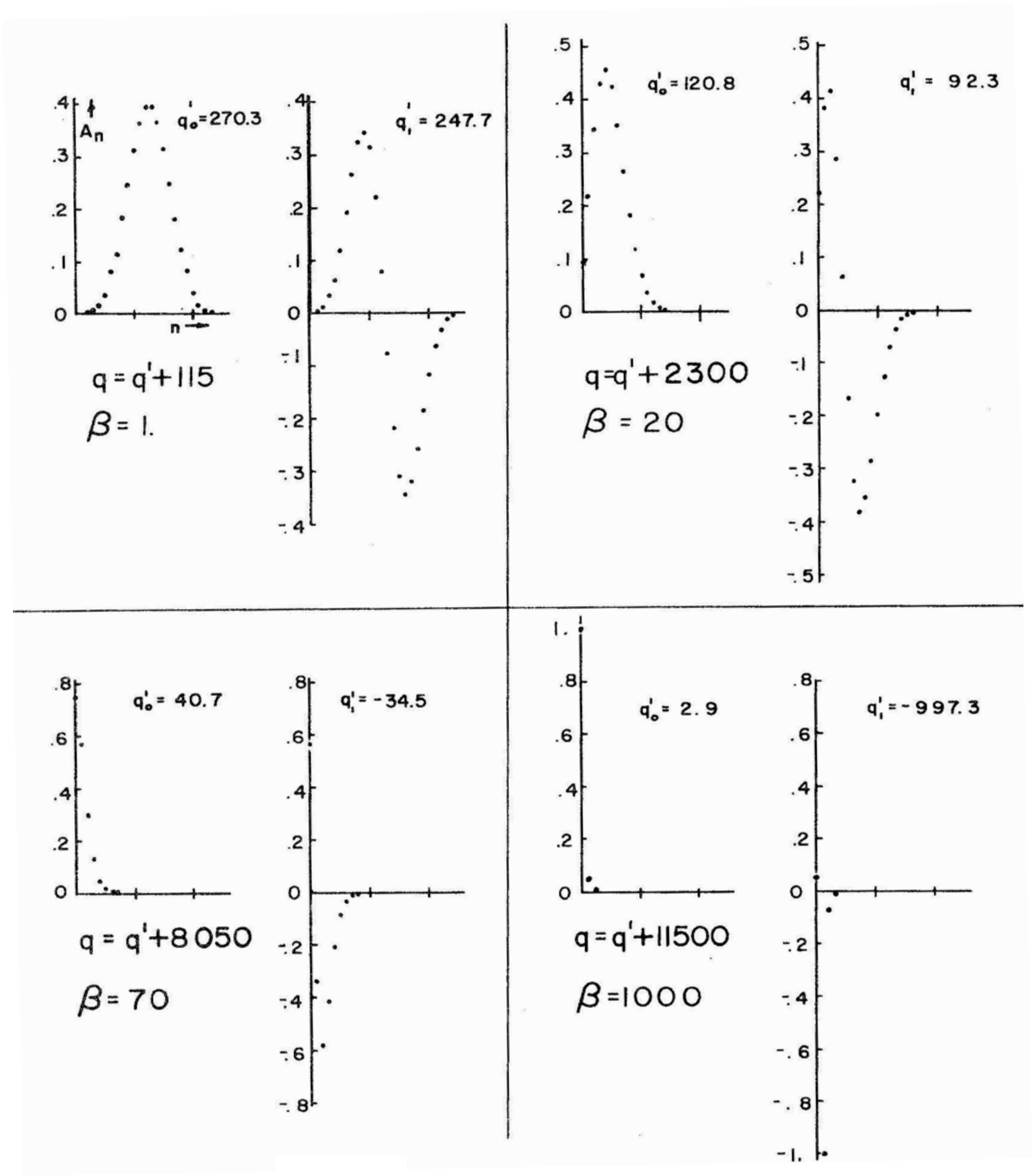

Figure 16: $A_n^{(12.5,127.5,j)}$, j=0, 1 and β≠0

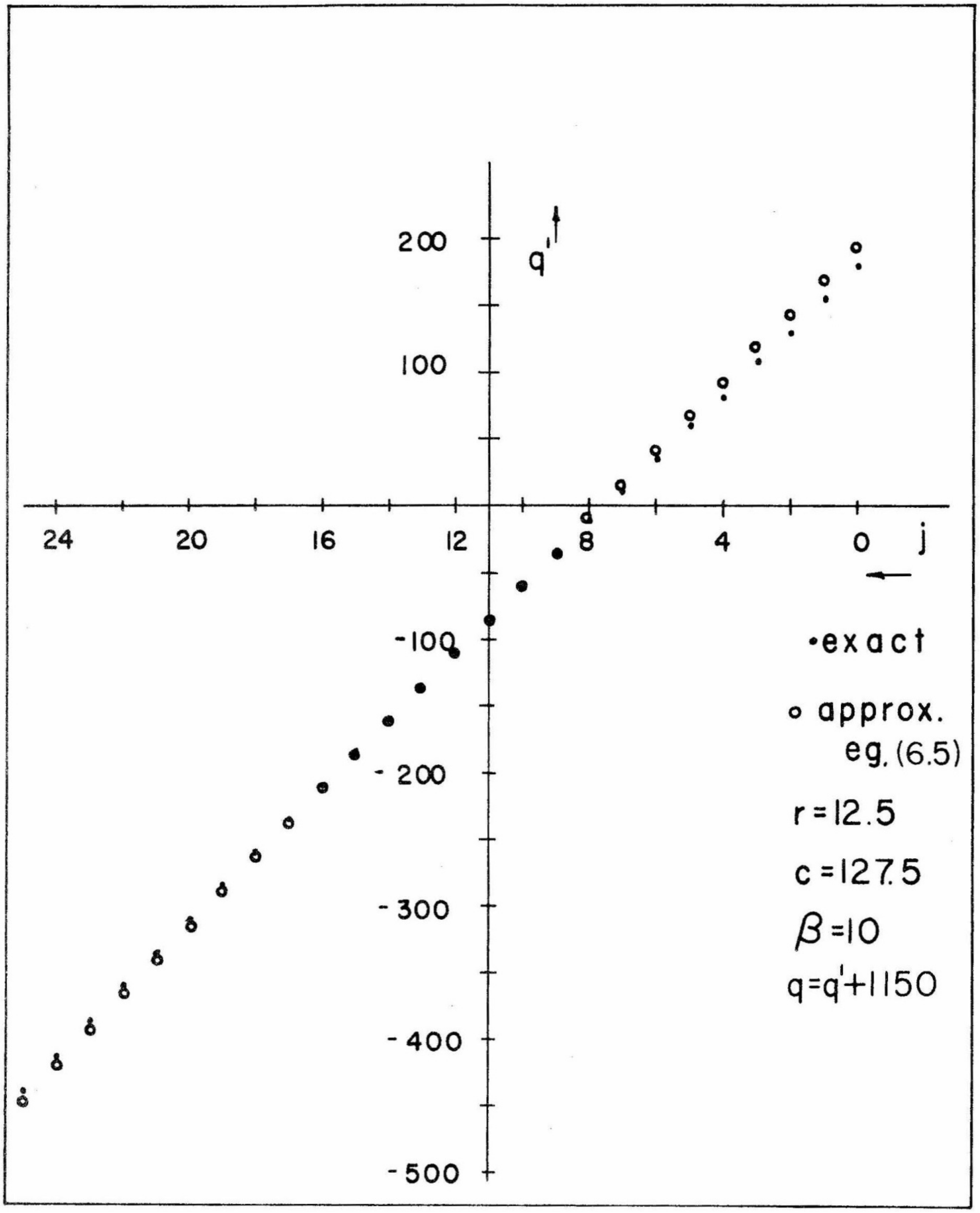

Figure 17: q vs. j for β≠0

becomes, within this approximation,

$$H = \omega c + \sum_{j=1}^{N} (\Omega - \omega) R_{3j} - \gamma \sqrt{n_o} \mathcal{L} R_{+j} - \gamma^* \sqrt{n_o} \mathcal{L}^\dagger R_{j-} \,. \tag{6.4}$$

The eigenvalues and eigenvectors are given by a treatment identical to that used in the "classical field approach" in Appendix B, i. e.,

$$\lambda_j = c - |(\Omega - \omega)|(r - j)\sqrt{1 + 4\left|\frac{\gamma}{\Omega - \omega}\right|^2 n_o} \tag{6.5}$$

and the states $|\overline{r, c, \beta, \mathrm{j}} >_f$ are similar to those given by Eq. (B.22). The difference is that tan $2\theta$ = $2\left|\frac{\gamma}{\Omega-\omega}\right|\sqrt{n_o}$ and that the states $|\overline{r, c, \beta, \mathrm{j}} >_f$ are expanded in terms of the states |r, m> |c-m>. If the relative tuning parameter becomes very large, meaning here that

$$|\beta| = \left|\frac{\Omega - \omega}{\gamma}\right| \gg \sqrt{4 n_o} \tag{6.6}$$

then $\theta \to 0$ and only the term with $L'$ = j-2r will contribute significantly to the state $|\overline{r, c, \beta, \mathrm{j}} >_f$, and

$$|\overline{r, c, \beta, \mathrm{j}} >_f = |r, (r - j)\frac{\beta}{|\beta|} > |c - \frac{\beta(r - j)}{|\beta|} >, \tag{6.7}$$

i.e., the problem is uncoupled. A similar result is obtained for c<0 where

$$\lambda_j = \omega c - \left[\frac{r - c}{2}\right](\Omega - \omega) - |(\Omega - \omega)|\left(\frac{c + r}{2} - j\right)\sqrt{1 + 8\left|\frac{\gamma}{\Omega - \omega}\right|^2 r} \tag{6.8}$$

and where j = 0, 1⋯, c+r. Eigenstates may also be written for this case by replacing r by $\frac{r+c}{2}$ in the expression (B. 22) and by replacing the states |r,m> by the states |L> |r,c-L> . Also, tan $2\theta = 2\left|\frac{\gamma}{\Omega-\omega}\right|\sqrt{2r}$. Uncoupling occurs in this case when

$$|\beta| \gg \sqrt{8r}. \qquad (6.9)$$

In general, if $|\beta|$ is large, a second order perturbation technique is applicable to find the energy shifts. As $|\beta|$ becomes very large, these perturbative results are more accurate than those calculated from Eqs. (6.5) and (6.8). The reason that Eqs. (6.5) and (6.8) do not give such accurate results is the neglect of spontaneous emission for c > r and the neglect of terms as explained above Eq. (6.1) for c<< r. Eqs. (6.6) and (6.9), along with the numerical results, indicate that the N TLMs have an effective "half-width" of interaction of approximately $4\sqrt{2r}|\gamma|\ to\ 4\sqrt{c}|\gamma|$, respectively for c < 0 to c > r. The TLMs are essentially completely uncoupled from the radiation field for detuning much larger than the effective "half-widths," except that the doubling terms neglected in Eq. (3.1) may play an important role in this regime. As $\beta \to 0$ the energies (6.5) and (6.8) go over to the energies (5.36) and (5. 50).

Note that for all figures showing $A_n$, the phase factor $\varphi$ has been set to zero.

# CHAPTER VII: CONCLUSION

We have taken an idealized problem of N Two-Level Molecules interacting with a single mode, quantized radiation field and solved it exactly. By making approximations to this problem we have also obtained analytical solutions and compared these to the exact results. Three of these approximations agree quite well with the exact results under certain conditions. In one approximation, the "differential equation approach," a difference equation for the eigenvectors of the exact solution is replaced by a differential equation. As a result of this approach, the approximate solutions for the eigenvectors are similar to the eigenstates of the harmonic oscillator when $\beta = 0$. This approximation was found to be good for the ground state and first few excited states over all energy ranges considered. A second approximation, a "modified TLM approach," is heuristically equivalent to treating the problem as if the TLMs form a macroscopic molecule which interacts with each photon independently, i.e. , the photons are independent of one another. It is found that this approximation gives good agreement for all eigenvectors and eigenvalues, but that the region of validity for this approximation is that where the energy, c, available to the system must be negative ($c < 0$). In a laser problem, this is the region of positive temperature (no inversion of molecular energy levels). The last of these approximations, the "average field approach," considers the interacting molecules as independent of one another, in so far as they "see" each other only through an averaged electric field. Spontaneous emission effects are ignored completely in this approximation. This last approximation gives good agreement for all the eigenvalues and eigenvectors but is only valid in the high energy regime, i.e. , $c \geq 5r$. This means that the number of photons available to the system is much larger than r. However, the photon number need not be larger than the number of molecules; N, since r may in fact be much less than N (see Appendix D).

These approximations applied to our idealized problem are especially interesting because of the bearing they have on some of the current approximations being used in the theoretical study of laser phenomena. For example, a recent analysis of a quantum coherent device by Scully and Lamb[5] describes the system of N radiating molecules by a factored product of 2x2 matrices instead of the correct matrix of dimension $2^N$. This is equivalent to the "average field" approach discussed above. In addition to explaining the steady state operation of a laser, they also claim their method can be used to explain the build-up of coherent radiation. In the build-up of coherent radiation, the system starts from a state of m positive (inversion) and no photons, and while pumping, builds to a state where m may still be positive but where the number of coherent photons is very large; that is, the system starts in a regime where $0 < c < r$. Therefore, according to the validity ranges given above, it is unlikely that either the "average field" or the "modified TLM" approach is appropriate for this analysis. On the other hand, it is possible that the "differential equation" approach can be used. It is also of interest to note that the conditions of steady state may exclude an independent molecule approximation. Consider two typical systems, a gas laser and a maser, each operating at approximately $10^3$ watts. Assume that for each system the typical dwell time (T) of a photon within the cavity is $10^{-7}$ seconds. This is not physically unreasonable for either system. For example, a gas laser with length of one meter and transmittance of two percent has a dwell time of $1.7 \times 10^{-7}$ seconds; a maser with a Q of 1000 for a frequency of $10^{10}$ cps has a dwell time of $10^{-7}$ seconds. In each of these cases, the number of photons, n, in the cavity is given roughly by $n\hbar\omega = \tau$ (energy output of the system). Therefore, for the laser operating at $5 \times 10^{14}$ cps, a typical value of n would be $5 \times 10^8$ photons in one mode, while for a maser operating at $10^{10}$ cps, a typical value of n would be $1.4 \times 10^{13}$. Typical gas lasers and masers contain $10^{14}$ to $10^{15}$ molecules. In a gas laser, $r \cong \sqrt{\frac{N}{2}}$ *[4] since the inversion, m, for this case is nearly zero, while for an ammonia beam maser $\cong \frac{N}{2}$ . This means that for the gas laser, the independent molecule picture may indeed be valid for steady state conditions, $n > r \Rightarrow c > r$, while for the maser, n≤ r, so that in this case and independent molecule picture is probably incorrect.

The other three approximations considered in this dissertation, the "average TLM approach," the "classical field approach," and the "classical TLM approach," did not agree well with the exact solution of the problem. The validity of these approximations, when applied to more general problems, is not understood well. However, it is felt that it would be incorrect to ignore the cooperative effects of the TLMs as is done in the "classical TLM approach," which averaged out the molecules completely. It is noted that in an actual problem r, the cooperation number, and c, the average energy of the system (when $\beta$= 0), will probably not remain good quantum numbers and their variations may be one of the important aspects of the problem.

[4] See Appendix D

# APPENDIX A

## NUMERICAL TECHNIQUE FOR THE EXACT SOLUTION[5]

Equations (3.11), (4.2), (4.4), (4.5), and (4.6) were used to find a numerical solution for the eigenvalues and eigenvectors of the Hamiltonian. The characteristic equation

$$B_{r+c+1} = 0 = \sum_{l=0}^{t/2} (-1)^l \mathcal{S}_l^{(t-1)} \tag{A.1}$$

was generated on the computer through use of the recursion relations (4.5). The steps in this development are as follows: The polynomial $\mathcal{S}_o^n$ is formed from $\mathcal{S}_o^{n-1}$ by multiplication by the factor (q+$\beta$n). The entire polynomial set $\{\mathcal{S}_o^n\} - 1 \leq n \leq t-1$ is formed in this manner as seen by the examples (4.4b). The set $\{\mathcal{S}_1^n\}\, 1 \leq n \leq t-1$ is then determined from $\mathcal{S}_o^n$ using Eq. (4.5), the set $\{\mathcal{S}_o^n\}$ being replaced in core by $\{\mathcal{S}_1^n\}$ except for $\mathcal{S}_o^{t-1}$ . This process continues until all of the $\{\mathcal{S}_\ell^{t-1}\}$ are determined at which time Eq. (A.1) is generated by addition.

A simple Newton-Rathson iterative technique is used to obtain the relative eigenvalues "q". The order of the polynomial is reduced as each root is found and the reduced equation is used to find the next root, although for accuracy, the initial polynomial is used for the final iterations in finding each root. Since the coefficients of the characteristic equation often reach a magnitude of $10^{30}$ or greater, the $\mathcal{C}_m$ (defined after Eq. (4.4a)) are scaled down and the final roots multiplied by the square root of this scaling factor.

It is convenient to define $D_n$'s so that

$$A_n = \left(e^{-i\varphi}\right)^n D_n\,. \tag{A.2}$$

---

[5] Note added in revision: This section is of historical interest only. All results have been recalculated using Mathematica version 7 or 8. The formulation is straightforward and available upon request.

Eq. (3.11) may then be rewritten in the form

$$D_{n+1} = \frac{\left[(q+n\beta)D_n - \sqrt{n}C_{r,c-n}D_{n-1}\right]}{\sqrt{n+1}C_{r,c-n-1}} \tag{A.3}$$

For each eigenvalue $q_j$ the different non-normalized eigenvectors are generated by using Eq. (A.3). Normalization is provided in the usual way. The phasing is given by Eq. (A.2) and arbitrarily multiplying the entire vector by $\left(e^{-i\varphi}\right)^j$.

There has been some problem with round-off error in the calculation of the eigenvectors and eigenvalues for larger values of r and c. This problem has not been entirely eliminated, although a great deal of improvement has been possible by using double precision variables in the calculations of both eigenvalues and eigenvectors, and in the case of the eigenvectors by generating the vector only half-way with Eq. (A.3) and generating the other half by

$$D_{n-1} = \frac{\left[(q+n\beta)D_n - \sqrt{n-1}C_{r,c-n-1}D_{n+1}\right]}{\sqrt{n}C_{r,c-n}} \tag{A.4}$$

and matching one of the terms in the middle.

Values for $\langle m \rangle$ and $\langle n^2 \rangle - \langle n \rangle^2$ are also calculated. A sample of one version of the computer program used is included in this appendix for easy reference. Note that for large r and c, accurate estimates for the q's must be given before the root finder will converge to a solution.

```
// JOB
// FOR
*ONE WORD INTEGERS
*IOCS(CARD,TYPEWRITER)
*NAMEMIKE
      REAL NORM
      DIMENSION SC(18),REN(18),A(19),ROTS(18)
      CALL TELL
    3 CALL DATSW(5,K)
      IF(K-2)4,1,4
    1 CALL TAVIS(SC,REN,BETA,LP,AB,ITR,R,ITC,C,NORM)
      CALL AFIX(A,SC,REN,LP)
      CALL TOM(A,SC,REN,NORM,AB,ITR,ITC,LP,BETA)
      CALL FIND(A,ROTS,LP,NDEG,M,BETA,NORM,EPSLO)
      CALL ORD(ROTS,M)
      CALL AROT(A,ROTS,NORM,NDEG,EPSLO,LP)
      CALL DATSW(10,LL)
      IF(LL-2)3,5,3
    5 CALL AVECT(ROTS,SC,REN,NDEG,M,LP,ITR,R,C)
      GO TO3
    4 CALL EXIT
      END
// DUP
*STORE        WS   UA   MIKE
// FOR
      SUBROUTINE TELL
      WRITE(1,2)
    2 FORMAT(/' SW 5 UP STOPS PROGRAM ON NEXT DATA SET'/
     1'SW 2 UP STOPS PRINTING OF EVERY ITERATION IN ROOT FINDER AND PRIN
     1T OUT OF POLYNOMIAL'/'SW 4 UP STOPS PRINTING OF ITERATION COUNT AN
     1D ROOTS IN ROOT FINDER'/)
      WRITE(1,3)
    3 FORMAT('SW 3 UP STOPS PRINT OF EVERY ITERATION IN AROT'/
     1'SW 9 UP  STOPS PRINT OF ITERATION COUNT IN AROT'/'
     1SW 10 UP STOPS ENTRY INTO AVECT'/
     1'SET SWITCHES   PRESS START')
      PAUSE
      RETURN
      END
// DUP
*STORE        WS   UA   TELL
// FOR
*ONE WORD INTEGERS
      SUBROUTINE TAVIS(SC,REN,BETA,LP,AB,ITR,R,ITC,C,NORM)
      REAL NORM
      DIMENSION SQC(18),SC(18),REN(18)
```

```
      READ(2,2)ITR,ITC,BETA,JJK
    2 FORMAT(2I5,2X,E15.7,1X,I5)
      IF(JJK)16,16,700
   16 LP=ITR+1
      TR=ITR
      TC=ITC
      R=TR/2.
      C=TC/2.
      II=ITR+ITC
      MAXN=II/2
      IF(2*MAXN-II)700,11,700
   11 IF(MAXN-ITR)12,13,13
   12 LP=MAXN+1
   13 NGAM=0
      NGAM=0
      IF(ITC-ITR)14,14,15
   15 NGAM=(ITC-ITR)/2
   14 T=ITR-LP
      REN(1)=NGAM
      SQC(1)=0.
      SC(1)=0.
      SS=0.
      DO 3 I=2,LP
      TI=T+I
      REN(I)=NGAM+I-1
      SQC(I)=0.
       SC(I)=0.
      SQC(I)=TI*(TR-TI+1.)
       SC(I)=REN(I)*SQC(I)
    3 SS=SS+SC(I)
      SS=SQRT(SS)
      NORM=5.*SS
      AB=SQRT(NORM)
      BETA=BETA/AB
      DO 4 I=1,LP
      SC(I)=SC(I)/NORM
    4 REN(I)=REN(I)*BETA
      RETURN
  700 CALL EXIT
      END
// DUP
*STORE        WS  UA  TAVIS
// FOR
*ONE WORD INTEGERS
      SUBROUTINE AFIX(A,SC,REN,LP)
      DIMENSION SC(18),REN(18),S(190),SAVE(3,19),A(19)
```

```
      JR(K,I)=((K-1)*(K-2))/2+K-1+I
      LPO=LP+1
      DO 2 I=1,LPO
      DO 1 J=1,I
      K=JR(I,J)
    1 S(K)=0.
      DO 7 J=1,3
    7 SAVE(J,I)=0.
    2 A(I)=0.
C
C     CALCULATION OF S SUB ZERO
C
      S(1)=1.
      DO 4 I=2,LPO
      K=I+1
      KJ=I-1
      DO 3 J=1,KJ
      JJ=K-J
      KK=I-J
      IJK=JR(I,JJ)
      IKJK=JR(KJ,KK)
      IKK=IKJK+1
      IF(KJ-JJ)5,6,6
    6 S(IJK)=REN(KJ)*S(IKK)
    5 S(IJK)=S(IJK)+S(IKJK)
    3 CONTINUE
      IJK=JR(I,1)
      IKJK=JR(KJ,1)
    4 S(IJK)=REN(KJ)*S(IKJK)
C
C     CALCULATION OF THE OTHER S'S
C
      I=1
      K=LP-1
      IF(I-K)75,75,100
   75 DO 76 J=1,I
      SAVE(1,J)=0.
      KK=JR(I,J)
      SAVE(2,J)=S(KK)
      KKK=JR(I+1,J)
      SAVE(3,J)=S(KKK)
   76 S(KK)=0.
      J=I+1
      KKK=KKK+1
      SAVE(3,J)=S(KKK)
      DO 80 JJ=I,K
```

```
      IK=I+JJ
      KK=JJ+1
      JK=JJ+2
      KJ=JJ-I
      IKK=KJ+1
      KKK=IKK+1
      IF(JJ-I)88,78,88
   88 DO 77 J=1,KJ
      M=KKK-J
      KJK=IKK-J
      IJ=JR(KK,M)
   77 S(IJ)=SAVE(1,KJK)+REN(IK)*SAVE(1,M)+SC(IK)*SAVE(2,M)
   78 IJ=JR(KK,1)
      S(IJ)=REN(IK)*SAVE(1,1)+SC(IK)*SAVE(2,1)
      MK=JR(KK,KKK)
      S(MK)=0.
      DO 68 J=1,KK
      JMM=JR(KK,J)
      SAVE(1,J)=S(JMM)
      SAVE(2,J)=SAVE(3,J)
      JM=JR(JK,J)
   68 SAVE(3,J)=S(JM)
   80 SAVE(3,JK)=S(JM+1)
      IF(I+2-K)87,87,90
   87 I=I+1
      K=K-1
      GO TO 75
   90 IF(I-K)92,100,92
   92 I=I+1
      DO 95 J=1,I
      KJ=JR(I,J)
   95 S(KJ)=0.
  100 DO 110 J=1,LPO
      ONE=1.
      IJ=LPO-J+1
      DO 110 K=1,IJ
      KJ=JR(LPO-K+1,J)
C
C     FORMATION OF CHARACTERISTIC EQUATION
C
      A(J)=A(J)+ONE*S(KJ)
  110 ONE=-ONE
      RETURN
      END
// DUP
*STORE      WS  UA  AFIX
```

```
// FOR
*ONE WORD INTEGERS
      SUBROUTINE TOM(A,SC,REN,NORM,AB,ITR,ITC,LP,BETA)
      REAL NORM
      DIMENSION SC(18),REN(18),A(19)
      DO 5 I=1,LP
      SC(I)=SC(I)*NORM
    5 REN(I)=REN(I)*AB
      LPP =LP+1
      WRITE(1,6)ITR,ITC,BETA,(A(I),I=1,LP),A(LPP)
    6 FORMAT(//1H ,'2R=',I5,2X,'2C=',I5,2X,'BETA=',E15.7/1H ,'A='
     1/5(E18.7,3X))
      WRITE(1,7)LP,NORM,(SC(I),I=1,LP)
    7 FORMAT(/1H ,'LP=',I5,2X,'NORM=',E15.7/1H ,'SC='/5(E18.7,3X))
      WRITE(1,8) AB,(REN(I),I=1,LP)
    8 FORMAT(/1H ,'AB=',E15.7,2X/1H ,'REN='/5(E18.7,3X))
      RETURN
      END
// DUP
*STORE      WS  UA  TOM
// FOR
*ONE WORD INTEGERS
      SUBROUTINE FIND(A,ROTS,LP,NDEG,M,BETA,NORM,EPSLO)
C
C     THIS IS THE ROOT FINDER
C
      REAL NORM
      DIMENSION A(19),ROTS(18),BB(19)
      NDEG=LP
      JJK=0
      M=1
      MM=0
      DO 4 I=1,LP
      ROTS(I)=0.
    4 BB(I)=0.
      I=LP+1
      BB(I)=0.
      CALL DATSW(2,K)
      READ(2,1)EPSLO
    1 FORMAT(E15.7)
      WRITE(1,2) EPSLO
    2 FORMAT(/1H ,40HSTART OF PRINT FROM ROOT FINDER
     1/1H , 8HEPSLON= ,E15.7)
      IF(LP-2*(LP/2))205,199,205
  199 JKJ=1
      GO TO 220
```

```
205 JKJ=0
220 CONTINUE
    IF(BETA)400,402,400
402 NDEG=NDEG/2
    JJK=1
400 NCOEF=NDEG+1
    IF(BETA)405,406,405
406 DO 300 I=1,NCOEF
    J=2*I-JKJ
300 A(I)=A(J)
    GO TO 410
405 NORM=SQRT(NORM)
410 X=A(1)
    GAM=ABS(A(NCOEF)/A(1))**(1./NDEG)
    DO 3 I=1,NCOEF
    A(I)=A(I)/X
    BB(I)=A(I)
  3 X=X*GAM
    NORM=NORM/GAM
    NGOOD=0
 20 N1=NGOOD+1
    IF(M+1-NDEG)33,36,33
 33 IF(M-NDEG)34,35,750
 34 IF(JJK)35,36,35
 35 Z=-A(N1)/A(N1+1)
    GO TO 45
 36 B=A(N1+1)
    AC=SQRT(B*B-4.*A(N1)*A(N1+2))
    ABB=2.*A(N1+2)
    Z=(-B-AC)/ABB
 45 IF(K-2)31,32,32
 32 WRITE(1,30)Z
 30 FORMAT(1H ,3HZ= ,E15.7)
    WRITE(1,25)(A(I),I=N1,NCOEF)
 25 FORMAT(/1H ,'POLYNOMIAL'/1H ,5(E15.7,3X))
 31 IF(M-NDEG)60,120,60
 60 DO 50 I=1,500
    Z1=A(NCOEF)
    Z2=0.
    DO 40 J=N1,NDEG
    JJ=NDEG+N1-J
    Z1=Z*Z1+A(JJ)
    IF(MM)47,46,47
 47 A(JJ)=Z1
 46 J1=JJ+1
    JJJ=J1-N1
```

```
      AJJ=JJJ
   40 Z2=Z2*Z+A(J1)*AJJ
      DELZ=-Z1/Z2
      IF(MM)120,48,120
   48 RATIO=ABS(DELZ)/(ABS(Z)+1.E-8)
      IF(RATIO-EPSLO)100,100,101
  101 Z =Z+DELZ
      IF(K-2)50,102,102
  102 WRITE(1,130)I,Z,DELZ,Z1,Z2
  130 FORMAT(1H ,I5,3X 4(E15.7,3X))
   50 CONTINUE
      GO TO 700
  100 IF(MM)120,105,120
  105 Z=Z+DELZ
      CALL DATSW(4,KK)
      IF(KK-2)41,42,42
   42 WRITE(1,130)I,Z,RATIO,Z1
   41 MM=1
      GO TO 60
  120 MM=0
      A(N1)=Z1
      Z=Z*NORM
      IF(JJK)13,12,13
   13 Z=SQRT(Z)
   12 ROTS(M)=Z
      IF(M-NDEG)600,700,600
  600 M=M+1
      NGOOD=NGOOD+1
      GO TO 20
  700 IF(M-NDEG)750,800,750
  750 WRITE(1,780)
  780 FORMAT(1X,'PROBLEM WITH THE ROOT FINDER')
  800 IF(JJK)820,810,820
  820 IF(JKJ)810,830,810
  830 M=M+1
  810 WRITE(1,500)(ROTS(J),J=1,M)
  500 FORMAT(1H ,'ROOTS'/1H ,5(E15.7,3X))
       DO 950 I=1,NCOEF
  950 A(I)=BB(I)
      RETURN
      END
// DUP
*STORE       WS  UA  FIND
// FOR
*ONE WORD INTEGERS
      SUBROUTINE ORD(ROTS,N)
```

```
      DIMENSION ROTS(18)
      IF(N-1)2,2,3
    3 M=N-1
      DO 1 I=1,M
      II=I+1
      DO 1 J=II,N
      IF(ROTS(I)-ROTS(J))4,1,1
    4 A=ROTS(I)
      ROTS(I)=ROTS(J)
      ROTS(J)=A
    1 CONTINUE
    2 RETURN
      END
// DUP
*STORE      WS  UA  ORD
// FOR
*ONE WORD INTEGERS
      SUBROUTINE AROT(A,ROTS,NORM,NDEG,EPSLO,LP)
C
C     THIS ROUTINE DOES THE LAST ITERATIONS FOR THE ROOTS
C     WITH THE ORIGINAL POLYNOMIAL
C
      REAL NORM
      DIMENSION A(19),ROTS(18)
      WRITE(1,10)
   10 FORMAT(//'PRINT FROM ROT FIXER')
      NCOEF = NDEG+1
      N=1
      MJ=0
      NGOOD=0
   20 N1=NGOOD+1
      Z=ROTS(N)
      IF(LP-NDEG)700,2,1
    1 Z=Z*Z
    2 Z=Z/NORM
      DO 50 I=1,500
      Z1=A(NCOEF)
      Z2=0.
      DO 40 J=N1,NDEG
      JJ=NDEG+N1-J
      Z1=Z*Z1+A(JJ)
      J1=JJ+1
      JJJ=J1-N1
      AJJ=JJJ
   40 Z2=Z2*Z+A(J1)*AJJ
      DELZ=-Z1/Z2
```

```
      RATIO=ABS(DELZ)/(ABS(Z)+1.E-8)
      IF(RATIO-EPSLO)120,120,101
  101 Z=Z+DELZ
      CALL DATSW(3,K)
      IF(K-2)50,102,102
  102 WRITE(1,130)I,Z,DELZ,Z1,Z2
   50 CONTINUE
      GO TO 700
  120 CALL DATSW(9,KK)
      IF(KK-2)31,121,121
  121  WRITE(1,130)I,Z,RATIO,Z1
  130 FORMAT(1H ,I5,3X,4(E15.7,3X))
   31 Z=Z*NORM
      IF(LP-NDEG)700,4,5
    5 Z=SQRT(Z)
    4 ROTS(N)=Z
      IF(N-NDEG)600,400,600
  600 N=N+1
      GO TO 20
  700 WRITE(1,780)
  780 FORMAT(1H ,'PROBLEMS IN THIS ROOT FINDER CALLING EXIT')
      MJ=10
  400 WRITE(1,500)(ROTS(J),J=1,N)
  500 FORMAT(1H ,'ROOTS='/5(E15.7,3X))
      IF(MJ-10)425,450,425
  425 RETURN
  450 CALL EXIT
      END
// DUP
*STORE       WS  UA  AROT
// FOR
*ONE WORD INTEGERS
      SUBROUTINE AVECT(ROTS,SC,REN,NDEG,M,LP,ITR,R,C)
C
C     SHOULD BE DOUBLE PRECISSION
C
      DIMENSION REN(18),ROTS(18),SC(18),VECT(18)
      WRITE(1,30)
   30 FORMAT(/1H ,'IN VECTOR FORMER')
      KKK=LP
      LO=0
      IF(LP-15)2,2,1
    1 KKK=LP/2+1
    2 DO 100 I=1,M
      DO 120 J=1,LP
  120 VECT(J)=0.
```

```
      VECT(1)=1.
      AB=SQRT(SC(2))
      VECT(2)=(ROTS(I)+REN(1))/AB
      IF(LP-2)125,125,3
    3 DO 200 J=3,KKK
      AB1=SQRT(SC(J))
      VECT(J)=((ROTS(I)+REN(J-1))*VECT(J-1)-AB*VECT(J-2))/AB1
      AB=AB1
  200 CONTINUE
      IF(LP-15)125,125,5
    5 AVE = VECT(KKK)
      IF(ROTS(I))10,12,10
   12 IF(KKK-2*(KKK/2))10,13,10
   13 AVE = -AB*VECT(KKK-1)/SQRT(SC(KKK+1))
      LO=10
C
C     VECTOR FROM THE OTHER END
C
   10 VECT(LP)=1.
      IF(I-2*(I/2))6,7,6
    7 VECT(LP)=-1.
    6 AB=SQRT(SC(LP))
      VECT(LP-1)=(ROTS(I)+REN(LP))*VECT(LP)/AB
      KKK1=KKK-1
      IF(LP-2*(LP/2))14,15,14
   15 KKK1=KKK1-1
   14 DO 800 J=2,KKK1
      LK=LP-J+1
      AB1=SQRT(SC(LK))
      LKK=LP-J
      VECT(LKK )=((ROTS(I)+REN(LK))*VECT(LK)-AB*VECT(LK+1))/ AB1
      AB=AB1
  800 CONTINUE
      LK=LP-KKK1
      BVECT=VECT(LK)
      IF(LO-10)16,17,16
   17 LK=LP-KKK1+1
      BVECT=VECT(LK)
   16 IF(LP-2*(LP/2))820,810,820
  820 ABCD=BVECT/AVE
      KKKK=KKK
      GO TO 840
  810 ABCD=BVECT/AVE
      KKKK=KKK1
  840 DO 850 J=1,KKKK
      LK=LP-J+1
```

```
  850 VECT(LK)=VECT(LK)/ABCD
  125 AVENS=0.
      SUM=0.
      AVER=0.
      DO 220 J=1,LP
      AC=VECT(J)*VECT(J)
      SUM=SUM+AC
      RJ = ITR-LP+J+1
      TM=R-RJ+1.
      AVER=AVER+AC*TM
      AVENS=AVENS+AC*(C-TM)**2
  220 CONTINUE
      AVER=AVER/SUM
      AVENS=AVENS/SUM-(C-AVER)**2
      SUM=SQRT(SUM)
      DO 210 J=1,LP
  210 VECT(J)=VECT(J)/SUM
      WRITE(1,300) AVER,AVENS,SUM,ROTS(I),(VECT(J),J=1,LP)
  300 FORMAT(/1H ,'AVER=',E15.7,2X,'AVENS=',E15.7,2X,'SUM=',E15.7,2X,
     1'Q=',E15.7/1H ,'VECT='/5(E15.7,3X))
  100 CONTINUE
      RETURN
      END
// DUP
*STORE        WS   UA   AVECT
```

# APPENDIX B
# FURTHER APPROXIMATION TO THE EXACT PROBLEM

There are three further approximations to the Hamiltonian (3.4) for $\omega$=$\Omega$. These consist of averaging out the magnitude of the TLM part of the interaction terms, considering the TLM part of the Hamiltonian as being classical, and finally as considering the field part of the Hamiltonian as being classical.

## A. Average TLM Approach

As in the modified TLM approach, Eq. (3.4) is written as

$$H = R_3 + a^\dagger a - \kappa\sqrt{R^2 - R_3^2 + R_3}\Gamma_+ a - \kappa^* \Gamma_- \sqrt{R^2 - R_3^2 + R_3}a^\dagger \, . \tag{B.1}$$

Replace $\sqrt{R^2 - R_3^2 + R_3}$ by its average $\langle\sqrt{R^2 - R_3^2 + R_3}\rangle$ instead of replacing it by a modified operator. Then

$$H_a = R_3 + a^\dagger a - \bar{\kappa}\Gamma_+ a - \bar{\kappa}^* \Gamma_- a^\dagger \, , \tag{B.2a}$$

where

$$\bar{\kappa} = \kappa\langle\sqrt{R^2 - R_3^2 + R_3}\rangle \, . \tag{B.2b}$$

Again, r and c are good quantum numbers. Solutions to Eq. (B.2) may again be found in a manner similar to that used in finding the exact solution, that is, by forming a difference equation and solving using the exact same boundary conditions. The difference equation is of the form

$$e^{-i\varphi}\sqrt{n}T_{n-1} - q_a T_n + e^{i\varphi}\sqrt{n+1}T_{n+1} = 0 \, , \tag{B.3}$$

where

$$q_a = \frac{c - \lambda_a}{|\bar{\kappa}|} \, . \tag{B.4}$$

If $Z_n$ is defined as

$$T_n = \frac{Z_n\left(e^{-i\varphi}\right)^n}{\sqrt{n!}}, \tag{B.5}$$

Eq.(B.3) becomes

$$nZ_{n-1} - q_a Z_n + Z_{n+1} = 0. \tag{B.6}$$

The boundary conditions are

$$Z_{max(-1,c-r-1)} = Z_{r+c+1} = 0. \tag{B.7}$$

The effective eigenvalues are again obtained from the characteristic equations for

$$Z_{r+c+1} = 0, \tag{B.8}$$

and the eigenvectors obtained from these eigenvalues as explained in Appendix A. The values of $q_a$ may be compared with those of the exact solution by finding $\langle\sqrt{R^2 - R_3^2 + R_3}\rangle$with the calculated eigenvectors, $T_n$, and comparing the exact q's with $q_a$ $\langle\sqrt{R^2 - R_3^2 + R_3}\rangle$ . It is found that there does not exist very good agreement between the eigenvalues and especially between the eigenvectors for any particular range of c (in relation to r). Of interest is the fact that on redefining $Z_n$ as

$$T_n = \frac{Z_n\left(e^{-i\varphi}\right)^n}{2^{n/2}\sqrt{n!}} \ (see\ Eq. B.5), \tag{B.8}$$

Eq. (B.3) becomes

$$Z_{n+1} + 2q_a^{'} Z_n + 2nZ_{n-1} = 0 \ (see\ Eq. B.6). \tag{B.9}$$

where

$$q_a^{'} = \frac{q_a}{2}. \tag{B.10}$$

Eq. (B.9) is the recursion relation for the Hermite polynomial of order n. Therefore for c<r a solution to Eq. (B.3) would give for the un-normalized eigenvectors

$$T_n = \frac{\left(e^{-i\varphi}/\sqrt{2}\right)^n}{\sqrt{n!}} H_n(q_a^{'}), \tag{B.11}$$

where the effective eigenvalues $q_a^{'}$ are given by the zeros of $H_{r+c+1}(q_a^{'}) = 0$.

## B. Classical TLM Approach

If the Hamiltonian (2.37) is averaged over the TLMs in a classical manner such that $\Omega R_3$ is replaced by the average classical TLM energy and $R_{\pm}$ replaced by the classical analog, that is let

$$\langle\mu\rangle(R_+ + R_-) = \langle\mu\rangle_{class}.$$

Then Eq. (2.37) becomes

$$H = \omega a^{\dagger} a - \gamma a - \gamma^* a^{\dagger} + H_a, \tag{B.12}$$

where $\gamma$ now contains the magnitude of the total dipole moment of the TLM system (B.12) and $H_a$ is the average energy of the N-TLMs. Define b = a-$\frac{\gamma^*}{\omega}$ and $b^{\dagger} = (b)^{\dagger}$, then

$$H = \omega[(b^{\dagger} b) - |\kappa|^2] + H_a, \tag{B.13}$$

where $|\kappa|$=$|\gamma^*/\omega|$. Eq. (B.13) is recognized as the harmonic oscillator Hamiltonian for quasi-particles. The ground state is given by

$$|\kappa> = e^{-|\kappa|^2/2} \sum_{n=0}^{\infty} \frac{(\kappa^*)^n |n>}{\sqrt{n!}} \tag{B.14}$$

with ground state energy $\lambda_o = -\omega|\bar{\kappa}|^2 + H_a$ This is, of course the displaced ground state oscillator solution and energy discussed by Glauber [(6)]. The excited energy states are given by

$$|m, \kappa> = \frac{(b^{\dagger})^m}{\sqrt{m!}} |\kappa>, \tag{B.15}$$

and the corresponding eigenvalues are

$$\lambda_m = \omega(m - |\bar{\kappa}|^2) + H_a . \tag{B.16}$$

Equation (B.12) may also be obtained from Eq.(3.4) if the phase of the TLM raising and lowering operators is incorporated into $\gamma$ before the TLM average is taken.

Note that in this approximation r and c are no longer relevant quantum numbers, i.e., a symmetry type breaking has occurred. Essentially, the SU (2) grouping of the problem has been averaged over and thereby lost.

The average value of the positive frequency part of the field in the ground state is given by

$$\langle \kappa | a | \kappa \rangle = \bar{\kappa}^* \,. \tag{B.17}$$

and the dispersion in photon number by

$$\sigma^2 = \langle \kappa | n^2 | \kappa \rangle - (\langle \kappa | n | \kappa \rangle)^2 = |\bar{\kappa}|^2 = \bar{\text{n}} \,. \tag{B.18}$$

In the exact solution, the deviation from the non-interacting energies was of order $|\kappa|$ and the largest dispersion in photon number for the ground state was approximately n/2. In the present approximation, the energy change is of order $|\kappa|^2$ and the dispersion given by n. It is therefore felt that this approximation does not give good estimates for either the eigenstates or eigenvalues of Eq. (3.4) for $\omega = \Omega$. Perhaps in some ensemble average of the system in question the present approximation is valid; however, questions of this sort will not be treated here.

## C. Classical Field Approach

As in Section B, the electromagnetic field may be considered as classical in Eq. (2.37) yielding

$$H = \Omega R_3 - \gamma R_+ - \gamma^* R_- + H_{f\,,} \tag{B. 19}$$

where $H_{f\,,}$ is the average energy of the electromagnetic field and here $\gamma$ contains the magnitude of the total electric field. This is, of course, Eq.(3.4) ($\omega$=$\Omega$) averaged over the field after the phase of the field has been incorporated into $\gamma$. Again c is no longer a valid quantum number; however $R^2$ still commutes with Eq. (B.19). The solution to this Hamiltonian follows exactly the solution for the "average field approach given in Chapter V of this paper for c > r. That is Eq.(B.19) is broken down into the sum of individual TLMs with interaction and the eigenstate and eigenvalues constructed from the single TLM solutions so that r is a good quantum number. Towards this end, Eq.(B.19) becomes

$$H = \sum_{j=1}^{N}\left(\Omega R_{j3} - \gamma R_{j+} - \gamma^{*} R_{j-}\right) , \tag{B.20}$$

where the field energy has been neglected for convenience. The states |+> and |->, excited and ground states of one TLM, expressed in terms of |↑>= |1/2, 1/2> and |↓> = |1/2, -1/2> are

$$|+> = \left(\cos\theta\,|\uparrow> + \sin\theta\, e^{i(\pi-\varphi)}\,|\downarrow>\right), \tag{B.21 a}$$

with energy

$$\lambda_{+} = \frac{\Omega}{2}\sqrt{1+4|\kappa|^{2}} , \tag{B .21 b}$$

where $\tan 2\theta = 2|\kappa|$ and $\kappa = \frac{|\gamma|}{\Omega}$,

$$|-> = \left(\sin\theta\, e^{i\varphi}|\uparrow> + \cos\theta\,|\downarrow>\right), \tag{B.21 c}$$

with energy

$$\lambda_{-} = -\frac{\Omega}{2}\sqrt{1+4|\kappa|^{2}} . \tag{B.21 d}$$

States of the system are given by

$$|r,j> = \frac{1}{\sqrt{\frac{(2r)!}{j!\,(2r-j)!}}}\sum_{l=0}^{2r}\sum_{l'=down'}^{up'}\left[cos^{j-l'}\theta\, sin^{n_{-}+l'}\theta\right]$$
$$\times\left[\left(e^{\frac{i(l+l')\pi}{2}}\right) l!\,(2r-l)!\right] \div \left[\left(\frac{(l+l')}{2}\right)!\left(\frac{(l-l')}{2}\right)!\left(j-\frac{(l+l')}{2}\right)!\left(n_{-}-\frac{(l-l')}{2}\right)!\right] \tag{B.22}$$
$$\times e^{i(n_{-}-l)\varphi}\sqrt{\frac{(2r)!}{l!\,(2r-l)!}}\,|r,m>$$

where

$$m = \text{r-}\ l, \tag{B.23a}$$

and

$$n_{-} = 2\text{ r} - \text{j}, \tag{B.23b}$$

also

$$\text{down}' = \max(-l\,, l - 2\,n_) \text{ and } \text{up}' = \min(l,\; 2\,j - l) \qquad \text{(B.23c)}$$

The energy of this state is given by

$$\lambda_j = -(r-j)\Omega\sqrt{1+4|\kappa|^2}\ , \qquad \text{(B.24)}$$

where j = 0 represents the ground state and j = 2 r the most excited state. Only for the coupling constant $|\kappa|$ becoming very large do the eigenstates and eigenvalues approach the form given for the "average field approach;" therefore, for $\omega$=$\Omega$ this is again not considered a too relevant approximation. The average value of m is not equal to zero for this case except for $|\kappa| \to \infty$, in fact, for the ground state m = r ($\sin^2\theta$ - $\cos^2\theta$). This is easily seen since the average for one TLMs in the ground state is l/2 ($\sin^2\theta$ - $\cos^2\theta$) and the total ground state is just the produce of 2r of the single TLM in the ground state. This average is considerably more difficult to find when j$\neq$ 0 or j $\neq$ r.

The approximations given in Sections B and C may also be obtained by assuming that the eigenstates are of the form

$$| > = \sum_{n=0}^{\infty}\sum_{m=-r}^{r} A_n B_m |n > |r, m > \qquad \text{(B.25)}$$

finding the energy of Eq.(3.4) with this state, and performing variational calculations with respect to the field component, $A_n^*$, or the TLM component,$B_m^*$. The eigenstates are then products of the eigenstates found in B and C and the eigenvalues the sum of energies, with care taken not to include the same energy more than once. This is again not considered appropriate and will not be discussed further.

# APPENDIX C

In this appendix a second order perturbation calculation is performed on the cubic term dropped from Eq.(5. 12). In order to perform this calculation, full use of the analogy between the harmonic oscillator equation and Eq. (5.12) is utilized. The equation for the harmonic oscillator is

$$\frac{d^2\psi}{dx^2} + \frac{2\mu}{\hbar^2}\left(E_n - \frac{1}{2}\mu\omega^2 x^2\right)\psi = 0, \qquad \text{(C .1)}$$

where $\omega$ is the circular frequency of the oscillation and μ is the mass of the particle. Let ħ→ 1 and μ→ 1 for convenience; therefore,

$$\frac{d^2\psi}{dx^2} + (2E_n - \omega^2 x^2)\psi = 0. \qquad \text{(C.2)}$$

This is to be compared to Eq.(5.12)

$$\frac{d^2E}{dn^2} + \left[\frac{\alpha_1 - q^2}{q^2} - \frac{\alpha_2}{q^2}(n - n_o)^2\right]E = 0. \qquad \text{(C.3)}$$

The cubic term which has been dropped is of the form $-\frac{1}{q^2}(n - n_o)^3$. Using the correspondence between the two equations (C.2) and (C.3) requires that

$$2E_n = \frac{\alpha_1 - q^2}{q^2}, \qquad \text{(C.4)}$$

and

$$\omega^2 = \frac{\alpha_1}{q^2}. \qquad \text{(C.5)}$$

It is well known that[12]

$$E_n = \omega(n + 1/2). \qquad \text{(C.6)}$$

If a cubic term of the form $bx^3$ is added to Eq. (C.2) then a correction to the energy (C.6) to second order is given by perturbation theory to be[(l9)]

$$E_n = \omega(n + 1/2) - \frac{b^2\left(\frac{1}{2\omega}\right)^3(30n^2 + 30n + 11)}{\omega}, (\hbar \to 1, \mu \to 1). \quad \text{(C.7)}$$

Substituting for $\omega$, b, and $E_n$

$$\frac{\alpha_1 - q^2}{2q^2} = \frac{\alpha_2^{1/2}}{q}\left(n + \frac{1}{2}\right) - \frac{1}{8\alpha_2^2}(30n^2 + 30n + 11). \quad \text{(C.8)}$$

Solving for the effective eigenvalues gives $q_n$ of the form

$$q_n = \frac{1}{1 - \frac{1}{4\alpha_2^2}(30n^2 + 30n + 11)}\left\{-\alpha_2^{1/2}\left(n + \frac{1}{2}\right) + \sqrt{\alpha_2\left(n + \frac{1}{2}\right)^2 + \alpha_1\left[1 - \frac{1}{4\alpha_2^2}(30n^2 + 30n + 11)\right]}\right\}$$

(C.9)

This perturbation calculation is valid only as long as

$$(30n^2 + 30n + 11) < 4\alpha_2^2 . \quad \text{(C . 10)}$$

The values calculated from Eq. (C.9) are larger than those calculated from Eq.(5.16). This difference is small (lying in the third significant figure) for those values of n for which the harmonic oscillator approach is valid. From Figure 10 it can be seen that the larger values of q calculated from Eq. (C.9) do not agree with the exact q's as well as the q's calculated with Eq. (5.16) .

# APPENDIX D

## PROOF OF SEVERAL RELATIONS INVOLVING "COOPERATION NUMBER"

Consider a system of N molecules in thermal equilibrium with one another. The states of the system |r, m>[7] have an energy given by mE where E is the energy level separation of one two-level molecule. The states also have a degeneracy given by[7]

$$P(r) = \frac{N!\,(2\,r+1)}{\left(\frac{1}{2}\,N+r+1\right)!\left(\frac{1}{2}\,N-r\right)!} \tag{D.1}$$

The values of m, r are limited by

$$|m| \leq r \;\leq \frac{N}{2}. \tag{D.2}$$

Dicke states that:

1. The average value of m for thermal equilibrium is

$$\overline{m} = \text{-NE/4kT} \tag{D.3}$$

2 That the mean square deviation from the mean is

$$\frac{N}{4} - \frac{\overline{m}^2}{N} \tag{D.4}$$

3 That for a definite value of m the mean value of r(r + 1) is

$$m^2 + \frac{N}{2} \tag{D.5}$$

4 and that the mean square deviation from the mean of Eq. (D.5) is

$$\frac{N^2}{4} - \,m^2 \tag{D.6}$$

Proof of (1). Let $\beta = \frac{E}{kT}$. Then

$$\bar{m} = \frac{\sum_{r=0,\frac{1}{2}}^{\frac{N}{2}} \frac{N!\,(2\,r+1)}{\left(\frac{N}{2}+\ r+1\right)!\left(\frac{N}{2}-\ r\right)!} \sum_{m=-r}^{r} m\, e^{-m\beta}}{\sum_{r=0,\frac{1}{2}}^{\frac{N}{2}} \frac{N!\,(2\,r+1)}{\left(\frac{N}{2}+\ r+1\right)!\left(\frac{N}{2}-\ r\right)!} \sum_{m=-r}^{r} e^{-m\beta}} = \frac{-d}{d\beta}\, ln\left(\sum_{r} P(r) \sum_{m=-r}^{r} e^{-m\beta}\right) \quad \text{(D.7)}$$

Now

$$\sum_{m=-r}^{r} e^{-m\beta} = e^{r\beta} \sum_{m=0}^{2\,r} e^{-m\beta} = \frac{e^{r\beta}\left[1 - e^{-2(r+1)\beta}\right]}{1-\ e^{-\beta}} = \frac{Sinh\left(r+\frac{1}{2}\right)\beta}{Sinh\ \frac{\beta}{2}}.$$

Doing the sum over r gives

$$\frac{\sum_{r=0,\frac{1}{2}}^{\frac{N}{2}} P(r) Sinh\left(r+\ \frac{1}{2}\right)\beta}{Sinh\ \frac{\beta}{2}} = \frac{2\frac{-d}{d\beta}\sum_{r=0,1/2}^{N/2} \frac{N!\,Cosh\left(r+\ \frac{1}{2}\right)\beta}{\left(\frac{N}{2}+\ r+1\right)!\left(\frac{N}{2}-\ r\ \right)!}}{Sinh\ \frac{\beta}{2}} \quad \text{(D.8)}$$

Expand the Cosh into the sum of exponential terms

$$\frac{e^{\left(r+\frac{1}{2}\right)\beta} + e^{-\left(r+\frac{1}{2}\right)\beta}}{2}$$

And rearrange the resulting sum (D.8), obtaining

$$\frac{\frac{-d}{d\beta}\left[\sum_{r=-(N+1)/2}^{N/2} \frac{N!Exp\left(r+\frac{1}{2}\right)\beta}{\left(\frac{N}{2}+r+1\right)!\left(\frac{N}{2}-r\right)!} - \frac{(N+1)!}{\left(\frac{N+1}{2}!\right)^2}\right]}{Sinh\,\frac{\beta}{2}} = \frac{\frac{-d}{d\beta}\left[\left(e^{\frac{\beta}{2}}+e^{-\frac{\beta}{2}}\right)^{N+1} - \frac{(N+1)!}{\left(\frac{N+1}{2}!\right)^2}\right]}{(N+1)Sinh\,\frac{\beta}{2}} = \frac{2^N(N+1)Sinh\frac{\beta}{2}\,Cosh^N\,\frac{\beta}{2}}{(N+1)Sinh\frac{\beta}{2}} = \quad \text{(D.9)}$$

$$2^N Cosh^N\,\frac{\beta}{2}$$

In the above the constant term in the square brackets is only added when N is odd. From Eq. (D.7)

$$\bar{m} = \frac{-d}{d\beta}\, ln\left(\sum_r P(r) \sum_{m=-r}^{r} e^{-m\beta}\right) = \frac{-d}{d\beta}\, ln\left(2^N Cosh^N \frac{\beta}{2}\right)$$

$$= \frac{2^N \frac{N}{2} Cosh^{N-1}\frac{\beta}{2} Sinh\frac{\beta}{2}}{2^N Cosh^N \frac{\beta}{2}} = -\frac{N}{2} Tanh\frac{\beta}{2} \tag{D.10}$$

If $\beta$ is small, using the definition of $\beta$ gives

$$\bar{m} = -\frac{N}{4}\,\beta = -\frac{NE}{4kT} \qquad \text{Q.E.D.}$$

Proof of (2):

$$\sigma^2(m) = \langle m^2 - \langle m\rangle^2\rangle \tag{D.11}$$

Using a formulation similar to (D.7)

$$\langle m^2\rangle = \frac{\frac{d^2}{d\beta^2}\left(\sum_r P(r) \sum_{m=-r}^{r} e^{-m\beta}\right)}{2^N Cosh^N \frac{\beta}{2}} = \frac{2^N \frac{N}{2}\frac{d}{d\beta}\left(Cosh^{N-1}\frac{\beta}{2} Sinh\frac{\beta}{2}\right)}{2^N Cosh^N \frac{\beta}{2}}$$

$$= \frac{N\left(Cosh^N\frac{\beta}{2} + \left((N-1)Cosh^{N-2}\frac{\beta}{2} Sinh^2\frac{\beta}{2}\right)\right)}{4Cosh^N\frac{\beta}{2}} \tag{D.12}$$

$$= \frac{N}{4} + \frac{N(N-1)}{4} Tanh^2\frac{\beta}{2}$$

Using (D.10)

$$\sigma^2(m) = \langle m^2 - \langle m\rangle^2\rangle = \frac{N}{4} + \frac{N(N-1)}{4} Tanh^2\frac{\beta}{2} - \frac{N^2}{4} Tanh^2\frac{\beta}{2}$$

$$= \frac{N}{4} - \frac{N}{4} Tanh^2\frac{\beta}{2} = \frac{N}{4} - \frac{\bar{m}^2}{N}, \tag{D.13}$$

This is the results for 2.

Proof of (3): $\overline{r(r+1)}$ for fixed m:

$$\overline{r(r+1)} = \frac{\sum_{r=m}^{N/2} P(r) r(r+1)}{\sum_{r=m}^{N/2} P(r)} \tag{D.14}$$

The normalization is given by:

$$\sum_{r=m}^{N/2} P(r) = \sum_{r=m}^{N/2} \frac{N!\,(2\,r+1)}{\left(\frac{N}{2}+r+1\right)!\left(\frac{N}{2}-r\right)!}$$

$$= 2\,N! \left\{ \sum_{r=m}^{N/2} \frac{\left(\frac{N}{2}+r+1\right)}{\left(\frac{N}{2}+r+1\right)!\left(\frac{N}{2}-r\right)!} \right.$$

$$\left. - \left(\frac{N+1}{2}\right) \sum_{r=m}^{N/2} \frac{1}{\left(\frac{N}{2}+r+1\right)!\left(\frac{N}{2}-r\right)!} \right\}$$

$$= 2\,N! \left\{ \sum_{r=m}^{N/2} \frac{1}{\left(\frac{N}{2}+r\right)!\left(\frac{N}{2}-r\right)!} \right.$$

$$\left. - \left(\frac{N+1}{2}\right) \sum_{r=m}^{N/2} \frac{1}{\left(\frac{N}{2}+r+1\right)!\left(\frac{N}{2}-r\right)!} \right\}$$

$$= 2\,N! \left\{ \frac{1}{\left(\frac{N}{2}+m\right)!\left(\frac{N}{2}-m\right)!} \right.$$

$$+ \sum_{r=m}^{\frac{N}{2}-1} \frac{1}{\left(\frac{N}{2}+r+1\right)!\left(\frac{N}{2}-r-1\right)!}$$

$$\left. - \left(\frac{N+1}{2}\right) \sum_{r=m}^{N/2} \frac{1}{\left(\frac{N}{2}+r+1\right)!\left(\frac{N}{2}-r\right)!} \right\}$$

$$= 2\,N! \left\{ \frac{1}{\left(\frac{N}{2}+m\right)!\left(\frac{N}{2}-m\right)!} - \sum_{r=m}^{N/2} \frac{\left(r+\frac{1}{2}\right)}{\left(\frac{N}{2}+r+1\right)!\left(\frac{N}{2}-r\right)!} \right\}$$

Thus

$$\sum_{r=m}^{N/2} P(r) = \sum_{r=m}^{N/2} \frac{N!(2\,r+1)}{\left(\frac{N}{2}+r+1\right)!\left(\frac{N}{2}-r\right)!} = \frac{N!}{\left(\frac{N}{2}+m\right)!\left(\frac{N}{2}-m\right)!} \tag{D.15}$$

Perform the sum of r(r+1) over P(r):

$$\sum_{r=m}^{N/2} P(r)r(r+1) = \sum_{r=m}^{N/2} \frac{N!\,(2\,r+1)r(r+1)}{\left(\frac{N}{2}+r+1\right)!\left(\frac{N}{2}-r\right)!}$$

$$= 2\,N!\left\{\sum_{r=m}^{N/2} \frac{\left(\frac{N}{2}+r+1\right)r(r+1)}{\left(\frac{N}{2}+r+1\right)!\left(\frac{N}{2}-r\right)!} - \left(\frac{N+1}{2}\right)\sum_{r=m}^{N/2} \frac{r(r+1)}{\left(\frac{N}{2}+r+1\right)!\left(\frac{N}{2}-r\right)!}\right\}$$

$$= 2\,N!\left\{\frac{m(m+1)}{\left(\frac{N}{2}+m\right)!\left(\frac{N}{2}-m\right)!}\right.$$

$$\left. + \sum_{r=m}^{\frac{N}{2}-1} \frac{(r+1)(r+2)}{\left(\frac{N}{2}+r+1\right)!\left(\frac{N}{2}-r-1\right)!} - \left(\frac{N+1}{2}\right)\sum_{r=m}^{N/2} \frac{r(r+1)}{\left(\frac{N}{2}+r+1\right)!\left(\frac{N}{2}-r\right)!}\right\}$$

$$= 2\,N!\left\{\frac{m(m+1)}{\left(\frac{N}{2}+m\right)!\left(\frac{N}{2}-m\right)!} + 2\sum_{r=m}^{\frac{N}{2}-1} \frac{(r+1)\left(\frac{N}{2}-r\right)}{\left(\frac{N}{2}+r+1\right)!\left(\frac{N}{2}-r-1\right)!}\right.$$

$$\left. - \sum_{r=m}^{N/2} \frac{r(r+1)\left(r+\frac{1}{2}\right)}{\left(\frac{N}{2}+r+1\right)!\left(\frac{N}{2}-r\right)!}\right\}$$

Therefore

$$\sum P(r)r(r+1) = \frac{N!m(m+1)}{\left(\frac{N}{2}+m\right)!\left(\frac{N}{2}-m\right)!} + 2N!\sum_{r=m}^{\frac{N}{2}-1} \frac{(r+1)\left(\frac{N}{2}-r\right)}{\left(\frac{N}{2}+r+1\right)!\left(\frac{N}{2}-r-1\right)!}$$

In a similar manner

$$2N!\sum_{r=m}^{\frac{N}{2}-1} \frac{(r+1)\left(\frac{N}{2}-r\right)}{\left(\frac{N}{2}+r+1\right)!\left(\frac{N}{2}-r-1\right)!} = \frac{N!}{\left(\frac{N}{2}+m\right)!\left(\frac{N}{2}-m-1\right)!},$$

so that

$$\overline{r(r+1)} = \frac{\sum P(r)r(r+1)}{\sum P(r)} = \frac{\left\{\frac{N!\,m(m+1)}{\left(\frac{N}{2}+m\right)!\left(\frac{N}{2}-m\right)!} + \frac{N!}{\left(\frac{N}{2}+\ m\right)!\left(\frac{N}{2}-m-1\right)!}\right\}}{\frac{N!}{\left(\frac{N}{2}+m\right)!\left(\frac{N}{2}-m\right)!}} \tag{D.16}$$

$$= m(m+1) + \left(\frac{N}{2} - m\right) = \ m^2 + \ \frac{N}{2} \quad Q.E.D.$$

Proof of (4):

In the same manner as in Part 3

$$\sum P(r)r^2(r+1)^2 = \frac{N!\,m^2(m+1)^2}{\left(\frac{N}{2}+m\right)!\left(\frac{N}{2}-m\right)!} + \frac{2(m+1)^2 N!}{\left(\frac{N}{2}+\ m\right)!\left(\frac{N}{2}-m-1\right)!} + \frac{2N!}{\left(\frac{N}{2}+\ m\right)!\left(\frac{N}{2}-m-2\right)!}, \tag{D.17}$$

So that

$$\langle r^2(r+1)^2\rangle - \ \langle r(r+1)\rangle^2$$

$$= \ m^2(m+1)^2 + \ 2(m+1)^2\left(\frac{N}{2} - \ m\right)^2 + \ 2\left(\frac{N}{2} - \ m\right)\left(\frac{N}{2} - \ m - 1\right) \tag{D.18}$$

$$-\left(m^2 - \frac{N}{2}\right)^2 = \ \frac{N^2}{4} - \ m^2 \quad Q.E.D.$$

# APPENDIX E.

## VALIDITY OF APPLYING A PERTURBATION CALCULATION TO THE DOUBLING TERMS

A quick estimate of the validity of a perturbative treatment in estimating the changes to the eigenvalues and eigenfunctions is that the value of $q_0|\kappa|$ must be less than the order of unity. The reason for this is that the doubling terms connect states with the quantum number , c, differing by 2, therefore, if $q_0|\kappa|$, the energy difference from c for the system in a state |r, c, j>, were of order 2, a perturbation treatment would probably be invalid.

Standard Second-Order Perturbation Theory[12] gives a second-order correction to the energy of the state |r,c,j> for β=0 of the form

$$W_2 = |\kappa|^2 \sum_{c',j'}{}^{'} \frac{\left|\langle r,c,j\middle|a^\dagger R_+ + aR_-\middle|r'.c',j'\rangle\right|^2}{c-|\kappa|q_{rcj} - \left(c'-|\kappa|q_{r'c'j'}\right)} \tag{E.1}$$

The prime on the summation indicates that the sum does not include the term for c'=c when j'=j. Also r=r' since $R^2$ commutes with the entire Hamiltonian (2.37). The states |r,c,j > and |r,c',j'> are expanded in terms of |n> |r,m> states in order to obtain understandable results.

$$\begin{aligned} W_2 &= |\kappa|^2 \left\{ \sum_{c',j'}{}^{'} \left| \sum_{n=\max(0,c-r)}^{c+r} \sum_{n'=\max(0,c'-r)}^{c'+r} A_n^{*r,c,j} A_{n'}^{r,c',j'} \left( \delta_{n,n'+1}\delta_{c-n,c'-n'+1}\sqrt{n'+1}\sqrt{r(r+1)-(c'-n)(c'-n+1)} \right.\right.\right. \\ &\left.\left.\left. + \delta_{n,n'-1}\delta_{c-n,c'-n'-1}\sqrt{n'}\sqrt{r(1+1)-(c'-n)(c'-n-1)} \right) \right|^2 \div \left(c-c'-|\kappa|\left[q_{rcj}-q_{rc'j'}\right]\right) \right\} \end{aligned} \tag{E.2}$$

Since the operators only connect states with c = c'+2, Eq. (E. 2) is rewritten as the sum of two terms

$$W_2 = |\kappa|^2 \left\{ \frac{\left[\sum_{j'=0}^{\min(2r,r+c+2)} \left( \left| \sum_{n'=\max(0,c-r+2)}^{r+c-2} A_{n'+1}^{*r,c,j} A_{n'}^{r,c+2,j'} \sqrt{n'+1}\sqrt{r(r+1)-(c-n'-2)(c-n'-1)} \right|^2 \right)\right]}{\left(2-|\kappa|\left[q_{r,c,j}-q_{r,c-2,j'}\right]\right)} - \frac{\left[\sum_{j''=0}^{\min(2r,r+c)} \left( \left| \sum_{n=\max(0,c-r)}^{r+c} A_{n}^{*r,c,j} A_{n+1}^{r,c+2,j'} \sqrt{n+1}\sqrt{r(r+1)-(c-n)(c-n+1)} \right|^2 \right)\right]}{\left(2-|\kappa|\left[q +_{r,c+2,j'} - q_{r,c,j}\right]\right)} \right\} \tag{E.3}$$

In order that perturbation theory be valid it is necessary that every term in the sum j'or j" be less unity, i. e.

$$|\kappa|^2 \left| \sum_{n'} A_{n'+1}^{*r,c,j} A_{n'}^{r,c+2,j'} \sqrt{n'+1}\sqrt{r(r+1)-(c-n'-2)(c-n'-1)} \right|^2 < 2-|\kappa|\left[q_{r,c,j}-q_{r,c-2,j'}\right] \tag{E.4}$$

The term on the left for j = j' may be thus easily recognized by inspection of the difference Eq. (3.4) for $\beta$ = 0 as $\frac{1}{4}|\kappa|^2 q_{r,c,j}^2$ the largest value of which is $\frac{1}{4}|\kappa|^2 q_{r,c,0}^2$. Therefore, for the calculation of $W_2$ to be valid

$$|\kappa|^2 q_{r,c,0}^2 < 8 - 4|\kappa|\left[q_{r,c,0} - q_{r,c-2,0}\right]$$

Or since

$$q_{r,c,0} \cong q_{r,c-2,0}$$

$$|\kappa| q_{r,c,0} \lesssim 2 \tag{E.5}$$

Once Eq. (E.5) is satisfied, the (E.4) is satisfied for all j' $\neq j$.

The validity of a perturbative treatment of the doubling terms for all cases is not understood well. For instance, the value of $q_o$ ranges from (c+r) $\sqrt{2r}$. to 2r$\sqrt{c}$ for c+r=$\epsilon$<<r to c>>r, respectively. From Appendix D it is noted that the average value of r(r+1) is approximately N/2 for m=0. On the other hand, if m=N/2, then r=N/2. For some laser applications, m is very near zero and slightly positive so that r $\cong \sqrt{N/2}$ while for the ammonia beam maser, nearly all the ammonia molecules entering the cavity are in the upper state. Therefore, treating the doubling terms perturbatively must be examined carefully for each case since the number of molecules involved in the system may be greater than $1/|\kappa|$.

# APPENDIX F.
# ENSEMBLE AVERAGES FOR $E^-$ AND $E^-E^+$

The ensemble average of the field operators $E^-(t)$ and $E^-E^+(t)$ may be found in the usual way[20]

$$\langle E^-(t)\rangle = -\left(\frac{\gamma}{\mu}\right)\sum_{n=0}^{\infty}(n+1)^{1/2}\langle n\left|\rho_f(t)\right|n+1\rangle, \tag{F.1a}$$

and

$$\langle E^-E^+(t)\rangle = \left|\frac{\gamma}{\mu}\right|^2\sum_{n=0}^{\infty} n\langle n\left|\rho_f(t)\right|n\rangle, \tag{F.1b}$$

where only one mode of the field is excited, $\gamma$ is the complex coupling constant, and $\mu$ the dipole moment of the TLM with which the field is interacting. The element of the field density matrix is given by the trace over the TLM states

$$\langle n\left|\rho_f(t)\right|n'\rangle = \sum_{r,m} P(r)\,\langle n|\langle r,m|\rho(t)|r,m\rangle|n'\rangle \tag{F.2}$$

where

$$P(r) = \frac{N!\,(2r+1)}{\left(\frac{N}{2}+r+1\right)!\left(\frac{N}{2}-r\right)!} \tag{F.3}$$

and $\rho(t)$ is given by a unitary transformation of the density operator at time $t_o=0$ when it is assumed that the N-TLMs and radiation field are not interacting! Therefore

$$\rho(t) = U(t)\rho(0)U^{-1}(t), \tag{F.4}$$

and

$$U(t) = e^{iHt}, \tag{F.5}$$

where H is given by Eq. (3.4) for $\beta$ = 0.

Since the system is non-interacting at time zero, the density operator is a direct product of the field part and N-TLM part of the system.

$$\rho(0) = \rho_f \; x \; \rho_T \tag{F.6}$$

Therefore, expanding Eq. (F.2) in terms of |r,c,j> states,

$$\langle n|\rho_f(t)|n'\rangle = \sum_{r,m} P(r) \sum_{r',c',j'} \sum_{r'',c'',j''} \langle n \left| \langle r,m|e^{-iHt}|r',c',j'\rangle\langle r',c',j'|\rho(0)|r'',c'',j''\rangle \langle r'',c'',j'' \left| |e^{iHt}| \right| r,m\rangle \right| n'\rangle. \tag{F.7}$$

Consider the factors in Eq. (F.7) separately

$$< n|\langle r,m|e^{-iHt}|r',c',j'\rangle = \sum_{j=0}^{\min(2r,r+c)} A_n^{r,c,j} \langle r,c,j|e^{-iHt}|r',c',j'\rangle \tag{F.8}$$

$$= A_n^{r,c,j} e^{-it(c-|\kappa|q_{rcj})} \delta_{rr'} \delta_{cc'} \delta_{jj'}$$

where c=n+m. Similarly

$$\langle r'',c'',j''|e^{iHt}|r,m\rangle|n' >= A_{n'}^{*r,c',j'} e^{it\left(c'-|\kappa|q_{rc'j'}\right)} \delta_{r''r} \delta_{c''c} \delta_{j''j} \tag{F.9}$$

where c'=n'+m.

Also

$$\langle \boldsymbol{r},\boldsymbol{c},\boldsymbol{j}|\boldsymbol{\rho}(\boldsymbol{0})|\boldsymbol{r},\boldsymbol{c}',\boldsymbol{j}'\rangle = \sum_{n''=max[0,c-r]}^{c+r} \sum_{n'''=max[0,c'-r]}^{c'+r} A_{n''}^{*r,c,j} A_{n'''}^{r,c',j'} \langle \boldsymbol{r},\boldsymbol{c}-\boldsymbol{n}''|\langle \boldsymbol{n}''|\boldsymbol{\rho}(\boldsymbol{0})|\boldsymbol{n}'''\rangle|\boldsymbol{r},\boldsymbol{c}'-\boldsymbol{n}'''\rangle \tag{F.10}$$

Therefore Eq. (F.7) becomes

$$\langle n|\rho_f(t)|n'\rangle = \sum_{r,m} P(r) \sum_{j=0}^{min[2r,c+r]} \sum_{j'=0}^{min[2r,c'+r]} \sum_{n''=max[0,c-r]}^{c+r} \sum_{n'''=max[0,c'-r]}^{c'+r} \times A_{n}^{r,c,j} A_{n'}^{*r,c',j'} A_{n''}^{*r,c,j} A_{n'''}^{r,c',j'} \times e^{-it\left(n-n'-|\kappa|\left(q_{r,c,j}-q_{r,c',j'}\right)\right)} \langle r, c-n''|\langle n''|\rho(0)|n'''\rangle|r, c'-n'''\rangle \tag{F.11}$$

where c=n+m and $c'$=$n'$+m. Applying (F.11) to Eq. (F.1), the desired ensemble averages are obtained

$$\langle \boldsymbol{E^{-}(t)}\rangle = -\left(\frac{\boldsymbol{\gamma}}{\boldsymbol{\mu}}\right) \boldsymbol{e^{it}} \sum_{n=0}^{\infty}\sum_{r} \boldsymbol{P(r)} \sum_{m=-r}^{r} \sum_{j=0}^{min[2r,c+r]} \sum_{j'=0}^{min[2r,c+1+r]} \times \sum_{n''=max[0,c-r]}^{c+r} \sum_{n'''=max[0,c+1-r]}^{c+1+r} (\boldsymbol{n}+\boldsymbol{1})^{1/2} \boldsymbol{A_{n}^{r,c,j} A_{n'}^{*r,c+1,j'} A_{n''}^{*r,c,j} A_{n'''}^{r,c+1,j'}} \times \boldsymbol{e^{it|\kappa|\left(q_{r,c,j}-q_{r,c+1,j'}\right)}} \langle \boldsymbol{r, c-n''}|\langle \boldsymbol{n''}|\boldsymbol{\rho(0)}|\boldsymbol{n'''}\rangle|\boldsymbol{r, c-n'''}\rangle \tag{F.12}$$

Where we set $c'$=c+1 and again c=n+m. Also

$$\langle E^{-}E^{+}(t)\rangle = \left|\frac{\gamma}{\mu}\right|^2 \sum_{n=0}^{\infty}\sum_{r,m} P(r) \sum_{m=-r}^{r} \sum_{j=0}^{min[2r,c+r]} \sum_{j'=0}^{min[2r,c+r]} \times \sum_{n''=max[0,c-r]}^{c+r} \sum_{n'''=max[0,c-r]}^{c+r} (n)\, A_{n}^{r,c,j} A_{n'}^{*r,c,j'} A_{n''}^{*r,c,j} A_{n'''}^{r,c,j'} \times e^{it|\kappa|\left(q_{r,c,j}-q_{r,c,j'}\right)} \langle r, c-n''|\langle n''|\rho(0)|n'''\rangle|r, c-n'''\rangle \tag{F.13}$$

where this time c'=c.

As is seen, these equations are quite complicated and to use the general expressions would be prohibitive except for the simplest of cases. Both Eqs.

(F.12) and (F.13) may, of course, be rewritten in terms of sin's and cos's by taking advantage of the symmetry about "c" in the sums over j and j'. It is noted that some simplification is possible by retaining only terms which have the "slowest" time dependence. This is possible because of the orthogonality between states of different j and/or n subscripts.

For the particular case of no photons initially in the field and all of the TLMs in the excited state, the equations for $\langle E^- \rangle$ and $\langle E^- E^+ \rangle$ have a considerably simpler form. Since

$$\langle n|\langle r, c-n|\rho(0)|r, c'-n'\rangle|n'\rangle = \delta_{n',0}\delta_{c,r}\delta_{c,c'}\delta_{n',n} \tag{F.14}$$

the average of E(t) is equal to zero and

$$\langle E^- E^+(t)\rangle = \left|\frac{\gamma}{\mu}\right|^2 \sum_r P(r) \sum_{m=-r}^{r} \sum_{j,j'=0}^{2r} A_{r-m}^{r,r,j} A_{r-m}^{*r,r,j'} A_0^{*r,r,j} A_0^{r,r,j'} e^{it|\kappa|(q_{r,r,j}-q_{r,r,j'})}. \tag{F.15}$$